\def\frac#1#2{{{{#1}}\over{{#2}}}}
\documentclass[12pt]{article}

\usepackage{todonotes}
\usepackage{amsmath}
\usepackage{wasysym}    
\usepackage{textcomp}
\usepackage{pict2e}
 \usepackage[absolute]{textpos} 
 \usepackage{caption}

\usepackage{blindtext}

\usepackage{cite}
\usepackage{todonotes}

\usepackage{xcolor, cancel}
\usepackage[normalem]{ulem}
\usepackage{graphicx}
\usepackage{pdfpages}
\usepackage{cite}

\newsavebox{\ns}
\newsavebox{\dbrane}
\newsavebox{\dbshort}

\usepackage{epsfig}
\usepackage{amssymb}

\newcommand\ba{\begin{eqnarray}}
\newcommand\ea{\end{eqnarray}}
\newcommand{\bbb}{\begin{eqnarray}\begin{array}{c}}
\newcommand{\bbl}[1]{\ba\label{#1}\begin{array}{c}{c}}
\newcommand{\eee}{\end{array}\end{eqnarray}}
\def\xxx{\eee\bbb}

\def\een#1{\label{#1} \eee}
\def\xxn#1{\een{#1}  \bbb}

\def\eenn{\nonumber\eee}

\definecolor{DarkGreen}{rgb}{0,.64,0}
\definecolor{gunmetal}{rgb}{0.171875, 0.207031, 0.222656}
\definecolor{chartreuse}{rgb}{.49,.98,0}
\definecolor{amethyst}{rgb}{0.59375,0.398438,0.792969}
\definecolor{brownrust}{rgb}{0.6875, 0.316406, 0.242188}
\definecolor{Violet}{rgb}{0.5,0,1}
\definecolor{BurntOrange}{rgb}{0.792969,0.332031,0}
\definecolor{FreshEggplant}{rgb}{0.59375, 0., 0.414063}
\definecolor{salmon}{rgb}{0.996094,0.507813,0.410156}
 \definecolor{FrenchRose}{rgb}{0.96875, 0.292969, 0.5625}
\definecolor{Cabaret}{rgb}{0.808594, 0.242188, 0.46875}
\definecolor{Shamrock}{rgb}{0.242188, 0.808594, 0.582031}
\definecolor{RobinsEggBlue}{rgb}{0., 0.792969, 0.792969}
\definecolor{GuardsmanRed}{rgb}{0.792969, 0., 0.}
\definecolor{Sapphire}{rgb}{0.183594, 0.328125, 0.621094}
\definecolor{Sorbus}{rgb}{0.996094, 0.429688, 0.0273438}
\definecolor{Red}{rgb}{1,0,0}
\definecolor{Blue}{rgb}{0,0,1}
\definecolor{Black}{rgb}{0,0,0}
\definecolor{Green}{rgb}{0,1,0}
\definecolor{thistle3}{rgb}{0.800781, 0.707031, 0.800781}
\definecolor{thistle4}{rgb}{0.542969, 0.480469, 0.542969}
\definecolor{DarkTurquoise}{RGB}{0,206,209}
\definecolor{turquoise4}{RGB}{0,134,139}

\definecolor{Purple}{rgb}{0.808594, 0.242188, 0.46875}
\newcommand{\purple}[1]{{\color{Purple} {#1}}}

\def\emm#1{{\it #1}}

\newcommand{\TENSOREDIT}[1]{{}}

\def\SHCommentColor{Sorbus}

\newcommand{\shg}[1]{{\color{\SHCommentColor} \it NOTE: {#1} -- SH ~~}}
\newcommand{\sshg}[1]{}

\newcommand{\sisg}[1]{}

\def\Dslash{\,\,{\raise.15ex\hbox{/}\mkern-12mu D}}
\def\Dbarslash{\,\,{\raise.15ex\hbox{/}\mkern-12mu {\bar D}}}
\def\delslash{\,\,{\raise.15ex\hbox{/}\mkern-9mu \partial}}
\def\delbarslash{\,\,{\raise.15ex\hbox{/}\mkern-9mu {\bar\partial}}}
\def\pslash{\,\,{\raise.15ex\hbox{/}\mkern-9mu p}}
\def\calDslash{\,\,{\raise.15ex\hbox{/}\mkern-12mu {\cal D}}}

\newcommand{\hh}{{1\over 2}}

\renewcommand{\ll}{_}
\newcommand{\uu}{^}
\newcommand{\pp}{\partial}

\renewcommand{\exp}[1]{{\rm exp}\{#1\}}

\usepackage{hyperref}

\renewcommand{\d}{\delta}
\newcommand{\m}{\mu}

\renewcommand{\dag}{{}^\dagger{}}

\renewcommand{\m}{\mu}
\newcommand{\n}{\nu}
\newcommand{\s}{\sigma}
\renewcommand{\t}{\tau}
\newcommand{\G}{\Gamma}
\newcommand{\g}{\gamma}
\renewcommand{\a}{\alpha}
\renewcommand{\r}{\rho}

\renewcommand{\o}{\omega}
\newcommand{\e}{\epsilon}
\renewcommand{\O}{\Omega}

\newcommand{\sqd}{^2}

\renewcommand{\hh}{{1\over 2}}
\renewcommand{\gg}{\nabla}

\renewcommand{\th}{\theta}
\renewcommand{\t}{\tau}

\def\D{\Delta}

\newcommand{\llsk}{\hskip .5in}

\newcommand{\IZ}{\relax\ifmmode\mathchoice
{\hbox{\cmss Z\kern-.4em Z}}{\hbox{\cmss Z\kern-.4em Z}}
{\lower.9pt\hbox{\cmsss Z\kern-.4em Z}} {\lower1.2pt\hbox{\cmsss
Z\kern-.4em Z}}\else{\cmss Z\kern-.4em Z}\fi} \font\cmss=cmss10
\font\cmsss=cmss10 at 7pt
\newcommand{\inbar}{\,\vrule height1.5ex width.4pt depth0pt}
\newcommand{\IC}{{\relax\hbox{$\inbar\kern-.3em{\rm C}$}}}
\newcommand{\IQ}{{\relax\hbox{$\inbar\kern-.3em{\rm Q}$}}}
\newcommand{\IP}{\relax{\rm I\kern-.18em P}}

\renewcommand{\k}[1]{{k_{#1}}}

\newcommand{\ed}{\dot{e}}

\renewcommand{\k}{\kappa}
\renewcommand{\l}{\lambda}

\newcommand{\cc}{{\cal I}}

\renewcommand{\cc}{{c_1}}

\renewcommand{\o}{\omega}

\renewcommand{\cc}{c}

\newcommand{\pr}{^\prime}

\newcommand{\cl}{{\cal L}}

\newcommand{\IR}{\relax{\rm I\kern-.18em R}}
\def\blfootnote{\xdef\@thefnmark{}\@footnotetext}

\renewcommand{\cc}[1]{\cite{#1}}

\newcommand{\ee}[1]{\ba {#1} \ea}

\newcommand{\upp }[1]{^{({#1})}{}}

\newcommand{\co}{{\cal O}}

\newcommand{\rr}[1]{(\ref{{#1}})}

\newcommand{\prpr}{^{\prime\prime}}

\newcommand{\heading}[1]{\begin{center}\it \blue{#1} \rm \end{center}}

\def\uth{^{\rm{\underline{th}}}}

\newcommand{\us}[2]{^{({#1}{#2})}}

\def\BeginItemize{\begin{itemize}}
\def\ei{\end{itemize}}

\def\qb{\bar{q}}

\def\ed{\end{document}}

\def\cc{{\cal I}}

\renewcommand{\rr}[1]{(\ref{#1})}

\def\cc{\,}

\def\r{\rho}

\def\uv{ultraviolet~}

\def\cc{\,}

\newcommand{\lrm}[1]{_{{\rm {#1}}}}
\newcommand{\urm}[1]{^{{\rm {#1}}}}
\newcommand{\uprm}[1]{^{({\rm {#1}})}}

\def\sb{{\bar{s}}}

\newcommand{\ls}[1]{_{[{#1}]}}

\usepackage{fancyhdr}
\renewcommand{\eqref}[1]{\rr{#1}}
\usepackage{dsfont}

\def\outt#1{}

\newcommand{\redd}[1]{{\color{Red} {#1}}}

\definecolor{Purple}{rgb}{0.808594, 0.242188, 0.46875}

\def\bll#1{{\blue{{#1}}}}
\def\bbl#1{\bll{#1}}

\def\tb{\bar{t}}

\newcommand{\aaa}[1]{}

\def\be{\begin{eqnarray}}
\def\ee{\end{eqnarray}}

\newcommand{\WhiteOutWithNotification}[1]{ {  \color{Red}  (  A SECTION HAS BEEN WHITED OUT HERE. \cc   )  } }
\newcommand{\WhiteOut}[1]{}
\newcommand{\RoundIEdit}[1]{}
\newcommand{\RoundHEdit}[1]{}
\newcommand{\RoundGEdit}[1]{}
\newcommand{\RoundFEdit}[1]{}
\newcommand{\RoundEEdit}[1]{}
\newcommand{\EMVertEdit}[1]{}
\newcommand{\HugeEditB}[1]{}
\newcommand{\HugeEditA}[1]{}
\newcommand{\RoundDEdit}[1]{}
\newcommand{\RoundCEdit}[1]{}
\newcommand{\RoundBEdit}[1]{}
\newcommand{\RoundAEdit}[1]{}

\newcommand{\blue}[1]{{\color{Blue}{#1}}}

\def\redlowdash{{\color{Red}{\rule[-0.5ex]{2pt}{0.4pt}}}}
\def\redmiddash{{\color{Red}{\rule[+0.5ex]{2pt}{0.4pt}}}}

\def\cute{{\lower3.5pt\hbox{\sixly
  \kern-.21pt \char58 \kern-.21pt }}}

\def\midcute{{\lower-1.0pt\hbox{\sixly
  \kern-.21pt \char58 \kern-.21pt }}}

  \def\lowcute{{\lower3.5pt\hbox{\sixly
  \kern-.21pt \char58 \kern-.21pt }}}
  
  \def\redmidcute{{\color{Red} \midcute}}
  
  \def\redlowcute{{\color{Red} \lowcute}}
   
    \def\bluelowcute{{\color{Blue} \lowcute}}

   \def\swave{\bgroup \markoverwith \midcute \ULon} 
  
  \def\redswave{\bgroup \markoverwith \redmidcute \ULon} 
  
  \def\reduline{\bgroup \markoverwith \redlowdash \ULon}
  
   \def\blueuline{\bgroup \markoverwith \bluelowdash \ULon}
  
   \def\reduwave{\bgroup \markoverwith \redlowcute \ULon}
   
   \def\blueuwave{\bgroup \markoverwith \bluelowcute \ULon}
  
  \def\redsout{\bgroup \markoverwith \redmiddash \ULon}
  
   \def\bluesout{\bgroup \markoverwith \bluemiddash \ULon}
   
   \def\Irrel{\bgroup \markoverwith {{\color{Red} {\bf X}}} \ULon}

\newcommand{\eqirrel}[1]{\rdots}

\let\oldcancel\cancel
\renewcommand\cancel[1][black]{%
  \def\CancelColor{\color{#1}}%
  \oldcancel}

   \let\oldbcancel\bcancel
\renewcommand\bcancel[1][black]{%
  \def\CancelColor{\color{#1}}%
  \oldbcancel}

\def\bca{\begin{cases}}
\def\eca{\end{cases}}

\def\rdots{\redd{\circ\circ\circ}}

\def\dag{^\dagger}

\renewcommand{\upp }[1]{^{({#1})}}

\def\lsim{\mathrel{\lower0.3em\hbox{$\stackrel{\textstyle <}{\sim}$}}}
\def\gsim{\mathrel{\lower0.3em\hbox{$\stackrel{\textstyle >}{\sim}$}}}

\def\negspace{\kern -0.4em}

\def\dvec{\raise 0.3 em \hbox{$^\leftrightarrow$} \kern -0.77 em}

\def\omegahat{\hat%
	{\setbox0=\hbox{$\omega$}%
		\kern-.025em\copy0\kern-\wd0
		\kern.05em\copy0\kern-\wd0
		\kern-.025em\raise.0433em\box0}}

\def\pol#1{}

\def\us#1^{[{#1}]}

{\color{Purple}}%
{\color{Black}}
{\color{Blue}}%
{\color{Black}}
{\color{Red}}%
{\color{Black}}

\def\blue#1{{\color{Blue}{#1}}}

{\color{Red}}%
{\color{Black}}

\def\bbsk{\hskip-.5in}

\usepackage{mathtools}

\def\tb{{\bar{\tau}}}

\def\JJM{\purple{{\cal J}}}

\makeatletter
\newcommand*{\Relbarfill@}{\arrowfill@\Relbar\Relbar\Relbar}
\newcommand*{\xeq}[2][]{\ext@arrow 0055\Relbarfill@{#1}{#2}}
\makeatother

\makeatletter
\newcommand*{\rxeq}[2][]{{\color{Red} \ext@arrow 0055\Relbarfill@{#1}{#2}}}
\makeatother

\def\sb{{\bar{\sigma}}}

\def\ordinary#1{#1}

\def\xxnn{\nonumber\xxx}

\def\frac#1#2{{ {#1}\over{#2}}}

\def\G{{\cal G}}

\def\cl{{\,\rm cl}}
\def\lambdabar{\bar\lambda}

\def\psibar{\bar\psi}

\def\new{{\scriptscriptstyle\rm new}}

\def\uX{\,\lower 1.2ex\hbox{$\sim$}\mkern-13.5mu X}
\def\uD{\,\lower 1.2ex\hbox{$\sim$}\mkern-13.5mu {\rm D}}

\def\uF{\,\lower 1.2ex\hbox{$\sim$}\mkern-13.5mu F}
\def\uW{\,\lower 1.2ex\hbox{$\sim$}\mkern-13.5mu W}
\def\uWbar{\,\lower 1.2ex\hbox{$\sim$}\mkern-13.5mu {\overline W}}
\def\uPhibar{\,\lower 1.2ex\hbox{$\sim$}\mkern-13.5mu {\overline \Phi}}

\def\uV{\,\lower 1.2ex\hbox{$\sim$}\mkern-13.5mu V}
\def\uv{\,\lower 1.0ex\hbox{$\scriptstyle\sim$}\mkern-11.0mu v}
\def\uPsi{\,\lower 1.2ex\hbox{$\sim$}\mkern-13.5mu \Psi}
\def\uPhi{\,\lower 1.2ex\hbox{$\sim$}\mkern-13.5mu \Phi}
\def\uchi{\,\lower 1.5ex\hbox{$\sim$}\mkern-13.5mu \chi}
\def\Psibar{\bar\Psi}
\def\uPsibar{\,\lower 1.2ex\hbox{$\sim$}\mkern-13.5mu \Psibar}
\def\upsi{\,\lower 1.5ex\hbox{$\sim$}\mkern-13.5mu \psi}
\def\psibar{\bar\psi}
\def\upsibar{\,\lower 1.5ex\hbox{$\sim$}\mkern-13.5mu \psibar}
\def\upsibarzero{\,\lower 1.5ex\hbox{$\sim$}\mkern-13.5mu \psibar^\zero}
\def\ulambda{\,\lower 1.2ex\hbox{$\sim$}\mkern-13.5mu \lambda}
\def\ulambdabar{\,\lower 1.2ex\hbox{$\sim$}\mkern-13.5mu \lambdabar}
\def\ulambdabarzero{\,\lower 1.2ex\hbox{$\sim$}\mkern-13.5mu \lambdabar^\zero}
\def\ulambdabarnew{\,\lower 1.2ex\hbox{$\sim$}\mkern-13.5mu \lambdabar^\new}

\def\Dslash{\,\,{\raise.15ex\hbox{/}\mkern-12mu \D}}
\def\Dbarslash{\,\,{\raise.15ex\hbox{/}\mkern-12mu {\bar\D}}}
\def\delslash{\,\,{\raise.15ex\hbox{/}\mkern-9mu \partial}}
\def\delbarslash{\,\,{\raise.15ex\hbox{/}\mkern-9mu {\bar\partial}}}

\def\uAcl{\,\lower 1.2ex\hbox{$\sim$}\mkern-13.5mu A^{}_{\cl}}
\def\uAbarcl{\,\lower 1.2ex\hbox{$\sim$}\mkern-13.5mu A_{\cl}^\dagger}

\def\uA{\,\lower 1.2ex\hbox{$\sim$}\mkern-13.5mu A}

\def\zero{{\scriptscriptstyle(0)}}

\def\bii{\begin{itemize}}

\def\cob{\overline{{\cal O}}}
\def\G{\Gamma}

\def\bbd{\bdol}
\def\bbd#1{${#1}$}

\usepackage{setspace}

\def\ddsc{\purple{\lambda}}
\def\DDSL{{\rm lim}_{ {{n\to\infty}\atop{\ddsc{\rm ~fixed}}}}}

\def\AlienEquation#1{\blue{(#1)}}
\def\aleq#1{\AlienEquation{#1}}
\def\AlienSection#1{\blue{#1}}
\def\alsec#1{\AlienSection{#1}}
\def\alfoo#1{\blue{#1}}
\def\WarningOut#1{}

\def\ModToAnharbridge{\purple{{\cal B}}}
\usepackage{slashbox}
\def\omit#1{}

\def\MacroW{\mathfrak{W}}
\def\mgkt{{\hbox{ref.~\cite{Grassi:2019txd}}}}
\def\smgkt{\hbox{\tiny{ref.~\cite{Grassi:2019txd}}}}
\def\movie{{\hbox{ref.~\cite{Bourget:2018obm}}}}
\def\smovie{\hbox{\scriptsize{ref.~\cite{Bourget:2018obm}}}}
\def\PREFAK{\mathfrak{P}[n,\t]}
\begin{document}

\begin{titlepage}
\begin{flushright}
\end{flushright}
\vspace{8 mm}
\begin{center}
  {\large \bf %
  Large R-charge EFT correlators in \(\mathcal{N} = 2\) SQCD
  }
\end{center}
\vspace{2 mm}
\begin{center}
{Simeon Hellerman$^1$ and Domenico Orlando$^{2,3}$%
}\\
\vspace{6mm}
{\it $^1$Kavli Institute for the Physics and Mathematics of the Universe \textsc{(wpi)}\\
The University of Tokyo \\
 Kashiwa, Chiba  277-8582, Japan\\
}
{$^1$simeon.hellerman.1@gmail.com\\}
\vspace{6mm}
{\it $^2$INFN sezione di Torino\\
  via Pietro Giuria 1, 10125 Torino\\}
\vspace{6mm}
{\it $^3$Albert Einstein Center for Fundamental Physics\\
  Institute for Theoretical Physics, University of Bern,\\
  Sidlerstrasse 5, CH-3012 Bern, Switzerland\\}
\end{center}
\vspace{-4 mm}
\begin{center}
{\large Abstract}
\end{center}
\noindent

We consider large-R-charge
Coulomb branch correlation functions 
in \bbd{{\cal N} = 2} superconformal QCD in D=4 dimensions, with gauge group \bbd{G=SU(2)} and
\bbd{N\lrm F = 4} hypermultiplets in the fundamental representation.  Using information from supersymmetric recursion relations, \bbd{S-}duality, and matching of EFT parameters with the double-scaling limit, we give an
exact formula for the massless Coulomb branch EFT contribution to the correlation function two-point
functions of the \bbd{n\uth} power of the chiral ring
generator, \bbd{G\ll{2n} \uprm{EFT}= 
{{2\uu{+4n}}\over{Z\ll{S\uu 4}[\t]}} \cc \G(2n + {5\over 2})\cc e\uu{A[\t] n + B[\t]},} with \bbd{A[\t]} and \bbd{B[\t]} given as explicit functions
of the coupling constant \bbd{\t} in closed form.  We note
the precise agreement of the EFT formula
with supersymmetric localization even at low values of \bbd{n}, and
discuss aspects of the post-EFT remainder contributed
by the macroscopic virtual propagation of massive particles.

\vspace{1cm}
\begin{flushleft}
\today
\end{flushleft}
\end{titlepage}

\tableofcontents
\newpage
\numberwithin{equation}{section}

\section{Introduction}\label{Introduction}

In previous work~\cite{Hellerman:2017sur, Hellerman:2018xpi, Hellerman:2020sqj} we have considered \bbd{{\cal N} = 2} superconformal field theories in \bbd{D=4} spacetime dimensions with one-dimensional Coulomb branch and whose low-energy degrees of freedom contain only a single \bbd{U(1)} vector multiplet and no charged hypermultiplets.
The chiral ring of the Coulomb branch is generated by a single BPS scalar primary operator \bbd{\co\ll\D,} of conformal dimension \bbd{\D} and 
\bbd{U(1)\lrm R} R-charge\footnote{We take the slightly nonstandard normalization for the \bbd{U(1)\lrm R}-charge that the supercharges \bbd{Q\ll\a\uu i} have
  \bbd{R-}charge \bbd{-\hh} and the free scalar component of a free vector multiplet has R-charge \bbd{+1}} \bbd{\D}.
One can study two-point functions of powers \bbd{\co\ll\D\uu n}
of the generator, and use analytic methods to estimate the behavior of the two-point function when the total R-charge \bbd{\JJM = n\D} of the operator insertion \bbd{\co\ll \D\uu n} is
large.
In~\cite{Hellerman:2017sur, Hellerman:2018xpi, Hellerman:2020sqj} we
used a supersymmetric version of the previously developed large-charge expansion\footnote{The large-charge expansion should itself
be thought of as a special case of a more general subject of 
the simplification of strongly-coupled quantum theories at
large quantum number, including large spin~\cite{Berenstein:2003gb, Alday:2007qf, Alday:2007mf, Alday:2013cwa, Hellerman:2013kba, Alday:2015eya, Fitzpatrick:2012yx, Komargodski:2012ek, Caron-Huot:2016icg} and more general high-energy~\cite{Son:1995wz, Srednicki, Deutsch, Cardy:1986ie, Hartman:2014oaa, Delacretaz:2020nit, Mukhametzhanov:2020swe} and
high-particle-number~\cite{Bachas:1991fd, Libanov:1994ug, Son:1995wz, Jaeckel:2018ipq, Monin:2018cbi, Khoze:2018mey, Dine:2020ybn} limits.}
~\cite{Hellerman:2015nra, Monin:2016jmo, Cuomo:2017vzg, Sharon:2020mjs, Gaume:2020bmp, Cuomo:2020rgt, Orlando:2019hte, Cuomo:2019ejv, Orlando:2020yii, Alvarez-Gaume:2016vff, Hellerman:2017veg, Hellerman:2017efx, Hellerman:2018sjf}
(see~\cite{Jafferis:2017zna, Alvarez-Gaume:2019biu, Kumar:2018nkf, Watanabe:2019adh,
Arias-Tamargo:2019xld, Arias-Tamargo:2019kfr, Arias-Tamargo:2020fow, Badel:2019oxl, Badel:2019khk, Giombi:2020enj, Antipin:2021akb, Cuomo:2021qws, Komargodski:2021zzy, Cuomo:2021ygt, Kravec:2018qnu, Favrod:2018xov, Kravec:2019djc, Orlando:2020idm, Hellerman:2020eff, Jack:2021ypd} for more recent developments) to determine the leading behavior of the correlation function at large \bbd{n,} and then used supersymmetric recursion relations~\cite{Papadodimas:2009eu, Baggio:2014sna, Baggio:2014ioa} to derive the higher \bbd{{1\over n}} corrections from the leading terms.  All this analysis uses only large \bbd{\JJM} as a control parameter; the analysis does
not require any weak-coupling limit of a marginal parameter of the CFT, or for that matter any marginal parameter at all. The large \bbd{R-}charge methods work equally
well for superconformal theories with no marginal coupling ({\it e.g.} the Argyres-Douglas theories~\cite{Argyres:1995jj, Argyres:1995xn, Xie:2012hs}) as for those with.

To describe the large-R-charge expansion
of the chiral primary correlation functions, write the correlation function as
\bbb
G\ll{2n} \equiv |x-y|\uu{2n\D}\cc \langle \co\ll\D\uu n(x) \overline{\co}\ll\D\uu n(y) \rangle
= {{2\uu{+4n}}\over{Z\ll{S\uu 4}}}\cc Z\ll n = {{2\uu{+ 4n}}\over{Z\ll{S\uu 4}}}\cc \exp{q\ll n} \ ,
\een{ZLowerNDefinition}
where \bbd{Z\ll{S\uu 4}} is the four-sphere partition function, and
\bbd{Z\ll n = \exp{q\ll n}} is the sphere partition function with \bbd{\co\ll \D\uu n} and its \bbd{\cob\ll\D\uu n} inserted at the two poles.  The result of the analysis of~\cite{Hellerman:2018xpi} is that
\bbd{q\ll n = {\rm Log}[Z\ll n]} has the decomposition
\bbb
q\ll n = q\ll n\uprm{EFT} + q\ll n\uprm{mmp}\ ,
\xxnn
q\ll n\uprm{EFT} =A n+ B +  {\rm log} \bigg [ \cc \G(\JJM + \a + 1) \cc \bigg ]\ , \llsk \JJM \equiv n\D\ ,
\xxnn
q\ll n\uprm{mmp} = [{\rm smaller~than~any~power~of~}\JJM]\ ,
\een{EFTPrediction}
where \bbd{A} and \bbd{B} are nonuniversal coefficients that may depend on the normalization of \bbd{\co\ll\D}, the renormalization scheme of the CFT path integral, and
the marginal parameters of the theory if any; \bbd{\a} is an anomaly coefficient describing the difference between the Weyl \bbd{a-}anomaly of the underlying CFT and
that of the EFT of an abelian vector multiplet and possibly neutral hypers; and \bbd{q\ll n\uprm{mmp}} is a set of corrections from degrees of freedom outside
the massless EFT,
describing macroscopic propagation of massive particles.  This last contribution, which we call the {\it massive macroscopic propagation \rm} (\textsc{mmp}) terms, 
is scheme-independent and independent of the normalization of the chiral ring generator, but may vary from theory to theory and may depend on the marginal parameters of the theory of any.

The \bbd{\a-}coefficient describing the anomaly mismatch, can be quantified as
\bbb
\a = 2 (a\uprm{CFT} - a\uprm{EFT})\lrm{units~of~\hbox{\cite{Anselmi:1997ys}}}\ , 
\eee
where the units expressing the \bbd{a-}anomaly are those given in~\cite{Anselmi:1997ys}.  This is a commonly used convention in which a \bbd{U(1)} vector multiplet has
\bbd{a-}anomaly \bbd{{5\over{24}}} and a massless hypermultiplet has \bbd{a-}anomaly \bbd{{1\over {24}}}.  Convention-independently, the formula for \bbd{\a} is

\bbb
\a = {5\over{12}}\cc \times {{ a\uprm{CFT} - a\uprm{EFT}}\over{a\uprm{U(1)~vector~multiplet}}}
\eee
A table of the values of \bbd{\a} for all known rank-one \bbd{{\cal N} = 2} superconformal theories, is given in the Appendix of~\cite{Hellerman:2017sur}.

Though our formula applies equally to theories with and without marginal couplings, a theory with a marginal coupling gives us the opportunity for explicit computations to
test our predictions and check the rapidity of convergence of the large-\bbd{\JJM} expansion to the exact answer.  The technique of supersymmetric localization
\cite{Nekrasov:2002qd, Pestun:2007rz} allows the computation of the four-sphere partition function, from which correlation functions can be calculated using the prescription
of~\cite{Gerchkovitz:2016gxx}. For instance the case of \bbd{{\cal N} = 4} super-Yang-Mills with gauge group \bbd{G = SU(2)} exactly obeys our asymptotic expansion, with correlators
\bbd{G\ll{2n}\upp{{\cal N} = 4} =  [{\rm Im}(\t)]\uu{-2n}\cc (2n+1)!} and \bbd{a-}anomaly mismatch coefficient \bbd{\a = 1}.

In the case of \bbd{{\cal N} = 2} superconformal QCD, with \bbd{G = SU(2)} and \bbd{N\lrm F = 4} hypermultiplets in the fundamental representation, the sphere
partition function \bbd{Z\ll{S\uu 4}[\t,\tb]} has a complicated functional form for the dependence on the marginal coupling \bbd{\t}, and there is no known
closed form for the correlation functions \bbd{G\ll{2n}[\t,\tb]}.  However one can compute correlation functions numerically using the construction of~\cite{Gerchkovitz:2016gxx},
and test the prediction \rr{EFTPrediction} against the numerical data.  This was done in~\cite{Hellerman:2018xpi, Hellerman:2020sqj}, with a confirmation of the prediction \rr{EFTPrediction} to
high precision.  Specifically, in~\cite{Hellerman:2018xpi} the authors compared the exact correlation functions with the EFT factor of the prediction for Coulomb-branch correlators in SQCD (with anomaly mismatch coefficient \bbd{\a = {3\over 2}}), with agreement sufficiently
precise that the nonperturbatively small correction \bbd{q\ll n\uprm{mmp}} is distinctly visible in the difference between the exact function \bbd{q\ll n} and
the massless EFT contribution \bbd{q\ll n\uprm{EFT}.}  The nonperturbatively small correction \bbd{q\ll n\uprm{mmp}} is expected to come from the breakdown of the
massless EFT description due to the macroscopic virtual propagation of massive BPS particles of mass \bbd{m} along a distance \bbd{\ell}, giving effects of size \bbd{(m\ell)\uprm{power}\cc \exp{- m\ell}.}  The numerical data for \bbd{q\ll n\uprm{mmp}} is in excellent agreement with such a functional form, with an exponent
\bbd{ -m\ell \propto -\sqrt{{n}\over{{\rm Im[\t]}}}} corresponding to the action \bbd{m\ell} for a massive hypermultiplet of mass \bbd{m} traversing the great circle of the \bbd{S\uu 3} time-slice of the cylinder, with length \bbd{\ell = 2\pi R} 
in the conformal frame describing radial quantization with the operator insertions at \bbd{\t = \pm\infty}.  In this conformal frame the magnitude of the vector multiplet scalar
is constant and the hypermultiplet receives an effective mass through its coupling to the vector multiplet, equal to \bbd{ m\lrm{hyper}\simeq {1\over R}\cc \sqrt{{\JJM\over{2\pi\cc {\rm Im}[\t]}}}  = {1\over R}\cc \sqrt{{n\over{\pi\cc {\rm Im}[\t]}}}} where \bbd{R} is the radius of the sphere.

In~\cite{Hellerman:2018xpi} the precision of comparison between theory and data was limited by the lack of knowledge of the theory-dependent functions of the coupling, \bbd{A} and \bbd{B}
in \bbd{q\ll n\uprm{EFT}}, forcing the authors to consider double-differences \bbd{q\ll{n+1} - 2 q\ll n + q\ll{n-1},} in which the theory-dependent terms \bbd{A n + B} drop out.  In
\cite{Hellerman:2020sqj} the authors solved for the full functional form \bbd{A[\t,\bar{\t}]} of the A-coefficient, using the S-duality~\cite{Gaiotto:2009we, Alday:2009aq, Grimm:2007tm, Dorey:1996bn} of superconformal SQCD (see also\cite{Billo:2013fi}) together with the supersymmetric recursion relations~\cite{Papadodimas:2009eu, Baggio:2014sna, Baggio:2014ioa} as theoretical inputs sufficient to determine \bbd{A} completely
up to a single \bbd{\t-}independent coefficient, which was matched to perturbation theory using the insights of~\cite{Bourget:2018obm} about the relationship between the large-\bbd{\JJM}
and weak-coupling limits.  This allowed the authors of~\cite{Hellerman:2020sqj} to make a theoretical prediction for the \emm{single} differences \bbd{q\ll{n+1} - q\ll n,} in which
only the \bbd{B-}coefficient drops out.  In the present work we will use the recursion relations and \bbd{S-}duality of \bbd{{\cal N} = 2} superconformal
SQCD again, to solve for the full \bbd{\t-}dependent
functional form of the \bbd{B-}coefficient, up to an overall constant factor; and then we shall
match the constant term in \bbd{B} with a semi-classical calculation in double-scaled perturbation theory~\cite{Grassi:2019txd} (which extends earlier work~\cite{Bourget:2018obm} to the
regime of strong double-scaled coupling).

The result will be as follows, written in terms of the coupling-dependent but scheme-independent combination \bbd{\tilde{B} \equiv B - {\rm log}[Z\ll{S\uu 4}]}.
 The exponentiated coefficient \bbd{\tilde{B},} whose determination is the main calculation of
this paper, is given by
\bbb
\exp{\tilde{B}}  = \g\lrm G\uu{+12}\cc e\uu{-1}\cc 2\uu{-{9\over 2}}\pi\uu{-{3\over 2}}\cc
{{ |\l(\s)|\uu{+{2\over 3}}\cc |1-\l(\s)|\uu{+ {8\over 3}}}\over{|\eta(\s)|\uu 8\cc [{\rm Im}(\s)]\sqd}}\cc \bigg [ \cc Z\lrm{{Pestun-}\atop{Nekrasov}}\cc \bigg ]\uu{-1}
\een{FinalResultSummaryEqFPrecap}
where \bbd{\l(\s)} is the modular lambda function, \bbd{\eta(\s)} is the Dedekind eta function, and \bbd{\g\lrm G} is the Glaisher constant \bbd{\g\lrm G \simeq 1.2824271291}.
The last factor is the reciprocal of the partition function \bbd{Z\lrm{{Pestun-}\atop{Nekrasov}}} as computed in the Pestun-Nekrasov scheme, by which we mean
as computed by localization using Pestun's one-loop determinant~\cite{Pestun:2007rz} and Nekrasov's \bbd{U(2)} (as opposed to \bbd{SU(2)}) instanton
factor~\cite{Nekrasov:2002qd}. 

The plan of the paper is as follows. In sec.~\ref{AAndBFunctions} we review the properties of the coupling-dependent functions \bbd{A[\t]} and \bbd{B[\t]}, and their 
transformation properties under scheme changes corresponding to holomorphic coupling reparametrization.  In sec.~\ref{SchemeDependenceSection} we review the properties of the \bbd{B[\t]}-function under shifts of the Euler-density counterterm, which acts on the log of the partition function as K\"ahler transformations of the K\"ahler potential of the Zamolodchikov 
metric.  In sec.~\ref{SolutionForBFunctionUpToOverallCoefficient}, we use these properties to determine the \bbd{\t-}dependence of the \bbd{B[\t]}-function (in any given scheme) up to an overall coupling-independent additive constant \bbd{{\bf C}}, which remains a free parameter in the EFT, even taking duality symmetries into account.  In sec.~\ref{DerivationOfConstantCoefficientBoldfaceC} we use EFT inputs for the behavior of the exponentially small
correction together with the recently derived solution~\cite{Grassi:2019txd} of the correlation functions in the double-scaling limit of~\cite{Bourget:2018obm}, to fix the final coupling-independent coefficient \bbd{{\bf C}}.
As a byproduct of this analysis we find a simple relationship between the "worldline instanton" piece
\bbd{F\uprm{inst}} of the double-scaled correlator in~\cite{Grassi:2019txd} on the one hand, and the non-EFT contribution to the correlation function on the other hand, which
we call the macroscopic massive propagation (\textsc{mmp}) function: We
show that the former is simply the double-scaling limit of the latter.
Having understood this relationship, we then go on in sec.~\ref{GeneralMMPSection} to discuss the expected behavior for the \textsc{mmp} function on  \textsc{eft} grounds, both
in the double-scaling limit and in the fixed-\bbd{\t} limit at large \bbd{\JJM}.  In particular, we calculate the expected \textsc{eft} value for the exponential in the
exponent of the function \bbd{F\uprm{inst}} of~\cite{Grassi:2019txd} at strong double-scaled coupling \bbd{\ddsc,} and we show that it matches the expected exponent
corresponding to a worldline instanton of a conformal hypermultiplet in the background Coulomb branch expectation value created by operator insertions of \bbd{R-}charge
\bbd{\JJM =2n}.  In sec. \ref{Comparison} we compare the completed EFT approximation with numerical results from localization by way of the
algorithm described in~\cite{Gerchkovitz:2016gxx}.  In sec~\ref{ConclusionsSection} we discuss our results and draw conclusions.
We also include several Appendices in which we give a comprehensive list of conventions used in this paper and in the closely adjacent literature (in sec.~\ref{ConventionAppendix}); and also weak-coupling
expansions of various quantities appearing in the paper (in sec.~\ref{WeakCouplingExpansions}), as well as a discussion of transformation laws of coupling-dependent functions
used in the paper, under the modular group and its quotient the anharmonic group (in sec.~\ref{ModularFormsAndAnharmonicForms}).

\section{Properties and transformation laws of the \bbd{A-} and \bbd{B-}functions}\label{AAndBFunctions}

\subsection{Recursion relations and their solution}

The correlation functions \bbd{G\ll {2n} = 2\uu{+4n}\cc e\uu{q\ll n[\t,\tb] - q\ll 0[\t, \tb]}} satisfy the recursion relation of a semi-infinite Toda lattice~\cite{Papadodimas:2009eu,Baggio:2014sna,Baggio:2014ioa}
\bbb
\pp\ll\t\pp\ll\tb q\ll n = \exp{q\ll{n+1}  - q\ll n} - \exp{q\ll n - q\ll{n-1}}\ , \llsk\llsk q\ll 0 \equiv {\rm log} \big [ \cc Z\ll{S\uu 4}[\t] \cc \big ]
\een{RecursionRelation}
These can be used to derive higher correlation functions from lower ones; alternately, by expanding \bbd{q\ll n} in a series at large \bbd{n} and applying the recursion relations
order by order, one can write equations for all the terms in the series.  Supplementing the recursion relations with additional information from the \textsc{eft} of the 
effective abelian vector multiplet on the Coulomb branch, and information about the coefficients of leading terms coming from explicit calculation of diagrams and matching
with known theories, we find the series is determined up to exponentially small corrections to be
\bbb
q\ll n \simeq q\ll n\uprm{eft} = A n + B + {\rm log} \bigg [ \cc \G(n\D+ \a + 1) \cc \bigg ]\ ,
\eee
with the corrections exponentially small in \bbd{\sqrt{n}} that
we shall call \bbd{q\ll n\uprm{mmp} \equiv q\ll n - q\ll n\uprm{eft}}.  In conformal SQCD, the generator of the Coulomb-branch chiral ring has \bbd{\D = 2},
and the anomaly-mismatch coefficient \bbd{\a = {3\over 2}.}  In that case,
\bbb
q\ll n\uprm{eft} = An + B + {\rm log} \bigg [ \cc \G(2n+ {5\over 2}) \cc \bigg ]\ , \llsk Z\ll n\uprm{eft} = e\uu{q\ll n\uprm{eft}} = \G(2n + {5\over 2})\cc e\uu{nA + B}\ .
\eee
In addition, the recursion relations imply that the \bbd{A,B} coefficients satisfy PDEs with respect to the gauge coupling \bbd{\t}:
\bbb
\pp\ll\t \pp\ll{\bar{\t}} A =8 \cc e\uu A\ , 
\xxn{PDEForAFunction}
 \pp\ll\t \pp\ll{\bar{\t}} (B - A) = 0\ .
\een{PDEForBFunction}

\subsection{Transformation of the \bbd{A-}function under holomorphic coupling reparametrizations}\label{AFuncHoloCouplingTransformation}

In~\cite{Hellerman:2018xpi} the authors fully solved for the form of \bbd{A} as a function of \bbd{\t} and \bbd{\bar{\t}}.  In doing this we used the S-duality properties of SQCD, and
it is useful to work in a different coupling coordinate \bbd{\s,} in which the S-duality acts more simply.  This coupling is sometimes referred to as the "infrared coupling"
as distinguished from the "ultraviolet coupling" \bbd{\t}; the relationship between the two is
\bbb
q\equiv \exp{2\pi i \t} = \l(\s) \ , 
\een{RelationBetweenUVAndIRCouplings}
where \bbd{\l(\s)} is the modular lambda function. (The weak-coupling expansion of the relationship between \bbd{\s} and \bbd{\t} is given to the first few orders in eqs. \rr{MLWCE1}-\rr{MLWCE3}).  In order for the recursion relations to be independent of the coordinate used on theory space, we have to transform
our \bbd{A} function among different holomorphic
coordinate systems with the transformation law of a Liouville field (see {\it e.g.\rm}~\cite{Seiberg:1990eb})  i.e., with an additive term that is the log of the norm-squared
of the holomorphic Jacobian, so that \bbd{\exp{A}} transforms as an object with one lower holomorphic and one lower antiholomorphic index:
\bbb
A\ls{\t\pr} \equiv A\ls \t + {\rm log} \bigg [ \cc \big | {{d\t}\over{d\t\pr}} \big |\sqd \cc \bigg ]\ , \llsk\llsk
\exp{A\ls{\t\pr}} = \big | {{d\t}\over{d\t\pr}} \big |\sqd \cc \exp{A\ls\t}\ ,
\een{TransformationLawForACoefficientUnderHolomorphicCouplingReparametrizations}
with the subscript in square brackets indicating the holomorphic coordinate with respect to which the \bbd{A}-field is defined.  With this transformation law, the equation
\rr{PDEForAFunction} is the same in any holomorphic coordinate.  

Note that the correlation functions themselves also transform under a change of holomorphic coordinate on theory space.  This is natural, since the generator
\bbd{\co\ll 2 = -4\pi i\cc {\rm tr}[\hat{\phi}\sqd]} is in the same superconformal multiplet as the chiral marginal operator \bbd{{\rm Tr}(F\ll +\sqd)}
multiplying the holomorphic gauge coupling \bbd{\t} in the Lagrangian, and marginal operators transform
as covector fields on theory space.  (Here by \bbd{F\ll\pm} we mean the self-dual and anti-self-dual parts of the gauge field strength.)  The correlation function \bbd{G\ll 2 = |x-y|\uu 4\cc  \langle \co\ls\t \overline{\co}\ls{\bar{\t}} \rangle} is therefore a tensor product of a holomorphic and
antiholomorphic covector field on theory space, and the correlation function transforms as
\bbb
G\ll {2[\t\pr]} = \big |{{d\t}\over{d\t\pr}} \big |\sqd \cc G\ll{2[\t]}\ ,
\eee
and more generally
\bbb
G\ll {2n[\t\pr]} = \big |{{d\t}\over{d\t\pr}} \big |\uu{2n} \cc G\ll{2n[\t]}\ ,
\eee
and the \bbd{q\ll n} transform as
\bbb
q\ll {n[\t\pr]} = q\ll{n[\t]} + n\cc{\rm log} \bigg [ \cc \big | {{d\t}\over{d\t\pr}} \big |\sqd \cc \bigg ]\ .
\eee
Note that this transformation law also follows from the fact that the partition function \bbd{Z\ll{S\uu 4}} transforms as a scalar, together with the recursion relations \rr{RecursionRelation}.

We note that these transformation laws also reflect the relationship between the Zamolodchikov metric on superconformal theory space, and the theory-space K\"ahler potential from
which it is derived~\cite{Gerchkovitz:2014gta, Gomis:2014woa}. The sphere partition function can be viewed as the exponential of the K\"ahler potential for the Zamolodchikov metric \bbd{g\ll{\t\bar{\t}}} on
theory space; the two-point function for a chiral primary, which is directly proportional by SUSY to the two-point function of the chiral marginal operator descended from
it, is proportional to the holomorphic times antiholomorphic second derivative of the K\"ahler potential:
\bbb
K = 12\cc\cc {\rm log}[Z]\ , 
\xxnn
g\ll{\t\bar{\t}} = {3\over 4}\cc G\ll {2[\t]} =  K\ll{,\tau\bar{\tau}} =  12\cc \pp\ll\t\pp\ll{\bar{\t}} \cc {\rm log}[Z]\ .  
\eee
So the transformation law of \bbd{G\ll 2} under holomorphic coupling reparametrization of \bbd{\t} follows from
its identification as \bbd{{4\over 3}} times the Zamolodchikov metric.  See the comments in sec.~\ref{SummaryOfNormalizationsOfTheKahlerPotential} of the
Appendix on the various normalization conventions used
in the relevant literature for the Zamolodchikov metric and its K\"ahler potential for \bbd{{\cal N} = 2} superconformal theories with
marginal operators, and their relationships to the normalizations of two-point functions.

\subsection{Transformation of the \bbd{A-}function under duality transformations}

In terms of the UV coupling \bbd{\t,} the action of the \bbd{SL(2,Z)} S-duality is not the familiar one acting by fractional linear transformations with integer coefficients~\cite{Gaiotto:2009we, Alday:2009aq, Grimm:2007tm, Dorey:1996bn}.
Rather, the familiar generators \bbd{S} and \bbd{T} of the S-duality group act on \bbd{q \equiv e\uu{2\pi i \t} = \l(\s)} as
\bbb
S: \llsk q\to 1-q\ ,
\llsk\llsk
T: \llsk q\to {q\over{q-1}}\ .
\een{ModularTransformationsOfq}
This is to be contrasted with the infrared coupling \bbd{\s}  that transforms in the familiar way by fractional linear transformations,
\bbb
\s\to {{a\s + b}\over{c\s + d}},~\llsk \left ( \begin{matrix} a & b \cr c & d \end{matrix} \right ) \in SL(2,\IZ)
\een{DualityTransformOfIRCoupling}
with the generators acting by
\bbb
S: \llsk \s\to - {1\over\s}\ ,
\xxnn
T: \llsk \s\to \s + 1\ .
\eee

\subsubsection{Duality transformations and solution for \bbd{A\ls\s}}

In~\cite{Hellerman:2020sqj} we used the duality properties of the \bbd{A} function to solve completely for its functional form, which is most easily
expressed in the \bbd{\s} coordinate, where \bbd{\exp{A\ls\s}} transforms as a modular form of weight \bbd{(2,2)}.  

The transformation law of \bbd{\exp{A\ls\s}} as a modular form of weight \bbd{(2,2)} follows from the underlying duality invariance of the theory.  Duality says that the
sphere partition function must be duality invariant up to a holomorphic times antiholomorphic factor.  From that property, it follows that the correlation functions 
\bbd{G\ll{2n[\s]} \equiv 2\uu{+ 4n}\cc e\uu{q\ll {n[\s]}- q\ll 0}},
which obey \rr{RecursionRelation} with respect to the coordinate \bbd{\s}, must transform as nonholomorphic modular forms of weight \bbd{(2n,2n)}.  From there,
we can express \bbd{A\ls\s} as
\bbb
\exp{A\ls\s} \equiv {\rm lim}\ll{n\to \infty}\cc {1\over{16\cc (2n + {7\over 2})(2n + {5\over 2})}}\cc {{G\ll{(2n+2)[\s]}}\over{G\ll{2n[\s]}}}\ ,
\een{ExtractionOfAFunction}
from which it follows immediately that \bbd{\exp{A\ls\s}} transforms as a modular form of weight \bbd{(2,2)}.

The simple transformation law under the duality group makes it simplest to express the solution for \bbd{A} in terms of the \bbd{\s-}coordinate.  In~\cite{Hellerman:2020sqj} we found
the solution
\bbb
\exp{A\ls\s} = {1\over{16\cc [{\rm Im}[\s]]\sqd}}\ .
\een{AFunctionInSigmaFrame}
This solution was uniquely determined, up to a single constant, by the requirement that \bbd{A\ls\s} transform as a nonholomorphic modular form of weights \bbd{(2,2)} under
the S-duality group acting on \bbd{\s}.  The single constant was then fixed by matching with perturbation theory, supplemented by the double-scaling analysis of~\cite{Bourget:2018obm}.

\subsubsection{Duality transformations and solution for \bbd{A\ls\t} in the \bbd{\t} variables}

We can use the transformation law to express the A-function \bbd{A\ls\t} in the \bbd{\t} frame, given the solution \rr{AFunctionInSigmaFrame} and the transformation
law \rr{TransformationLawForACoefficientUnderHolomorphicCouplingReparametrizations}.  We have
\bbb
\exp{A\ls\t} = |{{d\s}\over{d\t}}|\sqd \cc \exp{A\ls\s} =  {1\over {16}}\cc |{{d\s}\over{d\t}}|\sqd \cc [{\rm Im}[\s]|\uu{-2}  \ ,
\xxnn
A\ls\t = - 4\cc {\rm log}[2]- 2\cc {\rm log}[{\rm Im}[\s]] - 2\cc {\rm log}|{{d\t}\over{d\s}}|\ ,
\eee
with
\bbb
{{d\t}\over{d\s}} = {1\over{2\pi i}}\cc {{\l\pr(\s)}\over{\l(\s)}}\ ,
\eee
where \bbd{\l(\s)} is the modular lambda function.  Using an identity \rr{IdentityForTauToSigmaJacobian} for the \bbd{\t} to \bbd{\s} Jacobian, 
\bbd{\big |\cc {{d\t}\over{d\s}}\cc \big |\sqd =  2\uu{+{2\over 3}}\cc  |\l(\s)|\uu{-{2\over 3}}\times |1 - \l(\s)|\uu{+{4\over 3}} \times |\eta(\s)|\uu 8},
we can write the expression for the exponentiated \bbd{A-}coefficient in the \bbd{\t-}frame more simply as
\bbb
\bbsk
\exp{A\ls\t} =  2\uu{-{{14}\over 3}}\cc  |\l(\s)|\uu{+{2\over 3}}\times |1 - \l(\s)|\uu{-{4\over 3}} \times |\eta(\s)|\uu {-8} \times {1\over{[ {\rm Im}(\s)]\sqd}}
\llsk
\een{SimplerFormulaForExponentialOfACoefficientInTauFramePRecap}

It is far from transparent that \bbd{A\ls\t} as defined above, transforms appropriately under duality transformations.  To see this more clearly it is helpful
to discuss duality transformations directly in the \bbd{\t} variables.  Just as a holomorphic modular form of weight \bbd{k} is an object that transforms just like the \bbd{k/2} power of the \bbd{\s-}derivative of a modular-invariant function, we can define an {\it anharmonic form} of weight \bbd{w} as an object transforming like the \bbd{w/2} power of the \bbd{\t-}derivative of a modular-invariant function.  An anharmonic form \bbd{\phi\ll w} of weight \bbd{w} transforms as
\bbb
S:\llsk q\to 1-q\ , \llsk  \phi\ll w \to \big ( {{1-q}\over q} \big )\uu{w/2}\cc  \phi\ll w \ ,
\xxnn
STS: \llsk q \to + {1\over q}\ , \llsk  \phi\ll{w} \to (-1)\uu{w/2}\cc \phi\ll{w} \ .
\eee
for \bbd{w} an even integer.\footnote{We choose the name "anharmonic form" to indicate its simple
transformation law under the anharmonic group \bbd{S\ll 4 = SL(2,\IZ) / \G(2)} with \bbd{\G(2)} being the subgroup of the modular group \bbd{SL(2,\IZ)}
that leaves \bbd{q = \l(\s)} invariant.}

More generally, a not-necessarily-holomorphic anharmonic form of weights \bbd{(w,\tilde{w})} transforms as
\bbb
S:\llsk q\to 1-q\ , \llsk  \phi\ll{(w,\tilde{w})} \to \big ( {{1-q}\over q} \big )\uu{w/2}\cc \big ( {{1-\qb}\over \qb} \big )\uu{\tilde{w}/2}\cc \phi\ll{(w,\tilde{w})} \ ,
\xxnn
STS: \llsk q \to + {1\over q}\ , \llsk  \phi\ll{(w,\tilde{w})} \to (-1)\uu{(w - \tilde{w})/2}\cc \phi\ll{(w,\tilde{w})} \ .
\eee
for \bbd{w - \tilde{w}} an even integer.

The transformation law is chosen so that the derivatives \bbd{\pp\ll\t} and \bbd{\bar{\pp}\ll{\bar{\t}}} transform as anharmonic form of weights \bbd{(+2,0)} and \bbd{(0,+2)}, respectively.  It follows immediately
that the two-point function \bbd{G\ll{2[\t]}} of the chiral ring generator \bbd{\co \equiv \co\ls\t} whose descendant is the chiral marginal operator defined by differentiating the 
theory with respect to \bbd{\t,} is an anharmonic form of weights \bbd{(+2,+2)}.  We discuss these objects in more detail in Appendix~\ref{ModularFormsAndAnharmonicForms}.

\section{Scheme-dependence and duality properties of the sphere partition function and \bbd{B-}function}\label{SchemeDependenceSection}

\subsection{Scheme-dependence of the partition function and the \bbd{B-}function}

Just as we solved for the \bbd{A-}function using \bbd{S-}duality, we would like to use a similar strategy to solve for the coupling dependence of the \bbd{B-}function.  Unlike the \bbd{A}-function, the \bbd{B-}function is not defined with respect
to any coordinate system and does not transform under reparametrization of the gauge coupling.  Like the sphere partition function \bbd{Z\ll{S\uu 4},} the
\bbd{B-}function is a scalar with respect to coupling constant reparametrization.  

However also like the sphere partition function, the definition of the \bbd{B-}function \emm{does} have an renormalization-scheme ambiguity associated with the coupling dependence 
of the Euler-density counterterm~\cite{Gerchkovitz:2014gta, Gomis:2014woa}. At the level of the four-dimensional gauge theory, this counterterm is just a \bbd{c-}number multiplying
the Euler density of the background metric, but the dependence of this \bbd{c-}number counterterm on the gauge coupling gives a multiplicative renormalization by a positive
function of the coupling constant \bbd{\t}.

Absent further information, this scheme ambiguity would render the sphere partition function meaningless and completely arbitrary.  However superconformal symmetry
imposes constraints on the consistent SUSY-preserving regularization procedures and choice of counterterms, which reduces the ambiguity.  With the background metric
and marginal couplings assigned spurionic SUSY transformations and the counterterms chosen to preserve superconformal symmetry, the only counterterm ambiguities are in the form of a holomorphic plus antiholomorphic function of the marginal parameters multiplying the Euler density, so that the sphere partition function transforms under these ambiguities
into itself times a positive definite, holomorphically factorized function of the holomorphic coupling:
\bbb
Z\to |P(\t)|\sqd \cc Z
\eee
This scheme change can also be viewed as a K\"ahler transformation \bbd{K\to K +12\cc {\rm log}[ |P(\t)|\sqd]} by virtue of the discussion in~\cite{Gerchkovitz:2014gta, Gomis:2014woa}, reviewed at the end
of sec.~\ref{AFuncHoloCouplingTransformation}.

This scheme ambiguity clearly does not affect the correlation functions as derived from the recursion relations \rr{RecursionRelation}, nor does
it affect the \bbd{A-}function as can be seen from the relation \rr{ExtractionOfAFunction}:  Under a change of scheme / K\"ahler transformation, 
\bbb
G\ll{2n}\to G\ll{2n} \ , \llsk  A\to A\ , 
\eee
From these transformation properties and the definitions, we see 
that the \bbd{q\ll n} and the \bbd{B-}function:
\bbb
\exp{q\ll n} \to |P(z)|\sqd \cc \exp{q\ll n}\ , \llsk\llsk \exp{B}\to |P(z)|\sqd \cc \exp{B}\ .
\eee
In order to solve for the \bbd{B-}function, then, we must first recognize that there is no "the" \bbd{B-}function, and we must solve for the \bbd{B-}function in a given scheme.
Then we can reconstruct the \bbd{B-}function in any other scheme with the relation
\bbb
\exp{B\ll{{\rm scheme~}2}} = {{Z\ll{{\rm scheme~}2}}\over{Z\ll{{\rm scheme~}1}}}\cc \exp{B\ll{{\rm scheme~}1}} 
\een{SchemeCovarianceOfBFunction}
Note that the combination 
\bbb
\tilde{B} \equiv B - {\rm log}[Z]
\een{BTildeDefinition}
is scheme-independent, and can be described as the \bbd{n\uu 0} term in the large-\bbd{n} limit
of correlation functions:
\bbb
G\ll{2n} =  2\uu{+4n}\cc \exp{An +\tilde{B}}\times \G(2n + {5\over 2})\times \bigg [ 1 + ({\rm exponentially~small~in~}\sqrt{n})\bigg ]
\eee

Fortunately, the S-duality of SQCD, and its known action on the partition function in a particular scheme~\cite{Alday:2009aq}, will give us enough information to solve
for the \bbd{B-}function in any given scheme we like.

\subsection{Duality fixes the scheme uniquely up to an overall coupling-independent constant}

For a theory with an \bbd{S-}duality, the duality transformation properties of the sphere partition function may be seen to determine the Euler-counterterm choice completely up
to an overall, coupling independent constant, assuming that the counterterm choice is reasonable in the sense of having weak-coupling asymptotics, {\it i.e.\rm} 
growing no faster than a negative power of \bbd{q = \exp{2\pi i \t}} at weak coupling.  Suppose we have two different counterterm choices for the same theory at the same value of the coupling; the sphere partition
functions are \bbd{Z\lrm{scheme~1}} and \bbd{Z\lrm{scheme~2}}.  Under the assumption that both scheme choices respect superconformal symmetry, we know the ratio of
the two must be holomorphically factorized:
\bbb
Z\lrm{scheme~i} = |P\ll{ij}(\t)|\sqd \cc Z\lrm{scheme~j}\ , \llsk i,j \in\{1,2\}\ ,
\xxnn
|P\ll{ji}(\t)|\sqd = |P\ll{ij}(\t)|\uu{-2}\ .
\een{SchemeCovarianceRule}
Here we should pause to note we do not necessarily assume the holomorphic factorization is global, {\it i.e.} we do not necessarily assume \bbd{P\ll{ij}(z)} is single-valued;
we only assume the norm-squared is single valued.  For SQCD in particular also, the \bbd{|P\ll{ij}(\t)|\sqd} must also be nonsingular away from the weak-coupling
and dual-weak-coupling points \bbd{q = \exp{2\pi i \t} = 0,1},
as there are no other points in theory space where the dynamics become singular.

If both \bbd{Z\lrm{scheme~i}} transform the same way under \bbd{S-}duality and both have standard weak-coupling asymptotics, then the ratio \bbd{|P\ll{ij}(\t)|\sqd} is a positive definite, nonsingular harmonic function of \bbd{q = \exp{2\pi i \t}} that is invariant under the modular transformation \rr{ModularTransformationsOfq}.  By Liouville's
uniqueness theorem for modular-invariant harmonic functions, given in section~\ref{UniquenessTheoremProof} of the Appendix, 
 this implies \bbd{|P\ll{ij}(\t)|\sqd} must be a constant, independent of \bbd{\t}.  This requires only the assumption that
 \bbd{{\rm log}[Z\lrm{scheme~i}]} are both bounded by some power of \bbd{{\rm Im}[\t]} at weak coupling,
 and only a local, rather than global, holomorphic factorization of \bbd{|P\ll{ij}(\t)|\sqd}.

Once we know the duality transformation of \bbd{Z} in a given scheme, we can always construct \bbd{Z} in a duality-invariant scheme by multiplicatively averaging over
duality images of \bbd{Z}.  Note that this is far simpler than a Poincar\'e series for the log of \bbd{Z}, which would involve averaging over an infinite set of duality
images under \bbd{SL(2,\IZ)}.  The SQCD partition function is a function of \bbd{q = \exp{2\pi i \t},} on which the congruence group \bbd{\G(2)} 
acts trivially, and only the finite quotient \bbd{SL(2,\IZ) / \G(2)} acts nontrivially.  This quotient is of order \bbd{24,} and is isomorphic to the
symmetric group \bbd{S\ll 4} on four elements, sometimes called
the {\it anharmonic group \rm} in the role it plays here.  Of the \bbd{24-}element group, four elements fix \bbd{q} and so the duality orbit of \bbd{q} and any function depending on \bbd{q}, consists of only six objects.  So given
a partition function \bbd{Z[q]} in any scheme, we can always define
 \bbb
Z\uprm{inv}[q] \equiv \biggl [ \cc \prod\ll{q\lrm D\in\{\rm{six-element~duality~orbit~of~}q\}}
Z[q\lrm D] \biggl ]\uu{+{1\over 6}}\ .
\een{ConstructionOfDualityInvariantZByMultiplicativeAveraging} 
The partition function \bbd{Z\uprm{inv}[q]} is equivalent to the original partition function \bbd{Z[q]} up to a holomorphic plus antiholomorphic Euler counterterm in the action,
equivalently a K\"ahler transformation of \bbd{K \equiv 12\cc {\rm log}[Z]}, and is invariant under all duality transformations.  So, if we have any partition function
with a known duality transformation, we may always construct a duality-invariant version of the partition function by a calculable change
of the Euler-counterterm choice.

In the next section we shall use the AGT correspondence to write the duality transformation of the partition function in the scheme used by\cite{Alday:2009aq}, and relate
that scheme to others that will be of use to us.

\subsection{The AGT partition function~\cite{Alday:2009aq} and its duality transformation}

\subsubsection{General idea of~\cite{Alday:2009aq}}

In its role as a duality group of conformal SQCD, the \bbd{24-}element permutation group \bbd{S\ll 4} can be thought of as permuting the locations of
four local operators in a \emm{two-dimensional} CFT four-point function, which is an equivalent representation of the SQCD sphere partition function under the 4D/2D correspondence
known as AGT duality.~\cite{Alday:2009aq} In what follows we shall use this correspondence to fix the duality transformation of the partition function in the scheme 
in which ref.~\cite{Alday:2009aq} defines it, and use that definition to compare to other schemes as well.

The work of~\cite{Alday:2009aq} uses the duality ideas of~\cite{Gaiotto:2009we}
and the explicit calculations of~\cite{Nekrasov:2002qd, Pestun:2007rz} to make a correspondence
between a number of superconformal field theories in four dimensions with \bbd{SU(2)} gauge group on the one hand, and four-point functions in two-dimensional Liouville 
theory on the other hand.  This work, drawing on intuitions from the engineering of four-dimensional gauge theories by the compactification of the six-dimensional \bbd{(2,0)} theory
on Riemann surfaces, has had many applications and generalizations (see {\it e.g.\rm}~\cite{LeFloch:2020uop} for a recent review of the
state of the art), but for our purposes its main virtue is the fixing of the scheme-dependent
duality properties of the partition function of conformal SQCD on the four-sphere.

Schematically, the substance of the AGT correspondence for superconformal SQCD with \bbd{G = SU(2)} and four hypermultiplets in the fundamental representation, is 
the equality
\bbb
[{\rm partition~function~of~SQCD~on~}S\uu 4] = [{\rm four-point~function~of~Liouville~theory~on~}S\uu 2]\ ,
\een{SchematicEquivalence}
with the gauge coupling \bbd{\t} of SQCD is related to the holomorphic cross-ratio of the four-points:
\bbb
q = \exp{2\pi i \t} = {{z\ll{12} z\ll{34}}\over{z\ll{13} z\ll{24}}}\ .
\een{CrossRatioMapping}

\subsubsection{Map of parameters for the AGT correspondence for conformal SQCD}

In the case we are interested in, where the \bbd{S\uu 4} is round instead of squashed, and where the hypermultiplet masses vanish, the Liouville theory is the one with
asymptotic linear dilaton gradient \bbd{Q=2} and central charge \bbd{1 + 6Q\sqd = 25,} 
and the four vertex operators are those of the lowest-dimension normalizable states, \bbd{\a\ll{1,2,3,4} = h\ll{1,2,3,4} = \tilde{h}\ll{1,2,3,4} = +1}.  In terms
of AGT's parametrization of the four-dimensional geometry and mass parameters, these parameters correspond to\footnote{Some relevant comments on conventions for
mass parameters are given in~\cite{Okuda:2010ke}; see also~\cite{Billo:2013fi}.}
\bbb
 b = b\uu{-1} = \e\ll 1 = \e\ll 2 = +1\ ,  \llsk \e\ll 1 + \e\ll 2 = \e\ll + = Q = b + b\uu{-1} = +2\ , 
 \xxnn
 \tilde{m}\ll{0,1} = \hat{m}\ll{0,1} = 0\ , \llsk m\ll{0,1} = {Q\over 2} + \hat{m}\ll{0,1} = +1\ .
\een{ParameterMapAGTConformalSQCD}
(Here, both the \bbd{\e} and mass-parameter quantities are dimensionful with mass dimension \bbd{+1}; refs~\cite{Pestun:2007rz, Alday:2009aq} make up the mass dimensions by setting the radius of the four-sphere
equal to \bbd{1.})

\subsubsection{Definition[s] of CFT four-point function[s]}

At the schematic level we have stated it so far the equivalence \rr{SchematicEquivalence} is under-specified, since we have not quite stated what one means by "the" CFT
four-point function.  There are various, slightly different, definitions of a CFT four-point function, all related to each other by holomorphically factorized functions
of the cross-ratio, corresponding to holomorphically factorized functions of the exponentiated gauge coupling \bbd{q}.  Since it is precisely these kinds of holomorphically
factorized objects we wish to specify precisely, we need to give more information about how precisely the four-point function is defined.

There are various versions of the four-point function in the literature and the version of the four-point function equal to the four-sphere SQCD partition function under
the AGT correspondence, is none of the more commonly used versions.  

Taking all four operators to be scalars of equal dimension \bbd{\Delta = 2h}, the first
common normalization is just to define
\def\co{{\cal O}}
\bbb
\tilde{A}[q] \equiv {\rm lim}_{y\to\infty} |y|^{+4h} \tilde{A}[y,1,q,0]
\eee
where
\bbb
\tilde{A}[z_{1,2,3,4}] \equiv \langle \co(z_1) \co(z_2) \co(z_3) \co(z_4) \rangle
\eee
With this definition, \bbd{\tilde{A}[q]} has the transformation properties
\bbb
\tilde{A}[1-q] = \tilde{A}[q]
\eee
and
\bbb
\tilde{A}[ {1\over q} ] = |q|^{4h} \tilde{A}[q]
\eee
under the generators of the anharmonic group if the four scalar operators are identical.

Then there is another common definition of the four-point function that is commonly used in the bootstrap literature (see {\it e.g. \rm} the recent review~\cite{Poland:2018epd} and
references therein)
\bbb
g[q] \equiv |q|^{+4h} \tilde{A}[q]
\eee
for a four-point function with four scalar operators of equal dimension \bbd{\D = h + \tilde{h} = 2h = 2\tilde{h}}.

This definition is such that the four-point function
behaves more smoothly at small \bbd{q} for a unitary CFT with discrete spectrum.  For such a "nice" CFT, the limit \bbd{q\to 0} is dominated by the exchange of the lowest state, which is always the identity, in the intermediate channel~\cite{Poland:2018epd} so that \bbd{g[q]} scales as \bbd{q^0} instead of having a singularity.  The same is true
even for theories with continuous spectrum that are obtained as limits of unitary theories with discrete spectrum~\cite{Seiberg:1999xz}: Also in this case the identity is the lowest state, whether discretely below a continuum or at the bottom edge of a continuum starting at zero dimension.

For nonunitary CFT, or for CFT with continuous spectrum that are not the limit of CFT with discrete spectrum, the situation may be different and the lowest normalizable state
in the radial-quantization Hilbert space, may not correspond to the identity under the state-operator correspondence and may not have dimension zero.  In such cases, the
singularity of the four-point function \bbd{\tilde{A}[q]} at small \bbd{q} is not \bbd{|q|\uu{-4h}} but rather \bbd{|q|\uu{-4h +  \D\lrm{vac},}} where \bbd{\D\lrm{vac}} is the dimension
of the lowest normalizable state, with \bbd{\D\lrm{vac} = 2 h\ll 0 = 2 \tilde{h}\ll 0} for a parity-invariant CFT.

For these theories, the natural four-point function to define for four identical scalars with conformal weights \bbd{(h,h)}, is the generalized four-point function
\bbb
g\lrm{gen}[q] \equiv |q|\uu{-\D\lrm{vac}} \cc g[q] = |q|\uu{4h -  \D\lrm{vac}}\cc \tilde{A}[q] = 
|q|\uu{4h -  \D\lrm{vac}}\cc
{\rm lim}\ll{y\to\infty}\cc |y|\uu{4h}\cc \langle\co(y) \co(1)\co(q)\co(0)\rangle\ . \llsk
\een{DefinitionOfGeneralizedBootstrapFriendlyFourPointFunction}
This definition is chosen so that the power law in the singularity of \bbd{g\lrm{gen}[q]} is \bbd{q\uu 0,} although for theories with continuous spectrum, there may be corrections
in powers of \bbd{{\rm log}[q]} depending on the dimension of the continuum and the behavior of the spectral density near the vacuum.  For unitary theories with discrete spectrum, limits of such theories, or for any CFT whose ground state dimension is \bbd{\D\lrm{vac} \to 0,} the generalized four-point
function reduces to the usual bootstrap-friendly four-point function, \bbd{g\lrm{gen}[q] \to g[q]}.

\subsubsection{Modular transformations of the four-point function[s]}

For us what matters about these four-point functions is their modular properties, which are inherited directly from those of
\bbd{\tilde{A}[q]}, which we will explain here and summarize below in table~\ref{TypesOfLiouvilleAmplitudeTransformationTable}.

The unmodified four-point function \bbd{\tilde{{\cal A}}[q]}  for four identical scalar primaries with weights \bbd{(h,h)}, transforms as
\bbb
q\to 1-q: \llsk \tilde{A}[q]\to \tilde{A}[1-q] = \tilde{A}[q]\ ,
\xxnn
q\to {1\over q}: \llsk
\tilde{A}[q]\to \tilde{A}[{1\over q}] = |q|\uu{+4h}\cc \tilde{A}[q]\ ,
\eee
The "bootstrap-friendly" four-point function \bbd{g[q] \equiv |q|\uu{+4h}\cc \tilde{{\cal A}}[q]}  for four identical scalar primaries with weights \bbd{(h,h)}, transforms as
\bbb
q\to 1-q: \llsk g[q] \to g[1-q] = {{|1-q|\uu{+4h}}\over{|q|\uu{+4h}}}\cc g[q]\ ,
\xxnn
q\to {1\over q}:\llsk g[q]\to g[{1\over q}] =  |q|\uu{-4h}\cc g[q]\ .
\eee
The generalized bootstrap-friendly four-point function \bbd{g\lrm{gen}[q],} as defined in \rr{DefinitionOfGeneralizedBootstrapFriendlyFourPointFunction}, transforms as
\bbb
q\to 1-q: \llsk g\lrm{gen}[q] \to g\lrm{gen}[1-q] = {{|1-q|\uu{+4h - \D\lrm{vac}}}\over{|q|\uu{+4h - \D\lrm{vac}}}}\cc g\lrm{gen}[q]\ ,
\xxnn
q\to {1\over q}:\llsk g\lrm{gen}[q]\to g\lrm{gen}[{1\over q}] =  |q|\uu{2\D\lrm{vac} -4h}\cc g\lrm{gen}[q]\ .
\een{GeneralizedBootstrapFriendlyFourPointFunctionDualityTransformationLaw}

Finally we note the existence of another possible definition of the four-point function four-point which is basically never used, but which deserves attention in the context of the present
article.  We can define
\bbb
\bbsk
{\cal A}\uprm{inv}[q] \equiv |q|\uu{+{{4h}\over 3}}\cc |1-q|\uu{+{{4h}\over 3}}\cc \tilde{{\cal A}}[q] =  |q|\uu{-{{8h}\over 3}}\cc |1-q|\uu{+{{4h}\over 3}}\cc g[q] = |q|\uu{\D\lrm{vac}-{{8h}\over 3}}\cc |1-q|\uu{+{{4h}\over 3}}\cc g\lrm{gen}[q]\ .
\llsk
\een{InvariantFourPointFunctionDefined}
It is easy to check that this four-point function is invariant under \bbd{q\to 1-q} and \bbd{q\to {1\over q}} and therefore under the entire duality group with the anharmonic group \bbd{S\ll 4} acting on \bbd{q} as generated by \bbd{q\to 1-q} and \bbd{q\to {1\over q}}:
\bbb
q\to 1-q: \llsk {\cal A}\uprm{inv}[q]\to {\cal A}\uprm{inv}[1-q] = {\cal A}\uprm{inv}[q]\ ,
\xxnn
q\to {1\over q}: \llsk
{\cal A}\uprm{inv}[q]\to {\cal A}\uprm{inv}[{1\over q}] =  {\cal A}\uprm{inv}[q]\ ,
\eee
It is also straightforward
to see that the definition \rr{InvariantFourPointFunctionDefined} coincides with the definition of \bbd{{\cal A}\uprm{inv}[q] } as an explicit multiplicative average over its
orbit under the dualities:
\bbb
{\cal A}\uprm{inv}[q] \equiv \biggl [ \cc \prod\ll{q\lrm D\in\{\rm{six-element~duality~orbit~of~}q\}}
{\cal A}[q\lrm D] \biggl ]\uu{+{1\over 6}}
\eee

 We summarize their transformation laws in table~\ref{TypesOfLiouvilleAmplitudeTransformationTable}.

 \begin{table}
\begin{center}
\begin{tabular}{ |c||c|c|c|c| } 
 \hline\hline
\backslashbox{transformation}{object} & $\widetilde{{\cal A}[q]}$ & $ g[q]$ & $g\lrm{gen}[q]$ & ${\cal A}\uprm{inv}[q]$  \\ \hline\hline
$q\to 1-q$ &  $\widetilde{{\cal A}[q]}$  & ${{|1-q|\uu{+4h}}\over{|q|\uu{+4h}}}\cc g [q]$ & ${{|1-q|\uu{+4h - \D\lrm{vac}}}\over{|q|\uu{+4h - \D\lrm{vac}}}}\cc g\lrm{gen}[q]$ & ${\cal A}\uprm{inv}[q]$  \\ \hline
$q\to {1\over q}$ &$     |q|\uu{+4h}  \cc \widetilde{{\cal A}[q]}$              & $     |q|\uu{-4h}              \cc g[q] $& $ |q|\uu{2\D\lrm{vac} -4h}\cc\cc g\lrm{gen}[q]$  &  ${\cal A}\uprm{inv}[q]$ \\ \hline
\end{tabular}
\captionof{table}{Anharmonic-group transformations of types of four-point function in CFT, for four identical scalars of conformal weights \bbd{h\lrm L = h\lrm R = h}, and
the vacuum having a normalizable vacuum of total (left plus right) dimension \bbd{\D\lrm{vac}}  }
\label{TypesOfLiouvilleAmplitudeTransformationTable}
\end{center}
\end{table}

 \begin{table}
\begin{center}
\begin{tabular}{ |c||c|c|c|c| } 
 \hline\hline
\backslashbox{transformation}{object} & $\widetilde{{\cal A}[q]}$ & $ g[q]$ & $g\lrm{gen}[q]$ & ${\cal A}\uprm{inv}[q]$  \\ \hline\hline
$q\to 1-q$ &  $\widetilde{{\cal A}[q]}$  & ${{|1-q|\uu{4}}\over{|q|\uu{4}}}\cc g [q]$ & ${{|1-q|\uu{2}}\over{|q|\uu{2 }}}\cc g\lrm{gen}[q]$ & ${\cal A}\uprm{inv}[q]$  \\ \hline
$q\to {1\over q}$ &$     |q|\uu{+4}  \cc \widetilde{{\cal A}[q]}$              & $     |q|\uu{-4}              \cc g[q] $& $  g\lrm{gen}[q]$  &  ${\cal A}\uprm{inv}[q]$ \\ \hline
\end{tabular}
\captionof{table}{Anharmonic-group transformations of types of four-point function in CFT, for four identical scalars of conformal weights \bbd{h\lrm L = h\lrm R = h = +1}, and
the vacuum having a normalizable vacuum of total (left plus right) dimension \bbd{\D\lrm{vac}  =+2}.  This is the case of Liouville theory at \bbd{Q=2,~c=25} where the four
operators are the normalizable vacuum.  For these parameters, the amplitude \bbd{g\lrm{gen}[q]} corresponds under the AGT correspondence to the (round) \bbd{S\uu 4} partition function for conformal \bbd{{\cal N} = 2} SQCD
with \bbd{G=SU(2)} and \bbd{N\lrm F = 4}.  }
\label{TypesOfLiouvilleAmplitudeTransformationTableSpecificParameterChoiceOfInterestToUs}
\end{center}
\end{table}
 
For purposes of this paper, we are interested in the four-point function in the specific case of Liouville theory at \bbd{Q=2, c = 25,} which has \bbd{\D\lrm{vac} = {{Q\sqd}\over 2} = +2,}
and four identical external operators with \bbd{h = \tilde{h} = 1}.  For these values specifically, we summarize the transformations of the various four-point functions
in table~\ref{TypesOfLiouvilleAmplitudeTransformationTableSpecificParameterChoiceOfInterestToUs}.

\subsubsection{Precise definition of the \bbd{S\uu 4} partition function and Liouville four-point function in AGT correspondence}

\heading{\bbd{S\uu 4} partition function of \bbd{{\cal N} = 2} superconformal SQCD with \bbd{G = SU(2)} and \bbd{N\lrm F = 4}}

In the correspondence of~\cite{Alday:2009aq} the \bbd{S\uu 4} partition function for conformal SQCD is defined with the parameters \rr{ParameterMapAGTConformalSQCD}, but also in
a particular scheme that fixes the holomorphic plus antiholomorphic counterterm ambiguity of the Euler density term in the CFT action.  The authors define the
partition function using Pestun's~\cite{Pestun:2007rz} localization method, with one alteration.  While~\cite{Pestun:2007rz} uses Nekrasov's~\cite{Nekrasov:2002qd} \bbd{U(2)} instanton partition
function, the authors of~\cite{Alday:2009aq} factor out the effects of instantons that are in some sense purely inside the \bbd{U(1)} factor of the \bbd{U(2)} gauge group,
resolved to zero size by a deformation depending on parameters \bbd{\e\ll{1,2}}
that can be interpreted in terms of five-dimensional gauge theory or noncommutative geometry in four dimensions. This set of instantons gives a contribution that depends on the gauge coupling \bbd{\t} and on the \bbd{\e-} parameters \bbd{\e\ll{1,2}}
but is otherwise completely decoupled from the \bbd{SU(2)} gauge dynamics; its sole effect is a holomorphically factorized multiplicative contribution  \bbd{|1 - q|\uu{+4}}
to the \bbd{S\uu 4} partition function.  The authors of~\cite{Alday:2009aq}, in their eq. \aleq{3.9}, \emm{remove} this factor explicitly in defining their partition function:
\bbb
Z\lrm{AGT}[q] = |1-q|\uu{-4m\ll 0 (Q - m\ll 1)}\cc Z\lrm{{Pestun-}\atop{Nekrasov}}[q]\ , \llsk\llsk
Z\lrm{{Pestun-}\atop{Nekrasov}}[q] = |1-q|\uu{4m\ll 0 (Q - m\ll 1)} \cc Z\lrm{AGT}\ .
\eenn
For the specific case of interest to us, with parameters \bbd{m\ll 0 = m\ll 1 = 1, Q=2}, which correspond to conformal SQCD on the round \bbd{S\uu 4}, the relationship is
\bbb
Z\lrm{AGT}[q] = |1-q|\uu{-4}\cc Z\lrm{{Pestun-}\atop{Nekrasov}}[q]\ , \llsk\llsk
Z\lrm{{Pestun-}\atop{Nekrasov}}[q] = |1-q|\uu{+4} \cc Z\lrm{AGT}\ .
\een{PartitionFunctionRelationWithConformalValuesOfParameters}
Since this removal is positive definite and holomorphic times antiholomorphic, it can be absorbed completely into holomorphic plus antiholomorphic coupling
dependence of the Euler-density counterterm\footnote{We thank Zohar Komargodski for discussions on this point, and on the specific scheme choice of the instanton factor
of the SQCD partition function given in eq. \aleq{3.22} of~\cite{Gerchkovitz:2016gxx}}, and its presence or removal does not affect any observables or correlation functions.  However
it does affect the duality transformation of the AGT partition function, which is relevant for our considerations.  For us, it is important that the \bbd{S\uu 4} partition function
appearing in the equality \rr{SchematicEquivalence}, is the one whose instanton contributions have had their \bbd{U(1)} factor removed from the Nekrasov instanton sum~\cite{Nekrasov:2002qd}.

\heading{Liouville four-point function}

The Liouville four-point function that appears in the equivalence \rr{SchematicEquivalence}, is the one we have defined as the "generalized bootstrap-friendly" four-point
function \bbd{g\lrm{gen}[q]}; in the specific case of Liouville theory with screening charge \bbd{Q} and \bbd{c = 1 + 6Q\sqd,} we have \bbd{\D\lrm{vac} = {{Q\sqd}\over 2}}
and so the definition of \bbd{g\lrm{gen}} is
\bbb
\bbsk
g\lrm{gen}[q] = |q|\uu{-{{Q\sqd}\over 2}}\cc g[q] = |q|\uu{4h - {{Q\sqd}\over 2}}\cc \tilde{{\cal A}}[q] = |q|\uu{4h - {{Q\sqd}\over 2}}\cc {\rm lim}\ll{y\to\infty}\cc |y|\uu{+4h} \cc
\langle \co(y)\co(1)\co(q)\co(0)\rangle\ ,\llsk
\eee
For the specific parameters corresponding to conformal SQCD on the round four-sphere, \bbd{Q=2} and \bbd{h=1}, the relationship is
\bbb
g\lrm{gen}[q] = |q|\uu{-2}\cc g[q] = |q|\uu 2\cc \tilde{{\cal A}}[q] = |q|\uu 2\cc {\rm lim}\ll{y\to\infty}\cc |y|\uu{+4h} \cc
\langle \co(y)\co(1)\co(q)\co(0)\rangle\ ,
\eee

\heading{Precise correspondence and duality transformation of AGT's partition function}

So, specifically, the precise form of the correspondence \rr{SchematicEquivalence} can be found by comparing expressions \aleq{4.1} and \aleq{4.2} in~\cite{Alday:2009aq}
\bbb
Z\lrm{AGT}[q] = \n\cc g\lrm{gen}[q]\ ,
\een{PreciseVersionOfAGTRelationForConformalSQCD}
where \bbd{\n} is a \bbd{q-}independent constant which can be read off from eq. \aleq{4.2} of~\cite{Alday:2009aq} but which we need not record here.  For us, the key point is the duality transformation law of \bbd{Z\lrm{AGT}} which must match
that of \bbd{g\lrm{gen}[q],} which we have given in eq. \rr{GeneralizedBootstrapFriendlyFourPointFunctionDualityTransformationLaw} and in table~\ref{TypesOfLiouvilleAmplitudeTransformationTable}.  For our particular parameter choices \bbd{h=1, Q=+2, \D\lrm{vac} = 2,} the transformation law is: 
\bbb
q\to 1-q: \llsk g\lrm{gen}[q] \to g\lrm{gen}[1-q] = {{|1-q|\sqd}\over{|q|\sqd}}\cc g\lrm{gen}[q]\ ,
\xxnn
q\to {1\over q}:\llsk g\lrm{gen}[q]\to g\lrm{gen}[{1\over q}] =   g\lrm{gen}[q]\ .
\een{GeneralizedBootstrapFriendlyFourPointFunctionDualityTransformationLawForOurParameterChoices}

It follows from \rr{PreciseVersionOfAGTRelationForConformalSQCD} and \rr{GeneralizedBootstrapFriendlyFourPointFunctionDualityTransformationLawForOurParameterChoices} and that the AGT partition function transforms as 
\bbb
q\to 1-q: \llsk Z\lrm{AGT}[q] \to Z\lrm{AGT}[1-q] = {{|1-q|\sqd}\over{|q|\sqd}}\cc Z\lrm{AGT}[q]\ ,
\xxnn
q\to {1\over q}:\llsk Z\lrm{AGT}[q]\to Z\lrm{AGT}[{1\over q}] =   Z\lrm{AGT}[q]\ .
\een{AGTSQCDFourSpherePartitionFunctionDualityTransformationLawForOurParameterChoices}

\subsection{Duality transformations of four-sphere partition functions and \bbd{B-}coefficients in various schemes}

Having understood the duality transformations of the four-sphere SQCD partition function \bbd{Z\lrm{AGT}[q]} as defined with AGT's Euler-counterterm choice, we can now
use relationships to partition functions in other scheme choices, to find the duality transformations of the \bbd{S\uu 4} partition functions in those scheme choices.

As noted earlier in eq. \rr{PartitionFunctionRelationWithConformalValuesOfParameters},
the relationship between \bbd{S\uu 4} partition function for superconformal \bbd{G=2,~N\lrm F = 4} SQCD in AGT's scheme, and the partition function for
the same theory in the Nekrasov-Pestun scheme\footnote{This partition function is also the one whose expansion is given in eq. \aleq{3.23} of~\cite{Gerchkovitz:2016gxx}}
({\it i.e. \rm}using the \bbd{U(2)} instanton partition function without the \bbd{U(1)} factor removed), is
\bbb
Z\lrm{AGT}[q] = |1-q|\uu{-4}\cc Z\lrm{{Pestun-}\atop{Nekrasov}}[q]\ , \llsk\llsk
Z\lrm{{Pestun-}\atop{Nekrasov}}[q] = |1-q|\uu{+4} \cc Z\lrm{AGT}\ .
\een{PartitionFunctionRelationWithConformalValuesOfParametersRECAP}
It follows that the duality transformation is
\bbb
q\to 1-q: \llsk Z\lrm{{Pestun-}\atop{Nekrasov}}[q]\to Z\lrm{{Pestun-}\atop{Nekrasov}}[1-q] = {{|q|\sqd}\over{|1-q|\sqd}}\cc Z\lrm{{Pestun-}\atop{Nekrasov}}[q]\ ,
\xxnn
q\to {1\over q}:\llsk Z\lrm{{Pestun-}\atop{Nekrasov}}[q] \to Z\lrm{{Pestun-}\atop{Nekrasov}}[{1\over q}] 
= |q|\uu{-4}\cc Z\lrm{{Pestun-}\atop{Nekrasov}}[q] \llsk
\eee
We can also use construction \rr{ConstructionOfDualityInvariantZByMultiplicativeAveraging} starting either from \bbd{Z\lrm{AGT}} or \bbd{Z\lrm{{Pestun-}\atop{Nekrasov}},} to construct the partition function in a duality-invariant
scheme
\bbb
Z\uprm{inv}[q] \equiv \biggl [ \cc \prod\ll{q\lrm D\in\{\rm{six-element~duality~orbit~of~}q\}}
Z[q\lrm D] \biggl ]\uu{+{1\over 6}}\ ,
\een{DualityInvariantPartitionFunctionRecipe} 
which of course transforms as
\bbb
q\to 1-q: \llsk Z\uprm{inv}[q]\to Z\uprm{inv}[1-q] = Z\uprm{inv}[q]\ ,
\xxnn
q\to {1\over q}:\llsk Z\uprm{inv}[q] \to Z\uprm{inv}[{1\over q}] 
= Z\uprm{inv}[q] \llsk
\eee
We summarize the duality transformations of the sphere partition functions in various Euler-counterterm schemes, in table~\ref{DualityTransformationsOfSpherePartittionFunctionsTable}.

 \begin{table}
\begin{center}
\begin{tabular}{ |c||c|c|c|c| } 
 \hline\hline
\backslashbox{transformation}{object} &  $ {Z\uprm{inv}[q]} $ &  $  {Z\lrm{AGT}[q]} $&  $   {Z\lrm{{Pestun-}\atop{Nekrasov}}[q]}$   \\ \hline\hline
$q\to 1-q$ &   $ {Z\uprm{inv}[q]} $ & $ {{|1-q|\sqd}\over{|q|\sqd}}\cc  {Z\lrm{AGT}[q]} $ &   $ {{|q|\sqd}\over{|1-q|\sqd}}\cc  {Z\lrm{{Pestun-}\atop{Nekrasov}}[q]}$  \\ \hline
$q\to {1\over q}$ & $ {Z\uprm{inv}[q]} $            & $  {Z\lrm{AGT}[q]} $& $   |q|\uu{-4}\cc {Z\lrm{{Pestun-}\atop{Nekrasov}}[q]} $  \\ \hline
\end{tabular}
\captionof{table}{Duality transformations of the \bbd{S\uu 4} partition functions, in various schemes corresponding to different Euler-counterterm choices.}
\label{DualityTransformationsOfSpherePartittionFunctionsTable}
\end{center}
\end{table}

Working out the factors on the RHS of \rr{DualityInvariantPartitionFunctionRecipe}, explicitly, we find the relationships among all three scheme choices for the partition function which we have discussed so far:
\bbb
Z\uprm{inv}[q] =  |q|\uu{-{2\over 3}}\cc |1-q|\uu{+{4\over 3}}\cc Z\lrm{AGT}[q]
= |q|\uu{-{2\over 3}}\cc |1-q|\uu{-{8\over 3}}\cc Z\lrm{{Pestun-}\atop{Nekrasov}}[q]\ ,
\xxnn
Z\lrm{AGT}[q] = |1-q|\uu {-4}\cc Z\lrm{{Pestun-}\atop{Nekrasov}}[q] = |q|\uu{+{2\over 3}}\cc |1-q|\uu{- {4\over 3}}\cc Z\uprm{inv}[q]\ ,
\xxnn
Z\lrm{{Pestun-}\atop{Nekrasov}}[q] = |1-q|\uu{+4}\cc Z\lrm{AGT}[q]
= |q|\uu{+{2\over 3}}\cc |1-q|\uu{+{8\over 3}}\cc Z\uprm{inv}[q]\ ,
\een{TranslationsAmongAllPartitionFunctions}
We summarize these relationships in table~\ref{RatiosOfVariousPartitionFunctionsTable}.

 \begin{table}
\begin{center}
\begin{tabular}{ |c||c|c|c|c| } 
 \hline\hline
\backslashbox{denominator}{numerator} &  $  {Z\lrm{{Pestun-}\atop{Nekrasov}}[q]}  $ &  $  {Z\lrm{AGT}[q]} $&  $      {Z\uprm{inv}[q]} $   \\ \hline\hline
$  {Z\lrm{{Pestun-}\atop{Nekrasov}}[q]}  $ &   $1 $ & $ |1-q|\uu{-4}  $ &   $  |q|\uu{-{2\over 3}}\cc |1-q|\uu{-{8\over 3}} $  \\ \hline
$ {Z\lrm{AGT}[q]} $ & $ |1-q|\uu{+4} $          & $ 1 $& $   |q|\uu{-{2\over 3}}\cc |1-q|\uu{+{4\over 3}} $  \\ \hline
$  {Z\uprm{inv}[q]}    $ & $  |q|\uu{+{2\over 3}}\cc |1-q|\uu{+{8\over 3}} $            & $  |q|\uu{+{2\over 3}}\cc |1-q|\uu{- {4\over 3}} $& $  1 $  \\ \hline
\end{tabular}
\captionof{table}{Ratios of \bbd{S\uu 4} partition functions, in various schemes corresponding to different Euler-counterterm choices.  In a superconformally covariant
scheme, the counterterm is always holomorphic plus antiholomrphic in the complex gauge coupling, and so the ratio of the partition function in any two such schemes
is always the norm squared of a holomorphic function. }
\label{RatiosOfVariousPartitionFunctionsTable}
\end{center}
\end{table}

Together with the relations \rr{SchemeCovarianceOfBFunction}, eqs. \rr{TranslationsAmongAllPartitionFunctions} give the translations among the corresponding \bbd{B-}functions as computed from the corresponding partition functions {\it via\rm} the recipe of~\cite{Gerchkovitz:2016gxx}:
\bbb
\exp{B\uprm{inv}[q]} =  |q|\uu{-{2\over 3}}\cc |1-q|\uu{+{4\over 3}}\cc \exp{B\lrm{AGT}[q]}
= |q|\uu{-{2\over 3}}\cc |1-q|\uu{-{8\over 3}}\cc \exp{B\lrm{{Pestun-}\atop{Nekrasov}}[q]}\ ,
\xxnn
\exp{B\lrm{AGT}[q]} = |1-q|\uu {-4}\cc \exp{B\lrm{{Pestun-}\atop{Nekrasov}}[q]} = |q|\uu{+{2\over 3}}\cc |1-q|\uu{- {4\over 3}}\cc \exp{B\uprm{inv}[q]}\ ,
\xxnn
\exp{B\lrm{{Pestun-}\atop{Nekrasov}}[q]} = |1-q|\uu{+4}\cc \exp{B\lrm{AGT}[q]}
= |q|\uu{+{2\over 3}}\cc |1-q|\uu{+{8\over 3}}\cc \exp{B\uprm{inv}[q]}\ , \llsk
\een{TranslationsAmongAllExponentiatedFunctions}
or for the unexponentiated \bbd{B-}functions
\bbb
B\uprm{inv} = B\lrm{AGT} - {2\over 3}\cc {\rm log}|q| + {4\over 3}\cc {\rm log}|1-q| = B\lrm{{Pestun-}\atop{Nekrasov}} - {2\over 3}\cc {\rm log}|q| - {8\over 3}\cc {\rm log}|1-q| 
\xxnn
 B\lrm{AGT} = B\lrm{{Pestun-}\atop{Nekrasov}} - 4\cc {\rm log}|1-q|= B\uprm{inv} +  {2\over 3}\cc {\rm log}|q| - {4\over 3}\cc {\rm log}|1-q|
\xxnn
B\lrm{{Pestun-}\atop{Nekrasov}} = 
B\lrm{AGT} + 4\cc {\rm log}|1-q|
=  B\uprm{inv} + {2\over 3}\cc {\rm log}|q| + {8\over 3}\cc {\rm log}|1-q| \ .
\een{RelationsAmongUnexponentiatedBFunctions}
These relationships are summarized in table~\ref{DictionaryAmongExponentiatedBCoefficientsTable}.

 \begin{table}
\begin{center}
\begin{tabular}{ |c||c|c|c|c| } 
 \hline\hline
\backslashbox{denominator}{numerator} &  $  \exp{B\lrm{{Pestun-}\atop{Nekrasov}}[q]}  $ &  $  \exp{B\lrm{AGT}[q]} $&  $      \exp{B\uprm{inv}[q]} $   \\ \hline\hline
$  \exp{B\lrm{{Pestun-}\atop{Nekrasov}}[q]}  $ &   $1 $ & $ |1-q|\uu{-4}  $ &   $  |q|\uu{-{2\over 3}}\cc |1-q|\uu{-{8\over 3}} $  \\ \hline
$ \exp{B\lrm{AGT}[q]} $ & $ |1-q|\uu{+4} $          & $ 1 $& $   |q|\uu{-{2\over 3}}\cc |1-q|\uu{+{4\over 3}} $  \\ \hline
$  \exp{B\uprm{inv}[q]}    $ & $  |q|\uu{+{2\over 3}}\cc |1-q|\uu{+{8\over 3}} $            & $  |q|\uu{+{2\over 3}}\cc |1-q|\uu{- {4\over 3}} $& $  1 $  \\ \hline
\end{tabular}
\captionof{table}{Ratios of exponentiated \bbd{B-} functions, in various schemes corresponding to different Euler-counterterm choices for
the \bbd{S\uu 4} partition function.  The exponentiated \bbd{B-}function transforms exactly the same way as the partition function itself under
shift of the Euler-density counterterm, so the ratios of the exponentiated \bbd{B-}functions are exactly the same as the ratios of the corresponding \bbd{S\uu 4} partition functions.
}
\label{DictionaryAmongExponentiatedBCoefficientsTable}
\end{center}
\end{table}

The duality transformations of the exponentiated \bbd{B-}functions are then the same as those of the corresponding partition functions:
\bbb
q\to 1-q: \llsk \exp{B\lrm{AGT}[q]} \to \exp{B\lrm{AGT}[1-q]} = {{|1-q|\sqd}\over{|q|\sqd}}\cc \exp{B\lrm{AGT}[q]}\ ,
\xxnn
q\to {1\over q}:\llsk \exp{B\lrm{AGT}[q]}\to \exp{B\lrm{AGT}[{1\over q}]} =   \exp{B\lrm{AGT}[q]} \ .
\xxnn
q\to 1-q: \llsk \exp{B\lrm{{Pestun-}\atop{Nekrasov}}[q]}\to \exp{B\lrm{{Pestun-}\atop{Nekrasov}}[1-q]} = {{|q|\sqd}\over{|1-q|\sqd}}\cc \exp{B\lrm{{Pestun-}\atop{Nekrasov}}[q]}\ ,
\xxnn
q\to {1\over q}:\llsk \exp{B\lrm{{Pestun-}\atop{Nekrasov}}[q]} \to \exp{B\lrm{{Pestun-}\atop{Nekrasov}}[{1\over q}] }
= |q|\uu{-4}\cc \exp{B\lrm{{Pestun-}\atop{Nekrasov}}[q]} \llsk
\xxnn
q\to 1-q: \llsk \exp{B\uprm{inv}[q]}\to \exp{B\uprm{inv}[1-q]} =  \exp{B\uprm{inv}[q]}\ ,
\xxnn
q\to {1\over q}:\llsk \exp{B\uprm{inv}[q]} \to \exp{B\uprm{inv}[{1\over q}]} 
=  \exp{B\uprm{inv}[q]} \llsk
\een{DualityTransformationsOfExponentiatedBFunctions}
We summarize these transformations in table~\ref{DualityTransformationsOfExponentiatedBCoefficientTable}.

 \begin{table}
\begin{center}
\begin{tabular}{ |c||c|c|c|c| } 
 \hline\hline
\backslashbox{transformation}{object} &  $ \exp{B\uprm{inv}[q]} $ &  $  \exp{B\lrm{AGT}[q]} $&  $   \exp{B\lrm{{Pestun-}\atop{Nekrasov}}[q]}$   \\ \hline\hline
$q\to 1-q$ &   $ \exp{B\uprm{inv}[q]} $ & $ {{|1-q|\sqd}\over{|q|\sqd}}\cc  \exp{B\lrm{AGT}[q]} $ &   $ {{|q|\sqd}\over{|1-q|\sqd}}\cc  \exp{B\lrm{{Pestun-}\atop{Nekrasov}}[q]}$  \\ \hline
$q\to {1\over q}$ & $ \exp{B\uprm{inv}[q]} $            & $  \exp{B\lrm{AGT}[q]} $& $   |q|\uu{-4}\cc \exp{B\lrm{{Pestun-}\atop{Nekrasov}}[q]} $  \\ \hline
\end{tabular}
\captionof{table}{Duality transformations of the (exponentiated) \bbd{B-}coefficient, in various schemes corresponding to different Euler-counterterm choices
for the \bbd{S\uu 4} partition function. }
\label{DualityTransformationsOfExponentiatedBCoefficientTable}
\end{center}
\end{table}

In the next section we shall use the relations \rr{TranslationsAmongAllExponentiatedFunctions},\rr{RelationsAmongUnexponentiatedBFunctions} and the
transformations \rr{DualityTransformationsOfExponentiatedBFunctions} to completely determine the physically correct solution to the PDE \rr{PDEForBFunction} satisfied by
the \bbd{B-}function in any specific scheme, up to an overall coupling-independent constant, which we shall fix in sec.~\ref{DerivationOfConstantCoefficientBoldfaceC} using another set of considerations. 

\section{Explicit duality-covariant solution for the \bbd{B-}function  up to an overall coefficient \bbd{{\bf C}}}\label{SolutionForBFunctionUpToOverallCoefficient}
In this section we will solve for PDE \rr{PDEForBFunction} for the B-function, using duality to find the physically correct solution which will be uniquely determined up to an
overall, coupling-independent constant.

\subsection{General solution to the PDE for the \bbd{B-}function}

The PDE, which we recap here, is
\bbb
 \pp\bar{\pp} (B - A) = 0\ .
\een{PDEForBFunctionRecap}
This equation holds with respect to any choice of holomorphic coupling coordinate, and for any choice of scheme for the Euler counterterm.  Since any difference between
possible choices for those conventions changes B and A by additive holomorphic plus antiholomorphic terms, and rescales the LHS overall, the validity of the equation is
unaffected by any change in those choices.  The general solution is
\bbb
\exp{B} = {\cal H}\times \exp{A}\ ,
\eee
where \bbd{{\cal H}} is a positive definite, locally holomorphically factorized function of \bbd{\s} and \bbd{\bar{\s}} we have not yet determined.

Since \bbd{A} has a simple and manifestly duality-covariant form when expressed in terms of the infrared coupling \bbd{\s}, we will work in terms of that complex coupling parameter,
on which the \bbd{SL(2,\IZ)} duality transformations act in the familiar way as fractional linear transformations with integer coefficients:
\bbb
\s\to {{a\s + b}\over{c\s + d}}\ , \llsk a,b,c,d\in \IZ, \llsk ad-bc = 1\ .
\eee
In this coordinate frame, the expression for \bbd{A \equiv A\ls\s} was proven in~\cite{Hellerman:2020sqj} to be given by
\bbb
\exp{A\ls\s} = {1\over{16\cc [{\rm Im}(\s)]\sqd}}\ ,
\eee
which satisfies the Liouville equation \rr{PDEForAFunction} derived from the recursion relations at leading order in \bbd{n}.

So the expression for \bbd{\exp{B}} is given by
\bbb
\exp{B} = {{{\cal H[\s, \bar{\s}]}}\over{16\cc [{\rm Im}(\s)]\sqd}}\ ,
\eee
where \bbd{{\cal H}} is a positive definite, locally holomorphically factorized function of \bbd{\s} and \bbd{\bar{\s}}.

\subsection{Particular solutions for \bbd{B} in various Euler-counterterm schemes}

The choice of \bbd{{\cal H}} depends on the scheme in which \bbd{Z} and \bbd{B} are defined, with duality allowing us to narrow down the correct choice in any scheme up to
an overall constant.  Using duality is simplest in the infrequently\footnote{As far as we can tell actually never, until the present article.}-used but well-defined duality-invariant
scheme.  If the \bbd{B-}function on the LHS is taken to be the \bbd{B-}function  \bbd{B\lrm{inv}} defined in the duality-invariant scheme, the RHS must be duality-invariant
as well.  The function \bbd{[{\rm Im}(\s)]} transforms under \bbd{SL(2,\IZ)} as a nonholomorphic modular form of weight \bbd{(1,1)}, so we must choose \bbd{{\cal H}} to be
a locally holomorphically factorized, positive definite, nonholomorphic modular form of weights \bbd{(-2,-2)}.  

The Dedekind eta function \bbd{\eta(\s)} transforms as a modular form of weight \bbd{+\hh}, up to a phase:
\bbb
\eta(\s + 1) = \exp{{{\pi i}\over{12}}}\cc \eta(\s)\ , \llsk
\eta(- {1\over\s}) = (- i \s)\uu{+\hh}\cc \eta(\s)\ ,
\een{EtaFunctionTransformationsFromNotes}
and is nonvanishing everywhere in the upper half-plane, so \bbd{|\eta(\s)|} is a
positive definite, locally holomorphically factorized nonholomorphic modular form  of weight \bbd{(+{1\over 4}, +{1\over 4})} and its
\bbd{-8} power is a positive definite, locally holomorphically factorized nonholomorphic modular form of weight \bbd{(-2,-2)}.
By the uniqueness theorem~\ref{UniquenessTheoremProof}, such an object satisfying a weak version of weak-coupling asymptotics (growing no faster than exponentially
with \bbd{{\rm Im}(\s)} at infinity) is unique up to an overall constant factor.  So we find that the \bbd{B-}function \bbd{B\uprm{inv}} in the duality-invariant scheme, is given by
\bbb
\exp{B\uprm{inv}} = {{{\bf C}}\over{16\cc|\eta(\s)|\uu 8\cc [{\rm Im}(\s)]\sqd}}\ ,
\een{DualityInvariantBFunctionSolution}
where \(\mathbf{C}\) is a constant that we will compute in the following.

From the solution \rr{DualityInvariantBFunctionSolution} and the scheme relations \rr{TranslationsAmongAllExponentiatedFunctions},
we can immediately write down the \bbd{B-}functions as defined in the more widely-used AGT and Pestun-Nekrasov schemes:

\bbb
\exp{B\lrm{AGT}[q]}  = |q|\uu{+{2\over 3}}\cc |1-q|\uu{+{8\over 3}}\cc \exp{B\uprm{inv}[q]} =  {{{\bf C}\cc |\l(\s)|\uu{+{2\over 3}}\cc |1-\l(\s)|\uu{- {4\over 3}}}\over{16\cc|\eta(\s)|\uu 8\cc [{\rm Im}(\s)]\sqd}}\ .
\xxnn
\exp{B\lrm{{Pestun-}\atop{Nekrasov}}[q]} 
= |q|\uu{+{2\over 3}}\cc |1-q|\uu{+{8\over 3}}\cc \exp{B\uprm{inv}[q]} =  {{{\bf C}\cc |\l(\s)|\uu{+{2\over 3}}\cc |1-\l(\s)|\uu{+{8\over 3}}}\over{16\cc|\eta(\s)|\uu 8\cc [{\rm Im}(\s)]\sqd}}\ ,
\eee
where \bbd{\l(\s)} is the modular lambda function and we have used the relationship \rr{RelationBetweenUVAndIRCouplings} between the IR coupling \bbd{\s} and the UV coupling \bbd{\t}.

\subsection{Expression for the scheme-independent combination \bbd{\tilde{B}} and its expansion at weak coupling}

Finally, we can write down the fully duality-invariant and scheme-invariant combination \bbd{\exp{\tilde{B}}:}
\bbb
\exp{\tilde{B}} \equiv \exp{B} / Z = \exp{B\uprm{inv}} / Z\uprm{inv} = \exp{B\lrm{AGT}} / Z\lrm{AGT}  = \exp{B\lrm{{Pestun-}\atop{Nekrasov}}} / Z\lrm{{Pestun-}\atop{Nekrasov}}
\xxnn
=   {{{\bf C}\cc |\l(\s)|\uu{+{2\over 3}}\cc |1-\l(\s)|\uu{- {4\over 3}}}\over{16\cc|\eta(\s)|\uu 8\cc [{\rm In}(\s)]\sqd}}\cc \bigg [ \cc Z\lrm{AGT} \cc \bigg ]\uu{-1}
= {{{\bf C}\cc |\l(\s)|\uu{+{2\over 3}}\cc |1-\l(\s)|\uu{+ {8\over 3}}}\over{16\cc|\eta(\s)|\uu 8\cc [{\rm In}(\s)]\sqd}}\cc \bigg [ \cc Z\lrm{{Pestun-}\atop{Nekrasov}}\cc \bigg ]\uu{-1}
\een{FinalSolutionForExpTildeB}

\subsection{Weak-coupling expansion of \bbd{\tilde{B}}}

Now we would like to expand the scheme-invariant quantity \bbd{\tilde{B}} at weak coupling \bbd{\t}.

Since \bbd{\tilde{B}} is expressed above in terms of the IR coupling \bbd{\s}, we must first give the weak-coupling expansion of the nonholomorphic modular-invariant object \bbd{{1\over{|\eta(\s)|\uu 8\cc [{\rm Im}(\s)]\sqd }} } appearing in \bbd{\tilde{B},} in terms
of the UV coupling
\bbd{\t}.  We refer to the formulae from sec.~\ref{WeakCouplingRelationshipBetweenTauAndSigma} of the Appendix.
From Eq. \rr{ResultForSixteenTimesExponentiatedACoefficient} in the Appendix, which we recap here, we know
\bbb
{1\over{|\eta(\s)|\uu 8\cc [{\rm Im}(\s)]\sqd }} =  {{2\uu{+{2\over 3}}\cc |q|\uu{-{2\over 3}}}\over{  [{\rm Im}(\t) + {2\over \pi}\cc {\tt log}[2]]\sqd  }} + O ({{ |q|\uu{+{1\over 3}} }\over{ [{\rm Im}[\t]]\sqd}})\ .
\een{ResultForSixteenTimesExponentiatedACoefficientRecap}

We also have
\bbb
| \l(\s)|\uu{+{2\over 3}}\cc |1-\l(\s)|\uu{{\rm anything}} = |q|\uu{+{2\over 3}} + O(|q|\uu{+{5\over 3}})\ ,
\eee
so
\bbb
 \exp{B\lrm{AGT}}  = {{\bf C}\over{16}}\cc {{2\uu{+{2\over 3}}}\over{   [{\rm Im}(\t) + {2\over \pi}\cc {\tt log}[2]]\sqd  }}+ O ({{ |q| }\over{ [{\rm Im}[\t]]\sqd}})\ ,
 \eee
 and also
 \bbb
\exp{B\lrm{{Pestun-}\atop{Nekrasov} }} = {{\bf C}\over{16}}\cc {{2\uu{+{2\over 3}}}\over{   [{\rm Im}(\t) + {2\over \pi}\cc {\tt log}[2]]\sqd  }}+ O ({{ |q| }\over{ [{\rm Im}[\t]]\sqd}})\ ,
\eee
with different \bbd{ O ({{ |q| }\over{ [{\rm Im}[\t]]\sqd}})} terms.
So we have
\bbb
 \exp{B\lrm{AGT}} ,\cc\cc\cc
 \exp{B\lrm{{Pestun-}\atop{Nekrasov} }}  =  {{2\uu{-{{10}\over 3}}\cc {\bf C}}\over{   [{\rm Im}(\t) + {2\over \pi}\cc {\tt log}[2]]\sqd  }}+ O ({{ |q| }\over{ [{\rm Im}[\t]]\sqd}})
\eee

Finally, the weak-coupling asymptotics of the partition function itself in the Pestun-Nekrasov scheme (see {\it e.g.\rm} eq. \aleq{3.23} of~\cite{Gerchkovitz:2016gxx}, also given in eq. \rr{ExplicitFormulaForSpherePartitionFunctionInAppendix} of sec. \ref{ZS4WeakCouplingExpansionSec}
of our Appendix) and the AGT scheme is also the same up to
instanton corrections,
\bbb
Z\lrm{{Pestun-}\atop{Nekrasov} }  = |1 - q|\uu 4\cc Z\lrm{AGT} =    \frac{1}{4 \pi (\text{Im} \tau)^{3/2}}  \left[1 -  
   \frac{45 \zeta(3)}{16 \pi^2 (\text{In} \tau)^2} + \frac{525 \zeta(5)}{64 \pi^3 (\text{Im} \tau)^3}  + O([{\rm Im}[\t]\uu{-4} ])\right] \ .
\een{ExplicitFormulaForSpherePartitionFunctionOUTSIDEOfAppendix}
So the scheme-independent quotient \bbd{\exp{\tilde{B}} = \exp{B\lrm{AGT}} / Z\lrm{AGT}  = \exp{B\lrm{{Pestun-}\atop{Nekrasov}}[q]} / Z\lrm{{Pestun-}\atop{Nekrasov}}} is
\bbb
\exp{\tilde{B}}=   {{\exp{\tilde{b}} \cc \cc  [{\rm Im}(\t) ]\uu{{3\over 2}}}\over{   [{\rm Im}(\t) + {2\over \pi}\cc {\tt log}[2]]\sqd }}\left[1 +  
   \frac{45 \zeta(3)}{16 \pi^2 (\text{Im} \tau)^2} - \frac{525 \zeta(5)}{64 \pi^3 (\text{Im} \tau)^3}  \right] + O([{\rm Im}[\t]\uu{-{9\over 2}})
\eee
where we are defining
\bbb
\exp{\tilde{b}} \equiv (4\pi)\cc 2\uu{-{{10}\over 3}}\cc {\bf C} = 2\uu{- {4\over 3}}\cc \pi\cc {\bf C}\ .
\eee

So, duality and the recursion relations have totally fixed the functional form of \bbd{\tilde{B}} up to a single coupling-independent constant \bbd{{\bf C}}:
\bbb
\exp{\tilde{B}}=   {{\exp{\tilde{b}} \cc \cc  [{\rm Im}(\t) ]\uu{{3\over 2}}}\over{   [{\rm Im}(\t) + {2\over \pi}\cc {\tt log}[2]]\sqd }}\left[1 +  
   \frac{45 \zeta(3)}{16 \pi^2 (\text{Im} \tau)^2} - \frac{525 \zeta(5)}{64 \pi^3 (\text{Im} \tau)^3}  \right] + O([{\rm Im}[\t]\uu{-{9\over 2}})
\xxnn
{\rm with~~}\exp{\tilde{b}}  = 2\uu{- {4\over 3}}\cc \pi\cc {\bf C}\ .
\een{FinalResultForSchemeIndependentExpTildeBInTermsOfUnknownConstantBoldfaceC}

\section{Determination of the constant \bbd{{\bf C}} by matching with double-scaled perturbation theory at one loop at strong coupling}\label{DerivationOfConstantCoefficientBoldfaceC}

\subsection{The double-scaling limit versus the fixed-coupling, large-charge limit}\label{GeneralDoubleScalingDiscussion}

In order to fix this final constant \bbd{{\bf C}} in
the solution \rr{FinalSolutionForExpTildeB}, we must use perturbation theory as a boundary condition, as we did to
determine an unfixed constant in the expression
for the \bbd{A-}function in~\cite{Hellerman:2020sqj}.  For the
unfixed constant in the \bbd{B-}coefficient,
it is easiest to make contact with double-scaled perturbation theory in the limit of strong double-scaled coupling\footnote{In order to avoid a confusion of notations,
in this section we shall always use \bbd{\ddsc} to
refer to the double-scaled coupling constant and never
to the modular lambda function.  Also note that the normalization
convention used here for \bbd{\ddsc} is the same as that of
one of the two normalization conventions used for \bbd{\ddsc}
in~\cite{Grassi:2019txd}, which differs by a factor of \bbd{(4\pi)}
from the \emm{other} normalization convention used for \bbd{\ddsc} in 
\cite{Grassi:2019txd}, and differs by a factor of \bbd{(4\pi)\sqd}
from the normalization convention used in~\cite{Bourget:2018obm}. 
See sec.~\ref{ThreeLambdaNormalizationsWhyWhyWhyWhyWhyGodWhy}
of our Appendix.}
\bbb
\ddsc \equiv {n\over{4\pi\cc {\rm Im}[\t]}}
\een{DoubleScaledCouplingDef}

We consider the solution for correlation functions in the double-scaling limit in which \bbd{n} is
taken to infinity while \bbd{\ddsc} is held fixed.  This solution
was derived by Grassi, Komargodski, and Tizzano.  Their solution extends earlier work
\cite{Bourget:2018obm} that found the double-scaling limit of the correlators to several orders in \bbd{\ddsc} at small \bbd{\ddsc}.  Ref.~\cite{Grassi:2019txd} extended this to all orders
in \bbd{\ddsc} using matrix model methods.

The double-scaling limit of~\cite{Bourget:2018obm, Grassi:2019txd} is an example of a more general strategy for studying systems at large quantum number when
there is an additional quantum loop-suppressing parameter in the system.  This sort of limit has also been studied in less- and non-supersymmetric systems in double-scaling
limits involving large \bbd{N,}  and small \bbd{\e} in \bbd{4-\e} dimensions~\cite{Sharon:2020mjs, Arias-Tamargo:2019xld, Arias-Tamargo:2019kfr, Arias-Tamargo:2020fow, Alvarez-Gaume:2019biu, Watanabe:2019pdh, Badel:2019oxl, Badel:2019khk, Giombi:2020enj, Dondi:2021buw}. In all these cases
the double-scaling regime contains more complexity and more degrees of freedom than the fixed-coupling, large-charge regime, as the non-Nambu-Goldstone degrees of
freedom which are integrated out in the fixed-coupling-large-charge limits, stay at finite mass in the double-scaling limits.  

Qualitatively, the two
main features of the fixed-coupling large-charge limit are the parametric accuracy of the saddle-point approximation at large charge, and the use of effective field theory
to eliminate the non-Nambu-Goldstone degrees of freedom; the double-scaling limit is a 
conceptually interesting extension of the standard large-charge picture in that it disaggregates those
two ingredients, taking advantage of the first without the loss of information involved in the second.  In some qualitative sense one expects
the double-scaling limit to contain the fixed-coupling large-charge limit as a \emm{sub-} limit, in which the double-scaled coupling is 
taken strong and the non-NG modes become infinitely heavy.  This is not quite precise and indeed it is \emm{not} generally correct that the
large \bbd{\ddsc} limit of the double-scaling limit of a given quantity, is equal to the fixed-\bbd{\t,} large-charge limit of the same quantity.  However
in some broader sense the intuition is valid, and the information about the large-charge, fixed-coupling limit of certain quantities is contained in
the large-\bbd{\ddsc} limit of the double-scaling limit of the same quantities in a simple way.  In this section we will exploit that
relationship to calculate the coefficient \bbd{{\bf C}} which is essentially a threshold correction to the Euler density term in the action, coming from integrating out the massive hypers
and W-bosons.  

In order to fix the final undetermined constant \bbd{{\bf C}} of the preceding section, we will need to show exactly how the double-scaled expression of~\cite{Grassi:2019txd} for
the correlation function is related to
the massive macroscopic propagation contribution \bbd{q\ll n\uprm{mmp}} to the log of the correlator, and to its double-scaling limit
\bbd{ F\ls 0[\ddsc] \equiv \DDSL q\ll n\uprm{mmp}.}

\subsection{The MMP factor of the partition function}\label{MMPFactorDiscussion}

Now we will go beyond the EFT approximation to \bbd{Z\ll n}.  Thought of in terms of virtual particles, the EFT factor of \bbd{Z\ll n} contains only diagrams with explicit
contributions of
particles that are massless on the Coulomb branch.  It also contains \emm{implicit} contributions of massive particles whose trajectories are microscopic, parametrically
smaller than the infrared scale, so that their contributions can be absorbed into renormalized couplings.  So the leading contribution beyond the EFT factor should come from
processes containing at least one macroscopic worldline of a massive particle (a similar qualitative behavior is observed in the large-\(N\), large-charge double-scaling limit of the Wilson--Fisher point~\cite{Dondi:2021buw}):
\bbb
Z\ll n \equiv \exp{q\ll n} = 2\uu{- 4n}\cc Z\ll {S\uu 4}\cc G\ll{2n} = Z\ll n\uprm{EFT}\times Z\ll n\uprm{mmp}\ ,
\een{FactorizationThatIsEquivalentToADefinitionOfTheMMPFactor}
where \bbd{Z\ll n\uprm{EFT} \equiv e\uu{q\ll n\uprm{EFT}}} is the EFT factor
\bbb
Z\ll n\uprm{EFT} = \exp{n A + B}\times \G(2n + {5\over 2})\ ,
\eee
and \bbd{Z\ll n\uprm{mmp}} is a partition function exponentiating the sum over all connected configurations containing at least one massive macroscopic worldline instanton.
So
\bbb
Z\ll n\uprm{mmp} = \exp{ q\ll n\uprm{mmp}}
\eee
and \bbd{q\ll n\uprm{mmp}} is the sum over all connected configurations containing at least one macroscopic trajectory of a massive particle. 

Since we now have a closed form for the EFT factor, we can use
eq. \rr{FactorizationThatIsEquivalentToADefinitionOfTheMMPFactor}
as a \emm{definition} of the MMP factor, %
\bbb
Z\ll n\uprm{mmp} \equiv {{Z\ll n}\over{Z\ll n\uprm{EFT}}} = \exp{q\ll n\uprm{mmp}}\ ,
\xxnn
q\ll n\uprm{mmp} = q\ll n - q\ll n\uprm{EFT} = q\ll n - n A - B - {\rm log}\bigg [ \cc \G(2n + {5\over 2})\cc \bigg ]
\een{FactorizationThatIsEquivalentToADefinitionOfTheMMPFactor}
It was observed in~\cite{Hellerman:2018xpi, Hellerman:2020sqj} that the MMP
contribution \bbd{q\ll n\uprm{mmp}} has a well-defined double-scaling
limit in the sense of\footnote{Although we caution the reader that the normalization of the double-scaling parameter in~\cite{Bourget:2018obm} differs by powers of \bbd{4\pi} from each of the two distinct normalizations given for the same parameter in~\cite{Grassi:2019txd}.
See sec.~\ref{ThreeLambdaNormalizationsWhyWhyWhyWhyWhyGodWhy} of
the Appendix for a discussion of this lovely cornucopia of diverse normalization conventions.}\cite{Bourget:2018obm, Grassi:2019txd}.  (Also in the same sense of the double-scaling limits taken in 
several nonsupersymmetric examples recently~\cite{Sharon:2020mjs, Arias-Tamargo:2019xld, Arias-Tamargo:2019kfr, Arias-Tamargo:2020fow, Alvarez-Gaume:2019biu, Watanabe:2019pdh, Badel:2019oxl, Badel:2019khk, Giombi:2020enj})

 The existence
of a double-scaling limit for \bbd{q\ll n\uprm{mmp}} was theoretically
motivated from its interpretation in terms of macroscopic massive
propagation, with the action of a massive particle propagating
over a distance of order the infrared scale, being proportional to
\bbd{\ddsc\uu{+\hh}}.  In section~\ref{WorldlineInstantonActionSection} we will
make this argument more precise, giving a definite
physical prediction for the value of the worldline action.

The hypothesis of a well-defined double-scaling
limit for \bbd{q\ll n\uprm{mmp}} was borne out numerically
to high precision by numerical calculations of double-differences \bbd{q\ll {n+1}\uprm{mmp} - 2 q\ll n\uprm{mmp} + q\ll{n-1}\uprm{mmp}} in~\cite{Hellerman:2018xpi}, and
single differences \bbd{q\ll {n+1}\uprm{mmp} - q\ll n\uprm{mmp}}, in\cite{Hellerman:2020sqj}, from correlators computed with the
prescription of~\cite{Gerchkovitz:2016gxx}.  
So based on theoretical arguments well-supported by numerical
evidence, we can infer that
\bbb
\DDSL q\ll n\uprm{mmp} = ({\rm finite})  \equiv F\ls 0[\ddsc]\ ,
\een{ExistenceOfDoubleScalingLimitOfMMPContribution}
with the fixed-\bbd{\ddsc} large-\bbd{n} corrections coming in
an asymptotic series
\bbb
q\ll n\uprm{mmp} = \sum\ll{k\geq 0}n\uu{-k}\cc F\ls k[\ddsc]\ .
\eee

The worldline-instanton interpretation tells us more
than just the \emm{existence} of a limit \bbd{F\ls 0[\ddsc]}; it 
also tells us about the strong-coupling behavior of \bbd{F\ls 0[\ddsc]}.
Since we expect the large-\bbd{\ddsc} behavior of \bbd{F\ls 0[\ddsc]} 
to be given by the macroscopic propagation of a particle with
mass proportional to \bbd{\l\uu{+\hh},} we expect that 
\bbd{F\ls 0[\ddsc]} must vanish exponentially with \bbd{\l\uu{+\hh}}
in the limit \bbd{n\to\infty,{\rm Im}[\t]\to\infty} with \bbd{\l} fixed.

The existence of the limit \rr{ExistenceOfDoubleScalingLimitOfMMPContribution} is going to be useful to us for purposes of matching with~\cite{Grassi:2019txd} to find the value of \bbd{\tilde{b}}, because on \emm{physical grounds}
we know precisely the behavior of \bbd{F\ls 0[\l]} namely that \bbd{{\rm log}[F\ls 0[\ddsc]]} should go as \bbd{\propto \ddsc\uu{+\hh}} with a negative coefficient, so that \bbd{\DDSL\cc q\ll n =F\ls 0[\ddsc]} is
exponentially small at large \bbd{\ddsc.}

\subsection{Matching our \bbd{F\ls 0[\ddsc]} with the term \bbd{F\uprm{inst}} in~\cite{Grassi:2019txd}'s \bbd{\Delta C\ll 1}}\label{DetailsOfMatchingWithStrongDoubleScalingLimit}

Ref.~\cite{Grassi:2019txd} defines their double-scaling function, as~\cite{Bourget:2018obm} does, as a ratio between the full correlator and the \bbd{{\cal N} = 4} correlator. 
In ref.~\cite{Bourget:2018obm} the authors defined an object \bbd{F} which we shall superscript as\footnote{Because there are several objects denoted as "F" in the various literature
whose conventions we compare in the present paper , we need to avoid ambiguity by specifying the one we mean.} the "Oviedo quotient"
\bbd{F\uprm{Oviedo}} (referred to as \bbd{\Delta G} in~\cite{Grassi:2019txd}  ) as the ratio
\bbb
F\uprm{Oviedo} \equiv G\ll{2n} / G\ll{2n}\uu{{\cal N} = 4}
\een{OviedoRatioDefinition}
Then~\cite{Grassi:2019txd} takes the log, take the double-scaling limit, and call what comes out, \bbd{\Delta C\ll 1\uprm{\smgkt}[\ddsc]:}
\bbb
\bbsk
\Delta C\ll 1\uprm{\smgkt}[\ddsc] \equiv \cc \DDSL\cc {\tt log} \big [ \cc F\uprm{Oviedo} \cc \big ] =
\DDSL \cc {\tt log} \big [ \cc G\ll{2n} / G\ll{2n}\uu{{\cal N} = 4} \cc \big ]\llsk
\een{IdentificationOfGKTsDoubleScalingFunctionAsDoubleScalingLimitOfOviedoRatio}

Using the definitions \rr{ZLowerNDefinition}, \rr{BTildeDefinition}, and \rr{OviedoRatioDefinition}, 
\bbb
\bbsk
Z\uprm{mmp}\ll n = \exp{- n A - B} \cc {1\over{\G(2n + {5\over 2})}}\cc Z\ll n
=  \exp{- n A - \tilde{B}} \cc {1\over{\G(2n + {5\over 2})}}\cc 2\uu{- 4n}\cc G\ll {2n}\uprm{{\cal N} = 4} \cc F\uprm{Oviedo} \llsk
\eee
In this section, we will always be defining correlators
in the \bbd{\t} coordinate,
\bbb
Z\uprm{mmp}\ll {n} = \exp{- n A\ls{\t} - \tilde{B}} \cc {1\over{\G(2n + {5\over 2})}}\cc 2\uu{- 4n}\cc G\ll {2n}\uprm{{\cal N} = 4} \cc \ordinary{F\uprm{Oviedo}\ls\t}\ ,
\eee
but we will leave the subscript \bbd{{}\ls\t} implicit.

What we need to do now, is establish that this quantity on the RHS has a double-scaling limit in our approach.  Since
\bbd{ \ordinary{F\uprm{Oviedo}}}has a double-scaling limit, 
this is equivalent to the quantity 
\bbb
\PREFAK \equiv \exp{- n A\ls{\t} - \tilde{B}} \cc {1\over{\G(2n + {5\over 2})}}\cc 2\uu{- 4n}\cc G\ll {2n}\uprm{{\cal N} = 4}
\eee having a
double-scaling limit.  The quantity \bbd{\PREFAK}
should be thought of simply as the factor translating
the Oviedo ratio \bbd{F\uprm{Oviedo}} into the MMP function:
\bbb
Z\uprm{mmp}\ll n[\t] =\PREFAK \cc \ordinary{F\uprm{Oviedo}}[n,\t]
\een{MMPToOviedoQuotientInTermsOfFactor}

We recall that the expression for the \bbd{{\cal N} = 4} correlators is
\bbb
G\ll {2n}\uprm{{\cal N} = 4} =[{\rm Im}[\t]]\uu{-2n}\cc (2n+1)! =[{\rm Im}[\t]]\uu{-2n}\cc \G(2n+2)
\eee
We need to write the scalings of
the various other objects at large \bbd{n,} large \bbd{{\rm Im}[\t]} and fixed \bbd{\ddsc}.  And, in order to get things right, we are going to have to keep both leading and subleading terms, because we want to check
that the order \bbd{n\uu {+1}} terms in the double-scaling limit cancel as they need to do, but we also want to keep track of the order \bbd{n\uu 0} terms because these are
the things we want to match with our \bbd{F\ls 0[\ddsc]}.

In sec.~\ref{DoubleScalingExpansionForPrefactor} of the Appendix, 
we work out the large-\bbd{[{\rm Im}(\t)],~}large-\bbd{n} limit of the prefactor in eq. \rr{MMPToOviedoQuotientInTermsOfFactor} with \bbd{\ddsc} held fixed.  In eq. 
 \rr{FinalResultForDoubleScalingLimitOfPrefactorAsLimit} we find:
\bbb
\DDSL\cc \PREFAK = {{\exp{-\tilde{b}}}\over{\sqrt{8\pi \ddsc}}}
\cc \exp{16\cc {\rm log}(2)\cc\ddsc } \ .
\een{FinalResultForDoubleScalingLimitOfPrefactorAsLimitRecap}

So then going back to eq. \rr{MMPToOviedoQuotientInTermsOfFactor}, %
and using the expression \rr{FinalResultForDoubleScalingLimitOfPrefactorAsLimitRecap} for
the double-scaling limit of \bbd{\PREFAK} and
the definition \rr{IdentificationOfGKTsDoubleScalingFunctionAsDoubleScalingLimitOfOviedoRatio} of \mgkt's function \bbd{\Delta C\ll 1\uprm{\smgkt}[\l]} as the double-scaling limit of the Oviedo
ratio \bbd{F\uprm{Oviedo},} we have
\bbb
\DDSL Z\uprm{mmp}\ll n
= {{\exp{-\tilde{b}}}\over{\sqrt{8\pi \ddsc}}}\cc \cc \exp{16\cc {\rm log}(2)\cc\ddsc } \cc \exp{\Delta C\ll 1\uprm{\smgkt}}
\eee
Taking the logarithm
\bbb
\DDSL q\ll n\uprm{mmp}=-\tilde{b} - \hh \cc {\rm log}[8\pi \ddsc]  + 
16\cc {\rm log}(2)\cc\ddsc + \Delta C\ll 1\uprm{\smgkt}
\eee
At this point it is \emm{not at all obvious} there is a consistent
matching at all. We know on physical grounds based on
the WLI interpretation (which is strongly supported by numerical data~\cite{Hellerman:2018xpi, Hellerman:2020sqj})
 that \bbd{F\ls 0[\ddsc]} goes to zero exponentially at strong coupling (large \bbd{\ddsc}). Certainly the first three terms
on the RHS do not do that.  So our only hope is that those three would be canceled exactly by \bbd{\Delta C\ll 1\uprm{\smgkt}}, and then \bbd{\Delta C\ll 1\uprm{\smgkt}} would
add no other power law terms, or any terms larger than the size of the exponential of the negative of the worldline instanton action.  Indeed we will now see this is
precisely the case.

In eq. \bbd{4.20} of~\cite{Grassi:2019txd}, the expression for \bbd{\Delta C\ll 1\uprm{\smgkt}} is given.  Defining \bbd{\g\lrm G} to be the Glaisher constant \bbd{\g\lrm G = 1.28243}, they write
\bbb
\Delta C\ll 1\uprm{\smgkt} = 12\cc{\rm log}[\g\lrm G]-1 -{{{\rm log}(2)}\over 3} - 16\cc {\rm \ddsc}\cc {\rm log}[2]  + \hh\cc {\rm log}[\ddsc] + F\uprm{inst}[\ddsc]
\eee
where \bbd{F\uprm{inst}} is a worldline instanton piece going to zero as \bbd{\ddsc\uu{+{1\over 4}}\cc \exp{-  4\pi\sqrt{\ddsc}}}.

So we have
\bbb
\DDSL q\uprm{mmp}\ll n = \exp{- f\ll 0[\ddsc]} = -\tilde{b} - \hh \cc {\rm log}[8\pi \ddsc]  + 
16\cc {\rm log}(2)\cc\ddsc + \Delta C\ll 1\uprm{\smgkt}
\xxnn
= -\tilde{b} - \hh \cc {\rm log}[8\pi]  +12\cc{\rm log}[\g\lrm G]-1 -{{{\rm log}(2)}\over 3}  + F\uprm{inst}[\ddsc]
\eee

We know the LHS goes to zero exponentially at strong coupling, so total of the RHS must too; also, the WLI piece \bbd{F\uprm{inst}[\ddsc]} of~\cite{Grassi:2019txd}'s double scaling function
goes to zero at strong coupling as well.  Fortunately, the log and linear terms which would have made a consistent limit impossible, have canceled exactly, and in
the strong coupling limit we get the equalities:
\bbb
F\ls 0[\ddsc] \equiv \DDSL q\uprm{mmp}\ll n = F\uprm{inst}[\ddsc]
\cc \biggl |\ll{\hbox{ref.~\cite{Grassi:2019txd} }}
\een{MatchingTheDoubleScalingLimitOfOurMMPFunctionWithTheWorldlineInstantonFunctionInGKT}
and
\bbb
\exp{\tilde{b}} = \g\lrm G\uu{+12}\cc e\uu{-1}\cc 2\uu{-{{11}\over 6}}\pi\uu{-\hh} 
\een{PredictionForLittleTildeB}

Then using the definition of \bbd{\tilde{b},} \rr{FinalResultForSchemeIndependentExpTildeBInTermsOfUnknownConstantBoldfaceC}, which we recap here,
\bbb
\exp{\tilde{b}}  = 2\uu{- {4\over 3}}\cc \pi\cc {\bf C}
\ ,
\een{FinalResultForSchemeIndependentExpTildeBInTermsOfUnknownConstantBoldfaceCRecap}
we have
\bbb
{\bf C} = 2\uu{+{4\over 3}}\cc \pi\uu{-1}\cc \exp{\tilde{b}}
= 
\g\lrm G\uu{+12}\cc e\uu{-1}\cc 2\uu{-{1\over 2}}\pi\uu{-{3\over 2}}\ ,
\een{ExplicitFormulaForBoldC}
and
\bbb
{{{\bf C}\over{16}}} = \g\lrm G\uu{+12}\cc e\uu{-1}\cc 2\uu{-{9\over 2}}\pi\uu{-{3\over 2}}\ ,
\een{ExplicitFormulaForBoldCOver16}
so the full \bbd{\tilde{B}-}function \rr{FinalSolutionForExpTildeB} is given by
\bbb
\exp{\tilde{B}} 
=   \g\lrm G\uu{+12}\cc e\uu{-1}\cc 2\uu{-{9\over 2}}\pi\uu{-{3\over 2}}\cc
{{ |\l(\s)|\uu{+{2\over 3}}\cc |1-\l(\s)|\uu{- {4\over 3}}}\over{|\eta(\s)|\uu 8\cc [{\rm Im}(\s)]\sqd}}\cc \bigg [ \cc Z\lrm{AGT} \cc \bigg ]\uu{-1}
\eee
or equivalently
\bbb
\exp{\tilde{B}}  = \g\lrm G\uu{+12}\cc e\uu{-1}\cc 2\uu{-{9\over 2}}\pi\uu{-{3\over 2}}\cc
{{ |\l(\s)|\uu{+{2\over 3}}\cc |1-\l(\s)|\uu{+ {8\over 3}}}\over{|\eta(\s)|\uu 8\cc [{\rm Im}(\s)]\sqd}}\cc \bigg [ \cc Z\lrm{{Pestun-}\atop{Nekrasov}}\cc \bigg ]\uu{-1}
\een{FinalSolutionForExpTildeB}
For concrete expressions for \bbd{Z\lrm{{Pestun-}\atop{Nekrasov}}} and \bbd{Z\lrm{AGT}} at weak coupling, see section~\ref{ZS4WeakCouplingExpansionSec} of
the Appendix. 

Combining \rr{ExplicitFormulaForBoldC} with \rr{FinalResultForSchemeIndependentExpTildeBInTermsOfUnknownConstantBoldfaceC}
we have the explicit weak-cooupling expansion for \bbd{\exp{\tilde{B}}.}  We have
\bbb
\exp{\tilde{b}} = \g\lrm G\uu{+12}\cc e\uu{-1}\cc 2\uu{-{{11}\over 6}}\pi\uu{-{1\over 2}}
\een{ExplicitExpressionForExpTildeLowerCaseBWithBoldfaceCFilledIn}
and
\bbb
\bbsk
\exp{\tilde{B}}=   \g\lrm G\uu{+12}\cc e\uu{-1}\cc 2\uu{-{{11}\over 6}}\pi\uu{-{1\over 2}}\cc {{  [{\rm Im}(\t) ]\uu{{3\over 2}}}\over{   [{\rm Im}(\t) + {2\over \pi}\cc {\tt log}[2]]\sqd }}\left[1 +  
   \frac{45 \zeta(3)}{16 \pi^2 (\text{Im} \tau)^2} - \frac{525 \zeta(5)}{64 \pi^3 (\text{Im} \tau)^3}  \right] + O([{\rm Im}[\t]\uu{-{9\over 2}})
   \llsk
\een{FinalResultForSchemeIndependentExpTildeBWithBoldfaceCFilledIn}
\\
Also, the double-scaling limit \rr{FinalResultForDoubleScalingLimitOfPrefactorAsLimitRecap} of the prefactor is
\bbb
\DDSL\cc \PREFAK 
=  {{e\uu{+1}\cc 2\uu{+{1\over 3} }}\over{  \g\lrm G\uu{12} \ddsc\uu\hh }}
\cc \exp{16\cc {\rm log}(2)\cc\ddsc } 
\eee

\subsection{Summary: The exact EFT factor \bbd{Z\uprm{eft}}}

Building on previous results~\cite{Hellerman:2017sur, Hellerman:2018xpi, Hellerman:2020sqj} we have now solved completely for the form
of the correlation function of conformal SQCD to all orders in \bbd{n}
at fixed coupling \bbd{\t}; to recap the result,
\bbb
G\ll{2n} =  2\uu{+ 4n}\cc
\tilde{Z}\ll n\ ,
\xxnn
\tilde{Z}\ll n \equiv \tilde{Z}\ll n\uprm{eft}\cc Z\ll n\uprm{mmp}\ ,
\eee
where
\bbb
\tilde{Z}\ll n\uprm{eft} \equiv {1\over{Z\ll{S\uu 4}}}\cc Z\ll n\uprm{eft} =
 \exp{nA\ls\t + \tilde{B}}\cc \G(2n + {5\over 2})
\eee
with
\bbb
\exp{A\ls\t} =  (2\pi)\uu{+ 2}\cc
\big |\cc {{\l(\s)}\over{\l\pr(\s)}} \big |\sqd\cc
{1\over{16\cc [{\rm Im}[\s]]\sqd}}\ ,
\eee %
and
\bbb
\exp{\tilde{B}} =  {{{\bf C}\cc |\l(\s)|\uu{+{2\over 3}}\cc |1-\l(\s)|\uu{+ {8\over 3}}}\over{16\cc|\eta(\s)|\uu 8\cc [{\rm Im}(\s)]\sqd}}\cc \bigg [ \cc Z\lrm{{Pestun-}\atop{Nekrasov}}\cc \bigg ]\uu{-1}\ ,
\eee
where:
\bii
\item{ \bbd{\s} is the infrared coupling related
to the UV coupling by \bbd{q \equiv \exp{2\pi i \t} = \l(\s)} and \bbd{\l(\s)} is the modular lambda function,}
\item{\bbd{\eta(\s)} is the Dedekind eta function,}
\item{{\bbd{Z\lrm{{Pestun-}\atop{Nekrasov}}}} is the \bbd{S\uu 4} partition function as computed in Pestun's~\cite{Pestun:2007rz} scheme for
the one-loop localization integrand, and using Nekrasov's \bbd{U(2)} instanton partition
function\cite{Nekrasov:2002qd}}
\item{The constant \bbd{{\bf C}} is
\bbb
{\bf C} =  
\g\lrm G\uu{+12}\cc \exp{-1}\cc 2\uu{-{1\over 2}}\pi\uu{-{3\over 2}}\ ,
\eee
where \bbd{\g\lrm G} is the Glaisher constant \bbd{\g\lrm G \simeq 1.2824271291,}
and}
\item{The factor \bbd{Z\uprm{mmp}\ll n \equiv
\exp{q\ll n\uprm{mmp}}} is the macroscopic massive propagation contribution, for which \bbd{q\ll n\uprm{mmp}}
is exponentially small in \bbd{\sqrt{n}} at fixed \bbd{\t}.}
\ei

The first factor \bbd{\tilde{Z}\uprm{eft}\ll n} sums up
all diagrams of the massless Coulomb-branch
EFT, which for SQCD
contains only a single \bbd{U(1)} vector multiplet with
a free kinetic term and a supersymmetrized Wess-Zumino term for the Weyl \bbd{a-}anomaly, and no
other F-terms~\cite{Hellerman:2018xpi}.

At weak coupling, the behavior of \bbd{\tilde{B}} is given by
\bbb
\bbsk
\exp{\tilde{B}}=   \g\lrm G\uu{+12}\cc e\uu{-1}\cc 2\uu{-{{11}\over 6}}\pi\uu{-{1\over 2}}\cc {{  [{\rm Im}(\t) ]\uu{{3\over 2}}}\over{   [{\rm Im}(\t) + {2\over \pi}\cc {\tt log}[2]]\sqd }}\left[1 +  
   \frac{45 \zeta(3)}{16 \pi^2 (\text{Im} \tau)^2} - \frac{525 \zeta(5)}{64 \pi^3 (\text{Im} \tau)^3}  \right] + O([{\rm Im}[\t]\uu{-{9\over 2}})
   \llsk
\een{FinalResultForSchemeIndependentExpTildeBWithBoldfaceCFilledInRecap}

\section{The macroscopic massive propagation function \bbd{Z\uprm{mmp}}}\label{GeneralMMPSection}

\subsection{Physics of the massive macroscopic propagation factor \bbd{Z\ll n\uprm{mmp}}}\label{MMPPhysicsSection}

 We have now been able to write an exact expression for the EFT factor of the correlation function in closed form, with all
 previously undetermined coefficients fixed.  As a side benefit of fixing the overall normalization of the EFT factor
\bbd{Z\ll n\uprm{eft}} relative to the partition function \bbd{Z\ll{S\uu 4},} we have found the relationship
between our own massive macroscopic contribution
\bbd{q\ll n\uprm{mmp} = {\rm log} \big [ \cc Z\uprm{mmp}\ll n\cc \big ] =   {\rm log} \big [ \cc Z\ll n / Z\ll n\uprm{eft}\cc \big ]}
and the "[worldline] instanton term" \bbd{F\uprm{inst}[\ddsc]} of
ref.~\cite{Grassi:2019txd}: The latter is simply equal to the double-scaling
limit of the former.  Having done this, in this section we will shift our focus to the massive macroscopic
propagation partition function \bbd{Z\ll n\uprm{mmp}} itself.

This factor of the partition functions consists of a sum of all diagrams where each diagram contains at least one macroscopic worldline of a massive particle,
and its logarithm \bbd{q\ll n\uprm{mmp}} consists of a
sum of connected diagrams of this type.  As such,
it encodes a lot of interesting dynamical information
about the underlying degrees of freedom of the theory,
unlike the more universal EFT factor.  The dependence of this set of diagrams on \bbd{n} and \bbd{\t} is quite complex, and its many
interesting limits and behaviors are too intricate to explore
in any reasonable detail in this paper.  But in this section
we will briefly discuss some of its basic properties.

\subsubsection{Non-perturbative definition  of \bbd{Z\uprm{mmp}}}\label{SubSubSectionWhosePointICantRecallAtTheMeoment}

The most interesting property of the MMP partition function is that it is well-defined nonperturbatively
at all.  In general, sums of infinite subclasses of Feynman
diagrams only have {\it a priori\rm} definition as asymptotic
series in some loop-suppressing parameter, rather than
as well-defined functions.  In this case however, we
are able to take eq. \rr{FactorizationThatIsEquivalentToADefinitionOfTheMMPFactor} as a \emm{definition}
of \bbd{Z\uprm{mmp}\ll n} at any given \bbd{\t} and \bbd{n}; that is,
\bbb
Z\uprm{mmp}\ll n [\t]\equiv 2\uu{- 4n}\cc
{{Z\ll{S\uu 4}[\t]}\over{{Z}\uprm{eft}\ll n[\t]}} \cc G\ll{2n}[\t]\ .
\een{LiteralDefinitionOfMMPPartitionFunction}

As we have seen in sec.~\ref{DerivationOfConstantCoefficientBoldfaceC} (specifically eq. \rr{MatchingTheDoubleScalingLimitOfOurMMPFunctionWithTheWorldlineInstantonFunctionInGKT}), the \bbd{Z\ll n\uprm{mmp}} described
here also has a fixed limit at large \bbd{n} and fixed
\bbd{\ddsc}, where it becomes equal to the exponential of the~\cite{Grassi:2019txd}'s worldline-instanton function \bbd{F\uprm{inst}[\l]:}  
\bbb
\DDSL \cc Z\ll n\uprm{mmp}[\t]
= \exp{F\uprm{inst}[\ddsc]}\ ,
\eee
with the other terms on the RHS of~\cite{Grassi:2019txd}'s eq. \aleq{4.20}
canceled by factors in \bbd{2\uu{4n}\cc \tilde{Z}\ll n\uprm{eft},} the details of which we have seen
in sec.~\ref{DetailsOfMatchingWithStrongDoubleScalingLimit}.

\subsubsection{\bbd{S-}duality invariance of \bbd{Z\uprm{mmp}}}\label{DualityInvarianceOfMMPFunction}

The various objects \bbd{F\uprm{Oviedo}\ll n, (\Delta C)\ll 1} and \bbd{F\ls 0[\ddsc] = F\uprm{inst}[\ddsc]} are not
S-duality invariant.  The second and third objects, \bbd{ (\Delta C)\ll 1} and \bbd{F\ls 0[\ddsc] }, lack S-duality covariance because they are
double-scaled objects, and the double-scaling limit singles out a particular weak-coupling direction in coupling space, breaking S-duality from the beginning in the definition.
The Oviedo quotient \bbd{F\uprm{Oviedo}\ll n} is not itself a double-scaled object, but it is a quotient of two objects that are
separately duality-invariant, but only under separate and incompatibly defined S-duality operations.  That is, \bbd{F\uprm{Oviedo}\ll n} is invariant \emm{neither}
under the \bbd{{\cal N} = 4} S-duality \bbd{\t\to {{a\t + b}\over{c\t + d}},~\left ( \begin{matrix} a & b \cr c & d \end{matrix} \right ) \in SL(2,\IZ)}, \emm{nor} under
the SQCD \bbd{S-}duality \bbd{q\to 1-q} and \bbd{q\to {1\over q}}.  

By contrast, the object \bbd{Z\ll n\uprm{mmp}}
is invariant under the \bbd{S-}duality transformations
\bbd{q\to 1-q} and \bbd{q\to {1\over q}} acting on the
UV coupling \bbd{q\equiv \exp{2\pi i \t}}.  Note that unlike the full unnormalized correlator \bbd{Z\ll n,} the MMP factor \bbd{Z\ll n\uprm{mmp} = e\uu{q\ll n\uprm{mmp}}} does not get any nontrivial
tensorial prefactor under the duality transformation; it is literally strictly invariant under the entire duality group.

\subsubsection{\bbd{Z\uprm{mmp}} as a sum over diagrams with
macroscopic massive worldlines}

The function \bbd{Z\ll n\uprm{mmp}} has a behavior
consistent
with an interpretation as literally a sum of diagrams with massive particle trajectories propagating on the infrared
scale.  This can be seen clearly either at fixed \bbd{\t}
and large \bbd{n} or in the large \bbd{\ddsc} limit of
the fixed-\bbd{\ddsc} function when \bbd{n} is taken to
infinity first.  In either case the function \bbd{q\ll n\uprm{mmp} \equiv {\rm log}[Z\ll n\uprm{mmp}]} goes exponentially to zero, which it must do if interpreted as a literal sum of diagrams with at least one macroscopic massive worldline in each.  By contrast the logarithm
of \bbd{F\uprm{Oviedo}\ll n} contains additional
terms scaling as \bbd{\ddsc\uu 1, \ddsc\uu 0,} and \bbd{{\rm log}[\ddsc],} which are difficult to interpret
directly in terms of any specific dynamical field theory process.

\subsection{Geometry and \bbd{\JJM-}dependence of  worldline instanton effects}

Let us now discuss the macroscopic massive worldline interpretation
of \bbd{Z\ll n\uprm{mmp}} in the large-\bbd{n} limit,
either at fixed \bbd{\t} or in the closely
related limit of infinite \bbd{n} and fixed \bbd{\ddsc} with
a large-\bbd{\ddsc} limit taken subsequently.  Generally, the recent literature~\cite{Sharon:2020mjs, Arias-Tamargo:2019xld, Arias-Tamargo:2019kfr, Arias-Tamargo:2020fow, Alvarez-Gaume:2019biu, Watanabe:2019pdh, Badel:2019oxl, Badel:2019khk, Giombi:2020enj} on large-charge double-scaling limits tends to show those two limits 
have essentially the same behavior qualitatively.  These behaviors are similar but a quantitative discussion of
the two limits will help us distinguish the subtle differences between them in a concrete way. 

By "macroscopic" we simply mean that each connected 
diagram in the MMP contribution contains at least one closed worldline of a massive particle, whose length is of order
the infrared scale set by the size of the sphere.  Since
the theory is conformal, the effective mass \bbd{m} of any massive particles can only go as the expectation value of
the vector-multiplet scalar, which goes as \bbd{\sqrt{\JJM} = \sqrt{2n}}.

At large \bbd{n} and fixed \bbd{\t} the particles are very heavy and
their mechanical energy should dominate over interaction
energies with the massless degrees of freedom in the 
vector multiplet.  So at large \bbd{n} and fixed \bbd{\t} 
we expect the connected MMP path integral to be
a path integral for a single massive worldline of
\bbd{m} on an infinite cylinder \bbd{\IR\times S\uu 3} where the sphere has radius \bbd{R}.  We
have chosen the cylinder conformal frame to describe
the path integral in, because that is the conformal
frame in which the classical EFT solution has constant
magnitude \bbd{|\phi|} for the vector multiplet scalar (see
sec. \alsec{2.3} of~\cite{Hellerman:2017sur} for details
of the classical solution) and therefore
a space- and time-independent mass for the particle.

The connected
path integral for a free massive particle in a given background geometry
contains local terms coming from microscopic loops,
and also macroscopic loops.  The former are contained in
\bbd{q\ll n\uprm{EFT}} and are absent from \bbd{q\ll n\uprm{mmp}}, which contains only macroscopic massive worldlines.

Since the mass of the particle goes as \bbd{\sqrt{n}} in
units of the size of the sphere, the path integral for the
macroscopic contributions must be dominated by
a saddle point, which in the free limit would be a closed
geodesic of finite length. The only other possibility would be contributions coming from the boundary of the
sector of configuration
space defined by the cutoff on loop size; but these would be cutoff-dependent which would be subtracted as nonconformal counterterms in the process of renormalization, leaving
only saddle point contributions to dominate the MMP path integral at large \bbd{n} and fixed \bbd{\t}.

The only finite-length geodesics
on the cylinder are great circles of an \bbd{S\uu 3} section 
at fixed Euclidean time coordinate \bbd{t\lrm E}.  All
geodesics have the same length \bbd{2\pi R} and
so we would like to interpret the leading behavior
of \bbd{ -{\rm log}[q\ll n\uprm{mmp}] = -{\rm log}[{\rm log}[Z\ll n\uprm{mmp}]]} as the classical
worldline instanton action \bbd{  m \ell =  2\pi m R}.

To relate the mass of the lightest massive BPS particle to the
correlation function, define
\bbb
\MacroW [n, \t] \equiv - {\rm log}[q\ll n\uprm{mmp}] = 
 -{\rm log}[{\rm log}[Z\ll n\uprm{mmp}]]
\een{DefinitionOfLittleWSubLittleN}
The quantity \bbd{ {{\MacroW [n, \t]}\over{2\pi R}}} plays the role of
the effective mass of the lightest massive BPS particle
in the system in the background of a the Coulomb branch expectation value created by the insertion the operators \bbd{\co\uu n} and \bbd{\cob\ll n}. We express this quantity as a function of \bbd{n} and of the coupling \bbd{\t} because we expect the mass to depend on both.  The
dependence on \bbd{n}, because the mass of a
particle in the CFT must be proportional to the expectation value of the Coulomb branch field supported by the constant R-charge density on the cylinder.  The dependence on \bbd{\t,} because the mass formula for the BPS states depends on the value of \bbd{\t,} in units of the vector multiplet scalar \bbd{\phi\lrm{unit}}normalized to have unit kinetic term, the magnitude of whose expectation value in the classical solution controls
the magnitude of the R-charge density.

Then we can inquire about the actual behavior of these functions in various limits.

Since the
masses of the electrically charged hypers and \bbd{W-}
bosons go as \bbd{[{\rm Im}(\t)]\uu{-\hh}} at weak coupling
in units of the square root of the Coulomb-branch scalar with unit kinetic term, we also expect the weak-coupling
behavior of \bbd{{{\MacroW [n, \t]}\over{2\pi R\cc \sqrt{n}}}} will go as \bbd{[{\rm Im}(\t)]\uu{-\hh}}.   Let us now be slightly more quantitative about this.

\subsection{Particle mass from BPS formula}\label{BPSMassFormula}

First let us write the leading terms in the energies of the massless and massive degrees of freedom with some care paid
to the coefficients with a consistent set of normalizations.  The key point for us is to match the normalization of the scalar field in the vector multiplet as it appears in the BPS mass formula on the one hand, with the normalization of the vector multiplet scalar field as it appears in the bulk kinetic term, as the latter is relevant for the formula for
the R-charge density in the classical solution.

In section~\ref{RChargeAndVecMultScalarNormalizationConventions} of the Appendix we translate among various conventions in the literature for
the vector multiplet scalar \bbd{a}.  In the normalization specified in eq. \rr{ConventionForKineticTermAppendixVersionCopiedFromKineticTermNormalizationEquationFromDrivenRecursionRelationsV17} we use,
the kinetic term is
\bbb
{\cal L}\lrm{kin}  = {{{\tt Im}[\s]}\over{4\pi}}\cc |\pp a|\sqd\ .
\een{VMKineticTermNormalization}

In terms of the vector multiplet scalar normalized this way, 
the R-charge density is 
\bbb
\r\lrm J = \ordinary{+{i\over{4\pi}}} \cc {\tt Im}[\s] \cc \big [ \cc a\cc \dot{\bar{a}} - \bar{a} \cc \dot{a} \cc \big ]
\een{RChargeDensityBodyOfPaperVersion}
so the R-charge density of a helical solution with frequency \bbd{\o,} is given by 
\bbb
\r = +{{\o}\over{2\pi}}\cc {\rm Im}[\s]\cc |a|\sqd 
\eee
where the \bbd{R-}charge density is normalized as we've been doing, so that the complex vector multiplet scalar has R-charge \bbd{J = \pm 1}.

If the helical solution is homogeneous on a \bbd{3-}sphere of radius \bbd{R,} the total \bbd{U(1)} R-charge is
\bbd{\JJM =  {{\O\ll 3\cc R\uu 3\cc \o}\over{2\pi}}\cc{\rm Im}[\s]\cc |a|\sqd }
The area of the unit three-sphere \bbd{\O\ll 3 = 2\pi\sqd} and the frequency is \bbd{\o = {1\over R}} for a homogeneous helical solution for a conformally coupled free scalar on
an \bbd{S\uu 3} spatial slice of radius \bbd{R},
so we have
\bbb
\JJM =  \pi\cc R\sqd\cc \cc {\tt Im}[\s] \cc |a|\sqd\ , \llsk\llsk |a| = {1\over R}\cc \sqrt{{\JJM}\over {\pi\cc \cc {\tt Im}[\s] }}\ .
\eee

With the normalization \rr{VMKineticTermNormalization} for the vector multiplet scalar,
the mass of the hypermultiplet in the fundamental representation of \bbd{SU(2)} is given by
\bbb
m\lrm{hyper} =  |a|\ .
\een{MassOfHyperInTheFundamentalRepresentationBodyOfPaperVersion}
Using the BPS formula \rr{MassOfHyperInTheFundamentalRepresentation} for a fundamental hypermultiplet in the helical solution with \bbd{R-}charge \bbd{\JJM} then, the effective mass of a the hyper is given by
\bbb
m\lrm{hyper}  = {1\over R}\cc \sqrt{{\JJM}\over {\pi\cc \cc {\tt Im}[\s] }}\ .
\eee

\subsection{Worldline instanton action}\label{WorldlineInstantonActionSection}

Therefore the leading approximation to \bbd{q\ll n\uprm{mmp}} is given by the exponential of the negative of the classical worldline-instanton 
action,
\bbb
q\ll n\uprm{mmp}\sim e\uu{- S\lrm{WLI}}\ ,
\een{LeadingMMPFunctionIsExponential}
where \bbd{S\lrm{WLI}} is the action of a massive hyper circumnavigating a great circle of the \bbd{S\uu 3} spatial slice -- the only finite-length geodesic in the \bbd{S\uu 3 \times S\uu 1}
conformal frame -- is 
\bbb
S\lrm{WLI} = 2\pi R \cc m\lrm{hyper} =  \sqrt{{4\pi\cc \JJM}\over { \cc {\tt Im}[\s] }} =  \sqrt{{8\pi n}\over {  {\tt Im}[\s] }}\ .
\een{FormOfWLIAction}
We expect that the mass formula is \emm{exact} as a function of the gauge coupling, though the action can receive subleading contributions in \bbd{n} at fixed coupling,
coming from the effect of curvature terms and the gradient of the phase of the vector multiplet scalar \bbd{a}.

The point of this exercise was twofold: First, to check that the exponent of the exponentially small correction in the large-\bbd{\ddsc}-limit of~\cite{Grassi:2019txd}'s function \bbd{\Delta C\ll 1,}
is in fact given by the worldline instanton action.  Second, to illustrate the subtle difference between the behavior of \bbd{q\lrm n\uprm{mmp}[\t]} in the large-n limit at fixed \bbd{\ddsc} ob
the one hand, where it becomes identical with~\cite{Grassi:2019txd}'s function \bbd{F\uprm{inst}[\ddsc],} and the large-\bbd{n} limit at fixed \bbd{n}, where the exponentially small correction should
behave almost, but not quite, identically with the large-\bbd{\ddsc} behavior of~\cite{Grassi:2019txd}'s \bbd{F\uprm{inst}[\ddsc]}.

\subsubsection{Worldline instanton action in the double-scaling limit}

At large \bbd{n} and fixed \bbd{\ddsc,} the coupling \bbd{\t} goes to zero, so we can use the weak-coupling expansion \rr{ZeroInstantonApproximationToRelationBetweenCouplings} of the IR coupling
\bbd{\s} in terms of the UV coupling \bbd{\s:}
\bbb
\t 
\simeq  { \s\over 2} - {{2i}\over \pi}\cc {\tt log}[2] \ , 
\llsk\llsk \s \simeq 2\t + {{4i}\over \pi}\cc {\tt log}[2]\ ,
\xxnn
{\rm Im}[\t] = \hh \cc {\rm Im}[\s] - {2\over \pi}\cc {\tt log}[2] \ , 
\llsk
{\rm Im}[\s] = 2\cc {\rm Im}[\t] + {4\over \pi}\cc {\tt log}[2]\ ,
\een{ZeroInstantonApproximationToRelationBetweenCouplingsSortof}
with the \bbd{\simeq} indicating the omission only of exponentially small corrections.

In the double-scaling limit, the double-scaling behavior of the effective mass \bbd{{{\MacroW [n, \t]}\over{2\pi R}} = -{{ {\rm log}[q\ll n\uprm{mmp}]}\over{2\pi R}}} will be relevant; we will
use the definitions \rr{ExistenceOfDoubleScalingLimitOfMMPContribution} ,\rr{DefinitionOfLittleWSubLittleN}, which together give
\bbb
\DDSL \MacroW [n, \t] = - {\rm log}F\ls 0[\ddsc]
\eee 

So then at fixed \bbd{\ddsc =  {n\over{4\pi\cc {\rm Im}[\t]}}} we have
\bbb
{\rm Im}[\s] \simeq {n\over{2\pi \l}} + {4\over \pi}\cc {\tt log}[2] \ , \llsk {n\over{{\rm Im}[\s]}} = 2\pi\l + O(n\uu{-1})\ ,
\eee
and
\bbb
S\lrm{WLI} = \sqrt{{8\pi n}\over {  {\tt Im}[\s] }} = 4\pi\cc \sqrt{\l} + O(n\uu{-1})\ .
\eee
So taking the large-\bbd{n} limit at fixed \bbd{\ddsc} \emm{first,} and then taking the large \bbd{\ddsc} limit, we expect \bbd{\MacroW [n, \t] \equiv - {\rm log}[{\rm log}[Z\ll n\uprm{mmp}[\t]]] = - {\rm log}{[q\ll n\uprm{mmp}[\t]]}}
to go as \bbd{4\pi \cc \l\uu{+\hh}:}
\bbb
{\rm lim}\ll{\ddsc\to\infty}\cc \biggl [ \cc \DDSL\cc {{\MacroW [n, \t]}\over{\sqrt{\ddsc}}} \biggl ]=-{\rm lim}\ll{\ddsc\to\infty}\cc {{{\rm log}[F\ls 0[\ddsc]]}\over{\sqrt{\ddsc}}} = +4\pi\ .
\een{PhysicalExpectationForLargeLambdaLimitOfLargeChargeFixedLambdaLimit}

We can verify this prediction easily by using the equality derived earlier, between our \bbd{F\ls 0[\ddsc]} on the one hand, and ref.~\cite{Grassi:2019txd}'s \bbd{F\uprm{inst}[\ddsc]} on the other hand.  The latter has a large-\bbd{\ddsc} expansion that was already worked-out in~\cite{Grassi:2019txd}, with the result
\bbb
{\rm log}\bigg [ \cc F\uprm{inst}[\ddsc] \cc \bigg ] = -4\pi \ddsc\uu{+\hh} + O({\rm log}[\ddsc] )\ ,
\eee
in agreement with the prediction \rr{PhysicalExpectationForLargeLambdaLimitOfLargeChargeFixedLambdaLimit}.  This value is also in agreement with the
value inferred in~\cite{Hellerman:2018xpi} from numerical fitting of the exponent of the MMP factor in the double scaling limit.

This confirms our physical expectation for the worldline instanton action \bbd{S\lrm{WLI}} controlling 
the large-\bbd{\ddsc} limit of the large-\bbd{n} limit of the macroscopic massive propagation
function.  Taking the large-\bbd{n} double-scalimg limit first, the distinction between \bbd{\s} and \bbd{2\t} is washed out and the \bbd{{\cal N} = 2} threshold correction \bbd{{2\over \pi}\cc {\rm log}(2)}  to the relationship
between \bbd{\s} and \bbd{\t}, is irrelevant, along with the gauge instanton corrections.

\subsubsection{Worldline instanton action in the fixed-coupling, large-\bbd{\JJM} limit}

Next consider the behavior of \bbd{\MacroW [n, \t]} again, now at large \bbd{n} and fixed \bbd{\t}.  Here, we have no exact formula, but we can predict on physical grounds that 
the large \bbd{n} behavior of \bbd{\MacroW [n, \t]} at fixed \bbd{\t} should be dominated by the worldline of the lightest BPS dyon; in some range of \bbd{\t} with given
by \bbd{2\pi R} times the mass of the lightest BPS dyon for any given value of the coupling:
\bbb
{\tt lim}\ll{n\to\infty}\cc {1\over {\sqrt{n}}}\cc  \MacroW [n, \t] = (8\pi)\uu{+\hh}\cc {\rm min}\ll{n\ll e,n\ll m\in\IZ}\cc {{\bigl |  n\ll e + \s\cc n\ll m \cc \big |}\over{\sqrt{{\rm Im}[\s]}}}
\een{LightestDyonMassFormula}
The quantity \bbd{n\ll e + \s n\ll m} transforms as a holomorphic modular form of weight \bbd{-1} under \bbd{SL(2,\IZ)} %
so \bbd{|n\ll e+ \s n\ll m|\sqd}
transforms as a nonholomorphic modular form of weights \bbd{(-1,-1)} as
does \bbd{{\rm Im}[\s]},
so the RHS of \rr{LightestDyonMassFormula} is \bbd{SL(2,\IZ)-}invariant.  The expression
is also continuous, but not smooth; its first derivative is discontinuous at the locus where 
two distinct electromagnetic types of BPS dyons \bbd{(n\ll e, n\ll m)} are degenerate in mass.  

\subsubsection{Summary: The worldline instanton action}

So to summarize, we see that the fixed-coupling, large-\bbd{\JJM} limit of the massive macroscopic propagation function has at least three related differences from
the double-scaling limit, even at strong double-scaled coupling where they are most similar:
\bii
\item{The fixed-coupling, large-\bbd{n} limit is \bbd{S-}duality invariant whereas the double-scaling limit is not;}
\item{The fixed-coupling, large-\bbd{n} limit has continuous but non-smooth transitions in the coefficient of its leading term as a function of the coupling; and}
\item{Even at weak coupling, the formula for the exponent of the exponentially small correction contains specific subleading terms, starting with the \bbd{{\cal N}=2} threshold
correction, which have no analog in the exponent of~\cite{Grassi:2019txd}'s exponentially small worldline instanton contribution \bbd{F\uprm{inst}[\ddsc]} to the one-loop
double-scaling function \bbd{\Delta C\ll 1[\ddsc]}.}
\ei
The strong-coupling limit of the double-scaling limit of \bbd{\MacroW [n, \t] \equiv {\rm log}[q\ll n\uprm{mmp}]} matches the value of the worldline instanton action
as predicted from effective field theory,
\bbb
 \DDSL \MacroW [n, \t] = - \DDSL \cc {\rm log}[q\ll n\uprm{mmp}] = - {\rm log}[F\ls 0[\ddsc]] = - {\rm log}\big [ \cc F\uprm{inst}[\ddsc] \cc \big ]\ll{\hbox{ref.~\cite{Grassi:2019txd}}}
 \xxnn
 =  4\pi \cc \ddsc\uu{+\hh} + O({\rm log}[\ddsc])  = S\lrm{WLI}\cc\bigg |\lrm{double~scaling~limit} + O({\rm log}[\ddsc]) \ .
\eee
As for the  fixed-\bbd{\t}, large-\bbd{n} limit, we currently do not know any exact expression
for \bbd{q\ll n\uprm{mmp}} in this limit; it would be interesting to see if the features discussed here could be visible either numerically
in the data of correlation functions computed from localization data by the prescription in~\cite{Gerchkovitz:2016gxx} or else through an improved analytic understanding of
the extremal correlators.

\section{Numerical comparison}\label{Comparison}

\begin{figure}      
  \centering
  \begin{footnotesize}
  \begin{tabular}{c}
    \includegraphics[width=.7\textwidth]{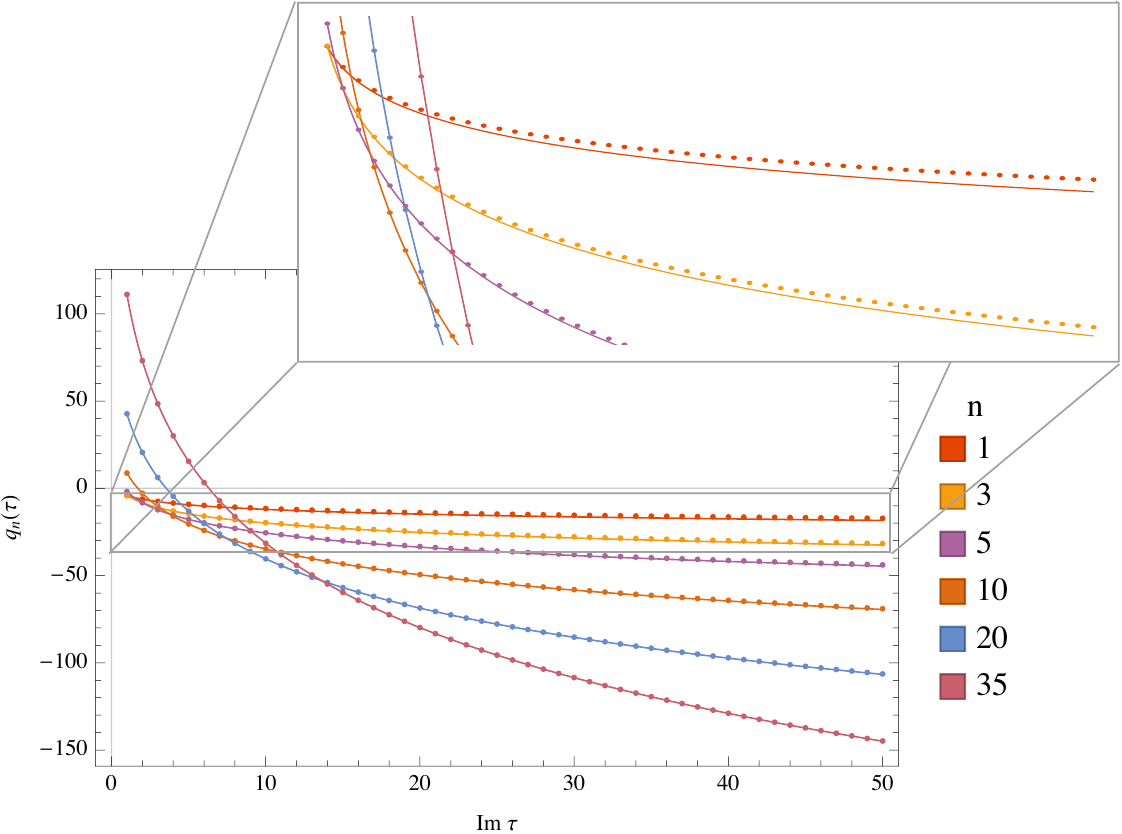} \\
  (a) \(q\) as function of \(\operatorname{Im}(\tau)\) at fixed \(n\). \\[2em]
  \includegraphics[width=.7\textwidth]{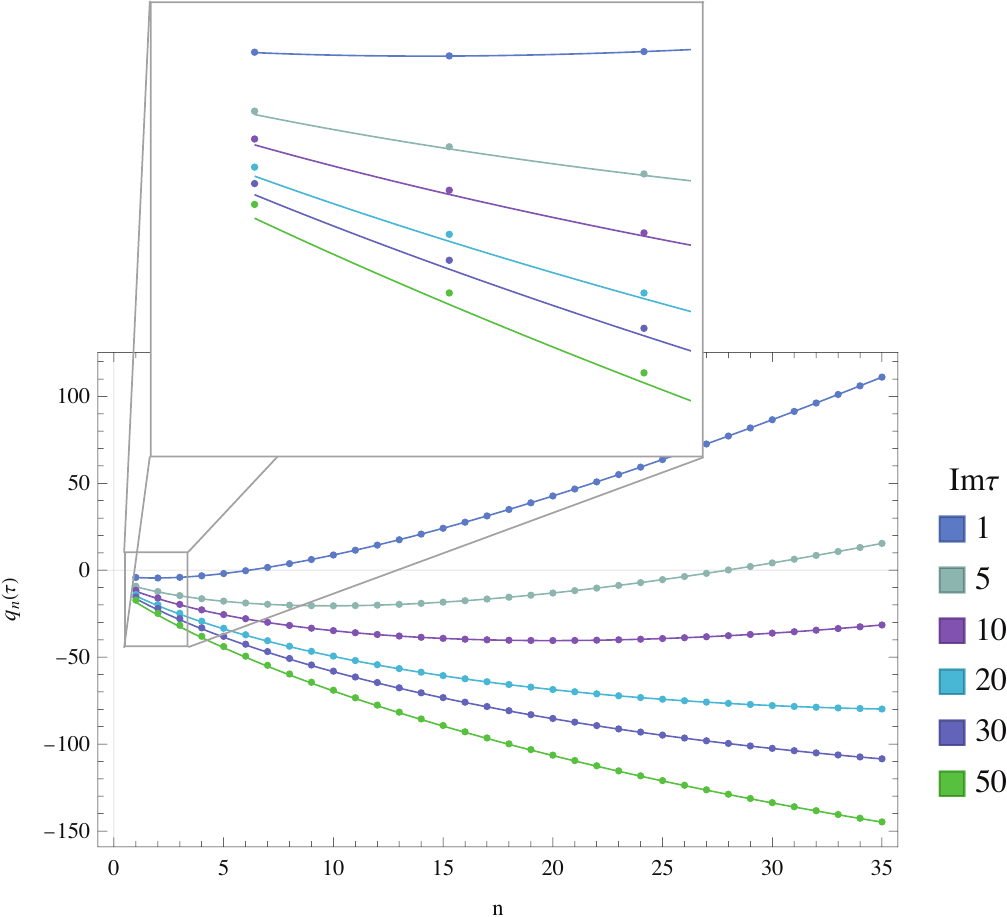} \\
  (b) \(q\) as function of \(n\) at fixed \(\operatorname{Im}(\tau)\).
  \end{tabular}
\end{footnotesize}
  \caption{Comparison with localization. Values of \(q_n\uprm{EFT}(\tau)\) from the prediction of Eq. \rr{EFTPrediction} (continuous lines) and numerical estimates from localization (dots) at fixed values of \(n\) (a) and fixed values of \(\operatorname{Im}(\tau)\). We need to zoom into the small-\(n\) region in order to have a visible discrepancy.}
  \label{fig:exploded-numerics}
\end{figure}

As we have stressed above, the correlation functions that we have studied can be computed numerically using the construction of~\cite{Gerchkovitz:2016gxx} and our predictions of large-\(n\) behavior compared to the exact numerical result as it had already been done in~\cite{Hellerman:2018xpi, Hellerman:2020sqj}.
In figure~\ref{fig:exploded-numerics} we show the value of \(q_n(\tau)\) both as function of \(\tau \) at fixed values of \(n\) and as function of \(n\) for fixed values of \(\tau\).
In both cases the agreement is quite remarkable, and we need to zoom on small values of \(n\) (under \(n=5\)) in order to see any discrepancy at all.

More quantitatively, at \(n = 1\), \(\tau = 2\) the relative error in the estimate of \(q_1(2)\) is of the order of one part on one hundred or better, and at \(n=30\), \(\tau = 2\) the relative error for \(q_{30}(2)\) is of the order of one part in ten millions (!), before taking into account any non-perturbative correction.
We find it quite remarkable that the agreement between the exact numerical results and the semiclassical approximation  remains very good even for small number \(n = O(1)\) of particles.
This seems to go against the common lore that a quatum system can be effectively approximated semiclassically only for a very large number of particles.\footnote{For example, a mesoscopic system where the semiclassical approximation is in line with the experimental results is a quantum dot made of \(10^5\) particles~\cite{richter2000semiclassical}.}
The complete EFT expression \bbd{q\ll n\uprm{EFT}} represents the maximum accuracy achievable
for the large-charge correlation function without additional data from the underlying CFT.  However with only the additional input of the mass of the lightest BPS particle as a function of the coupling \bbd{\t,} one may \cite{HellermanExponentialCorrectionsToAppear} give a further asymptotic series for the exponentially small correction itself with the expression \rr{LeadingMMPFunctionIsExponential},\rr{FormOfWLIAction} as the leading approximation.  This "hyperasymptotic" or "transseries"     
correction to the EFT estimate, similar in spirit to the recent results~\cite{Dondi:2021buw}
in the Wilson--Fisher critical \bbd{O(2N)} models, further improves upon the accuracy of the EFT estimate by several orders of magnitude.

\section{Conclusions}\label{ConclusionsSection}

\subsection{Concise and self-contained summary of the results}\label{ConciseSummaryInConclusions}

In the present paper we have completed the solution of the universal EFT factor \bbd{Z\ll n\uprm{eft}} of the correlation function of Coulomb-branch chiral primary
operators in superconformal SQCD with \bbd{G = SU(2)} and \bbd{N\ll f=4} massless 
hypermultiplets in the fundamental representation. Building on previous results~\cite{Hellerman:2017sur, Hellerman:2018xpi, Hellerman:2020sqj} at large \bbd{\JJM} and fixed gauge coupling, and also
using the strong-coupling limit of the exact formula ~\cite{Grassi:2019txd} for the double-scaling limit~\cite{Bourget:2018obm}
of the correlator to match an otherwise-undetermined constant, we have fully solved for the two coupling dependent
functions \bbd{A} and \bbd{B} in the expansion EFT factor of the correlator.
We now give a concise and self-contained statement of the result.

The correlation function is
\bbb
\langle\co\uu n(x) \cob\uu n(y) \rangle =  |x-y|\uu{-4n}\cc G\ll{2n[\co]} \ ,
\xxn{FinalResultSummaryEqA}
G\ll{2n[\co]} = G\ll{2n[\co]}\uprm{eft} \times Z\ll n\uprm{mmp}\ ,
\xxn{FinalResultSummaryEqB}
G\ll{2n[\co]}\uprm{eft} =  {\bf N}\uu n\ll\co \cc \exp{\tilde{B}}\cc \Gamma[2n + {5\over 2}] \ ,
\een{FinalResultSummaryEqC}
where
\bbb
{\bf N}\ll\co \equiv   \big |\cc {{\co}\over{\co\ls\t}}\cc \big |\sqd \cc 16\cc e\uu{A\ls\t} 
= \big |\cc {{\co}\over{\co\ls\t}}\cc \big |\sqd \cc {{\big |{{d\s}\over{d\t}}\big |\sqd }\over{ \big ( \cc {\rm Im}[\s]\big )\sqd}}\ ,
\een{FinalResultSummaryEqD}
Here the unsubscripted \bbd{\co} is the generator of the Coulomb branch chiral ring with any normalization at all, and \bbd{\co\ls\t} is the generator normalized so that
\bbb
\langle \co\ls\t(x) \cob\ls\tb(y)\rangle = {{16}\over{|x - y|\uu 4}}\cc \pp\ll\t\pp\ll{\bar{\t}}\cc {\rm log}[Z\ll{S\uu 4}]
\een{FinalResultSummaryEqE}
The coupling \bbd{\s} is the \bbd{SL(2,\IZ)}-covariant "infrared coupling" \bbd{\s}, which transforms under \bbd{SL(2,\IZ)} as \bbd{\s\mapsto {{a\s + b}\over{c\s + d}}} where
\bbd{a,b,c,d} are integers satisfying \bbd{ad-bc=1}.  The coupling \bbd{\t} can be taken to be the "ultraviolet coupling" related to \bbd{\s} by
\bbd{e\uu{2\pi i \t} = \l(\s)} where \bbd{\l(\s)} is the modular lambda function, but the formulae above are fully covariant with
respect to choice of holomorphic coupling constant \bbd{\t}.  The coupling-dependent coefficient \bbd{\tilde{B},} whose determination was the main calculation of
this paper, is given by
\bbb
\exp{\tilde{B}}  = \g\lrm G\uu{+12}\cc e\uu{-1}\cc 2\uu{-{9\over 2}}\pi\uu{-{3\over 2}}\cc
{{ |\l(\s)|\uu{+{2\over 3}}\cc |1-\l(\s)|\uu{+ {8\over 3}}}\over{|\eta(\s)|\uu 8\cc [{\rm Im}(\s)]\sqd}}\cc \bigg [ \cc Z\lrm{{Pestun-}\atop{Nekrasov}}\cc \bigg ]\uu{-1}
\een{FinalResultSummaryEqF}
where \bbd{\l(\s)} is the modular lambda function, \bbd{\eta(\s)} is the Dedekind eta function, and \bbd{\g\lrm G} is the Glaisher constant \bbd{\g\lrm G \simeq 1.2824271291}.
The last factor is the reciprocal of the partition function \bbd{Z\lrm{{Pestun-}\atop{Nekrasov}}} as computed in the Pestun-Nekrasov scheme, by which we mean
as computed by localization with a localization integrand using Pestun's one-loop determinant~\cite{Pestun:2007rz} and Nekrasov's \bbd{U(2)} (as opposed to \bbd{SU(2)} as in~\cite{Alday:2009aq}) instanton
factor~\cite{Nekrasov:2002qd}.  The detail of the \bbd{S\uu 4} partition function of \bbd{{\cal N} = 2, G=SU(2), N\ll f = 4} SQCD and its application
to the computation of Coulomb-branch correlators, is given in~\cite{Gerchkovitz:2016gxx}; its weak-coupling expansion is given there and also reviewed in sec.~\ref{ZS4WeakCouplingExpansionSec} of the
Appendix here.

\subsection{Other results and conclusions}

In addition to deriving the result \rr{FinalResultSummaryEqA}-\rr{FinalResultSummaryEqF}, we have also:
\bii
\item{Used S-duality-invariance together with appropriate covariance of the \bbd{B-}function under scheme changes, to determine it fully up to a single, coupling-independent coefficient;}
\item{Determined the remaining coefficient scaling the EFT factor out of the correlator and matching the large-\bbd{\JJM} limit
of the remaining factor \bbd{Z\uprm{mmp}\ll n} with the exponential
of~\cite{Grassi:2019txd}'s function \bbd{F\uprm{inst}}.}
\item{Discussed the physics of \bbd{Z\ll n\uprm{mmp}} which describes macroscopic propagation of massive BPS particles, and used this physical picture to predict precise behaviors
of \bbd{Z\ll n\uprm{mmp}} in the large-\bbd{n}, fixed-coupling limit, including a Stokes phenomenon of BPS dyon dominance exchange as the coupling is varied.}
\item{Compared our predictions, both for the EFT factor and the massive macroscopic propagation factor of the correlators, with numerical results for the same correlators using
supersymmetric localization, finding agreement with the EFT prediction for both factors to one part in ten million or smaller.}\ei

We have hope that the results presented here will stimulate further study of precision correlators at large R-charge.  More generally we would wish this paper to foster an improved appreciation and wider application of the unreasonable effectiveness of the large quantum number expansion.

\section*{Acknowledgments}%

The authors thank Marco Bill\'o for discussions on the S-duality properties
of the \bbd{S\uu 4} partition function
and Zohar Komargodski for discussions on the
scheme-dependence of the sphere partition function in \bbd{{\cal N} = 2} superconformal SQCD.  The work of S.H. is supported by the World Premier International Research
Center Initiative (\textsc{wpi} Initiative), \textsc{mext}, Japan; by the \textsc{jsps} Program for Advancing Strategic International Networks to Accelerate the Circulation of Talented Researchers;
and also supported in part by \textsc{jsps kakenhi} Grant Numbers \textsc{jp22740153, jp26400242}.
D.O. acknowledges partial support by the \textsc{nccr 51nf40--141869} ``The Mathematics of Physics'' (Swiss\textsc{map}).
We thank Susanne Reffert for valuable discussion and collaboration on closely related work.
The authors also thank the Simons Center for Geometry and Physics for hospitality during the program, ``Quantum Mechanical Systems at Large Quantum Number,'' during which this work was initiated.

\newpage

\appendix

\section{\blue{Appendix:} Conventions}\label{ConventionAppendix}

\subsection{Conventions for normalizations of the double-scaling parameter}\label{ThreeLambdaNormalizationsWhyWhyWhyWhyWhyGodWhy}

 Here are the various normalizations of \bbd{\ddsc}.  So far in the
 literature on double-scaling at large charge in \bbd{{\cal N} = 2} superconformal QCD, there are already three distinct conventions, two of them introduced by \mgkt ~in the same paper, both of which differ from the convention of \movie.

\bii
\item{\bbd{\ddsc\lrm{\smovie} = g\sqd n }; this normalization
is even given in the abstract of \movie. Now, \bbd{{\rm Im}[\t] = {{4\pi}\over{g\sqd}}} so \bbd{g\sqd = {{4\pi}\over{{\rm Im}[\t]}}}.  This means
\bbb
\ddsc\lrm{\smovie} = {{4\pi n}\over{{\rm Im}[\t]}}
\eee
}
\item{In \mgkt, as the authors \emm{explicitly note,} there are \emm{two different normalizations} for their \bbd{\ddsc}, a different one in the first half of their paper than the one
in the second half of their paper.  See the comment about this near their eq. \blue{(4.1)}. For the convention
in the earlier part of the paper, \mgkt~ writes (in their eq. \blue{(1.2)})  \bbd{\ddsc\lrm{\smgkt,~earlier~part~of~paper}  = {{g\lrm{YM}\sqd}\over{4\pi}}\cc n} and since \bbd{g\lrm{YM}\sqd = {{4\pi}\over{{\rm Im}[\t]}}}
we get
\bbb
\ddsc\lrm{\smgkt,~earlier~part~of~paper} = {n\over{{\rm Im}[\t]}}\ .
\eee
}
\item{Then, in the later part of their paper, in eq. \blue{(4.1)} they write 
\bbb
\ddsc\lrm{\smgkt,~later~part~of~paper} = {n\over{4\pi\cc {\rm Im}[\t]}}\ .
\eee
}
\item{Our own convention is that of \emm{the later part of} \mgkt
\bbb
\ddsc = \ddsc\lrm{here} = \ddsc\lrm{\smgkt,~later~part~of~paper} = {n\over{4\pi\cc {\rm Im}[\t]}}\ .
\eee
}
\ei

The relations among the various normalizations of the coupling appearing in the recent literature, are summarized in table~\ref{NormalizationsOfDoubleScaledCouplingTable}.

 \begin{table}
\begin{center}
\begin{tabular}{ |c||c|c|c|c| } 
 \hline\hline
\backslashbox{denominator}{numerator} &  $  \ddsc \equiv \ddsc\lrm{ {{this}\atop{paper}}}  $ &  $   \ddsc\ll{\hbox{\scriptsize ref.~\cite{Bourget:2018obm}}} $&  $     \ddsc\ll{ {{\rm earlier~part}\atop{\hbox{\tiny of ref.\cite{Grassi:2019txd} }}}}$  &  $     \ddsc\ll{ {{\rm later~part}\atop{\hbox{\tiny of ref.\cite{Grassi:2019txd} }}}}$ \\ \hline\hline
$ \ddsc\equiv \ddsc\lrm{this~paper} $ &   $1 $ & $ (4\pi)\uu{+2}  $ &   $ (4\pi)\uu{+1}   $   & 1 \\ \hline
$   \ddsc\ll{\hbox{{\scriptsize ref.~\cite{Bourget:2018obm}}}}     $ & $    (4\pi)\uu{-2} $          & $ 1 $& $ (4\pi)\uu{-1}     $  & $ (4\pi)\uu{-2} $ \\ \hline
$  \ddsc\ll{\hbox{{\scriptsize earlier part of ref.~\cite{Grassi:2019txd}}}}    $        & $ (4\pi)\uu{-1} $    & $ (4\pi)\uu{+1}  $& $  1 $  & $(4\pi)\uu{-1}$  \\ \hline
$  \ddsc\ll{\hbox{{\scriptsize later part of ref.~\cite{Grassi:2019txd}}}}    $ & $ 1 $            & $  (4\pi)\uu{+2}  $& $  (4\pi)\uu{+1} $  &  1 \\ \hline
\end{tabular}
\captionof{table}{Relationships among different conventions for the normalization of the double-scaled coupling \bbd{\ddsc,} one of which was introduced in~\cite{Bourget:2018obm} and the
second and third of which were both introduced in~\cite{Grassi:2019txd}.}\label{NormalizationsOfDoubleScaledCouplingTable}
\end{center}
\end{table}

\subsection{Conventions for the normalization of the two-derivative effective action of the \bbd{U(1)} vector multiplet}\label{RChargeAndVecMultScalarNormalizationConventions}

\subsubsection{The effective kinetic term}

In order to compute the action of the worldline instanton of the massive BPS hypermultiplet at fixed \bbd{R-}charge \bbd{\JJM = 2n} we need
to specify the conventions for the normalization of the two-derivative effective action for the scalar field \bbd{a} in the \bbd{U(1)} vector multiplet.
At the moment we are only going to compute the leading large-\bbd{\JJM} action for the worldline instanton, either at fixed gauge coupling or fixed 
double-scaled coupling, and so we do not need higher derivative terms coming from the Wess-Zumino term, which affect the result only at
subleading order in \bbd{\JJM}.

In the present paper we will use a commonly used normalization convention for the scalar \bbd{a} in the vector multiplet, matching the normalization convention of~\cite{Pestun:2007rz},
and also used in the detailed review~\cite{Tachikawa:2013kta} of \bbd{{\cal N} = 2} supersymmetric gauge theory in \bbd{D=4}. 
In this normalization the effective kinetic term is written in terms of the effective holomorphic prepotential \bbd{{\cal F} = \hh \cc \s\cc a\sqd} or the effective K\"ahler potential  on field space, with \bbd{K = {\rm Im}[\bar{a}\cc  {\cal F}\pr(a)] = {\rm Im}[\s] \cc |a|\sqd}:
\bbb
{\cal L}\lrm{{kin,}\atop{effective}} 
= \ordinary{{1\over{4\pi}}} \cc {\tt Im}\big [{\cal F}\prpr(a) \big ]\cc|\pp a|\sqd =  \ordinary{{1\over{4\pi}}} \cc K\ll{,a\bar{a}}\cc |\pp a|\sqd  
= {{{\tt Im}[\s]}\over{4\pi}}\cc |\pp a|\sqd\ .
\een{ConventionForKineticTermAppendixVersionCopiedFromKineticTermNormalizationEquationFromDrivenRecursionRelationsV17}
The kinetic term \rr{ConventionForKineticTermAppendixVersionCopiedFromKineticTermNormalizationEquationFromDrivenRecursionRelationsV17} should be
interpreted as a Wilsonian effective kinetic term for the \bbd{U(1)} Abelian vector multiplet of the unbroken gauge group on the vacuum manifold, 
and \bbd{\s} is the complexified Abelian gauge coupling that transforms as \bbd{\s\to {{a\s + b}\over{c\s + d}}} under the \bbd{SL(2,\IZ)} S-duality.  If expressed
in terms of the UV coupling \bbd{\t,} which is related to \bbd{\s} by \bbd{e\uu{2\pi i \t} = \l(\s)} where \bbd{\l(\s)} is the modular lambda function, then the effective
kinetic term is
\bbb
{\cal L}\lrm{{kin,}\atop{effective}} 
= \ordinary{{1\over{4\pi}}} \cc {\tt Im}\big [{\cal F}\prpr(a) \big ]\cc|\pp a|\sqd =  \ordinary{{1\over{4\pi}}} \cc K\ll{,a\bar{a}}\cc |\pp a|\sqd  
= {{{\tt Im}[\t]}\over{2\pi}}\cc |\pp a|\sqd + O(({\rm Im}[\t])\uu 0)\ .
\een{ConventionForKineticTermAppendixVersionCopiedFromKineticTermNormalizationEquationFromDrivenRecursionRelationsV17ExpressedInTermsOfTau}

With these conventions understood, the conjugate momenta are
\bbb
\Pi\ll a = \ordinary{ {1\over{4\pi}}} \cc {\tt Im}[\s]\cc \dot{\bar{a}}\ , \llsk\llsk
\Pi\ll {\bar{a}} = \ordinary{ {1\over{4\pi}}} \cc {\tt Im}[\s]\cc \dot{a}
\een{ConjugateMomentaToAScalarField}
and we will use them to compute the canonical expression for the $R$-current density.  

\subsubsection{The central charge and hypermultiplet mass}

In terms of the normalization choice \rr{ConventionForKineticTermAppendixVersionCopiedFromKineticTermNormalizationEquationFromDrivenRecursionRelationsV17} for the vector multiplet scalar,
the relationship between the vector multiplet scalar \bbd{a} and the central charge \bbd{Z} appearing
in the BPS mass formula, is
\bbb
m =   |Z|\ , \llsk Z = (n\ll e + \s\cc n\ll m)\cc a\ ,
\eee

In this convention the effective Abelian electric charge of the hypermultiplet in the fundamental representation
is \bbd{n\ll e = \pm 1} and so the hypermultiplet mass is
\bbb
m\lrm{hyper} =  |a|\ .
\een{MassOfHyperInTheFundamentalRepresentation}

\subsubsection{Relation to normalizations in the older literature}

Note that we are giving our formulae with the modern normalizations of the central charge \bbd{Z} and vector multiplet \bbd{a}.  There are differences by a power of \bbd{\sqrt{2}}
between the modern and older literature in this respect, with the older literature ({\it e.g.}~\cite{Seiberg:1994rs, Seiberg:1994aj}) using the formul\ae
\bbb
\bbsk
m = \sqrt{2} |Z\lrm{old}| \ , \llsk Z\lrm{old} = (n\ll e + n\ll m \s)\cc a\lrm{old}\ ,
\llsk
{\cal L}\lrm{kin} =  {1\over{2\pi}}\cc {\tt Im}[\s] \cc |\pp a\lrm{old}|\sqd\ .
\llsk
\eee
The normalizations are related by
\bbb
Z\lrm{old} = {1\over{\sqrt{2}}}\cc Z\lrm{here}\  , \llsk\llsk  Z\lrm{here} =  \sqrt{2}\cc Z\lrm{old} \ ,
\xxnn
a\lrm{old} ={1\over{\sqrt{2}}}\cc a\lrm{here} \ , \llsk\llsk a\lrm{here} =  \sqrt{2}\cc a\lrm{old} \ .
\eee
This is just a redefinition of variables and does not affect any relationships among physical quantities such as masses and R-charges.
\subsubsection{The microscopic Lagrangian and its vacuum modulus}

Now let us give the convention for the tree-level microscopic action and describe how the vacuum modulus is related to the nonabelian vector multiplet
at weak coupling.  In the normalization conventions of ~\cite{Pestun:2007rz}, with the notational difference \bbd{\ A\lrm{here} = (\Phi\ll 0\uu E + i \Phi\ll 9) }, we have
\bbb
{\cal L}\lrm{{{microscopic,}\atop{ Pestun}}} =  {1\over{g\sqd}}\cc {\rm tr}\lrm F\bigg [ \cc -\hh\cc \hat{F}\ll{\m\n}\sqd + (\gg\ll\m \hat{A})(\gg\ll\m \hat{A}\dag)
+ {1\over 4}\cc [\hat{A}, \hat{A}\dag]\sqd  \cc \bigg ]
\xxnn
+|\gg\ll\m \phi|\sqd + \hh\cc \phi\dag(\hat{A}\dag \hat{A} + \hat{A} \hat{A}\dag)\phi 
\xxnn
+ (\th-{\rm term}) + ({\rm fermions}) 
\eee
The gauge field and the scalar \bbd{\hat{A}} are traceless  \bbd{2\times 2}
matrix-valued fields, with the gauge connection being antihermitean and the scalar being complex.  The trace is taken in the fundamental representation.  The complex modulus \bbd{a} is embedded in the
vacuum solution space of the field \bbd{\hat{A}} as
\bbb
\hat{A} = i a \s\uu 3\ , \llsk\llsk a =  \sqrt{- \hh\cc {\rm Tr}\lrm F(\hat{A}\sqd)} 
\eee 
Then the mass of the hypermultiplets is \bbd{m\lrm{hyper} = 1\times |a|} and
the kinetic term for the vacuum modulus is 
\bbb
{\cal L}\lrm{  {{kinetic}\atop{tree}}} = {2\over{g\sqd}}\cc |\pp a|\sqd
\eee
The complex UV coupling is
\bbb
\t\lrm{UV} \equiv \t = {{4\pi i}\over{g\sqd}} + {{\th}\over{2\pi}}\ ,
\eee
so
\bbb
{\cal L}\lrm{  {{kinetic}\atop{tree}}} = {2\over{g\sqd}}\cc |\pp a|\sqd = {{{\rm Im}[\t]}\over{2\pi}}\cc |\pp a|\sqd \simeq {{{\rm Im}[\s]}\over{4\pi}}\cc |\pp a|\sqd
\eee
agreeing at leading order in the weak coupling expansion with the Wilsonian effective kinetic term \rr{ConventionForKineticTermAppendixVersionCopiedFromKineticTermNormalizationEquationFromDrivenRecursionRelationsV17ExpressedInTermsOfTau},
\rr{ConventionForKineticTermAppendixVersionCopiedFromKineticTermNormalizationEquationFromDrivenRecursionRelationsV17}, as expected.

\subsection{Normalization of the \bbd{U(1)} R-charge}

We also note that this present paper (and earlier work~\cite{Hellerman:2017sur, Hellerman:2018xpi, Hellerman:2020sqj} on the same subject) use a different convention for normalizing the \bbd{U(1)} R-charge
in an \bbd{{\cal N} = 2} superconformal theory than the one commonly used elsewhere.  In the commonly used convention the supercharges have R-charge \bbd{\mp 1} and
the scalar component of a free vector multiplet has R-charge \bbd{\pm 2}.  In our own convention the supercharges have \bbd{U(1)} R-charge \bbd{\pm\hh} and the R-charge
of the scalar component of a free vector multiplet has R-charge \bbd{\pm 1}.  The translation between our normalization convention for the R-charge and the
one used almost universally elsewhere is
\bbb
[U(1)~{\rm R-charge}]\ll{\hbox{\tiny here,~\cite{Hellerman:2017sur, Hellerman:2018xpi, Hellerman:2020sqj}}} =
 \hh\cc [U(1)~{\rm R-charge}]\lrm{ {{\rm everywhere}\atop{\rm else}}} \ , 
 \xxnn
  [U(1)~{\rm R-charge}]\lrm{ {{\rm everywhere}\atop{\rm else}}} = 2\cc [U(1)~{\rm R-charge}]\ll{\hbox{\tiny here,~\cite{Hellerman:2017sur, Hellerman:2018xpi, Hellerman:2020sqj}}}\ .
\eee

Now we  we will use the normalization \rr{ConjugateMomentaToAScalarField} of the conjugate momenta to the \bbd{a-}field, to compute the canonical expression for the $R$-current density.  In 
our normalization the R-charge, the total R-charge \bbd{\hat{J},} commutes with the vector multiplet scalar as
\bbb
[\hat{J}, a] = + \ordinary{1}\times a \ , \llsk\llsk [\hat{J}, \bar{a}] = - \ordinary{1}\times \bar{a}\ ,
\eee
so for the density we have
\bbb
\r\lrm J =  \ordinary{+i}\cc \big [ \cc a\cc \Pi\ll a - \bar{a} \cc \Pi\ll{\bar{a}} \cc \big ]
= \ordinary{+{i\over{4\pi}}} \cc {\tt Im}[\s] \cc \big [ \cc a\cc \dot{\bar{a}} - \bar{a} \cc \dot{a} \cc \big ]
\eee

\subsection{Conventions and notations for the normalizations
of correlation functions}\label{CorrelationFunctionNormalizationSection}

In the literature there are several similar but slightly different
notations for differently-normalized correlation functions of Coulomb-branch chiral primaries of \bbd{{\cal N} = 2} theories in \bbd{D=4}.  In
some cases the same notation is used differently in
different papers, to indicate correlators with
different normalizations.  Here we will give a summary of
the main conventions and notations that are relevant to
the computation of correlators here and in the literature we have
referred to.

\subsubsection{The notation correlators \bbd{G\ll{2n}}, the K\"ahler potential \bbd{K}, and the Zamolodchikov metric \bbd{g\ll{\t\bar{\t}}}}\label{ZamolodchikovMetricAndCorrelatorConventionsSection}

\heading{The two-point functions as normalized in~\cite{Gerchkovitz:2016gxx}}

Here, the Zamolodchikov metric is defined by eq. \aleq{1.8}, 
\bbb
g\ll{i\bar{j}}= |x-y|\uu{2 D}\cc \langle O\ll i(x) \overline{O}\ll{\bar{j}}
(y) \rangle
\eee
which in four dimensions means
\bbb
g\ll{i\bar{j}} = |x-y|\uu 8\cc \langle O\ll i(x) \overline{O}\ll{\bar{j}}
(y) \rangle
\een{ZamolodchikovMetricEqualsMarginalTwoPointFunction}

The capital-\bbd{G} metric defined in \aleq{1.19} is "the Hermitean metric on
the vector bundle" and is proportional to, but does not have the same normalization
as, the Zamolodchikov metric:
\bbb
\langle \co\ll I(x) \bar{\co}\ll{\bar{J}}(y) \rangle = |x-y|\uu{-2\D\ll I}\cc
G\ll{I\bar{J}}
\een{DefinitionOfAMetricSimilarToTheZamolodchikovMetric}

We also have the formula, from their \aleq{3.3}:
\bbb
G\ll{2n} \equiv
1\times |x-y|\uu{4n}\cc \big \langle \co\ll n(x)\overline{\co\ll n}(y) \big\rangle
\eee
and then they have the formula
\bbb
G\ll 2 = 16\cc \pp\ll\t\bar{\pp}\ll{\bar{\t}}\cc {\rm log}\big [ Z\ll{S\uu 4} \big ]
\eee

The best way to \emm{define} the normalization of the Zamolodchikov metric -- in any given convention -- is to do it in a way that is completely independent of the
normalization with which the operators enter the Lagrangian.  In~\cite{Gerchkovitz:2016gxx} we have (in four dimensions specifically)
\bbb
g\ll{\t\uu i \bar{\t}\uu j} 
 =2\uu{10}\cc \pi\uu 4\cc  |x - y|\uu 8\cc \bigg \langle {{\d S}\over{\d \t\uu i(x)}} \cc {{\d S}\over{\d \bar{\t}\uu j(y)}} \bigg \rangle
\eee

Then the way ref.~\cite{Gerchkovitz:2016gxx} is normalizing the marginal operator \bbd{C\ll i} entering the action can be understood by comparing their \aleq{1.16} with their \aleq{2.1}, which gives
\bbb
{{\d S}\over{\d \t\uu i}} = {1\over{32\pi\sqd}} \cc C\ll i\ ,
 \llsk\llsk C\ll i \cc  =  32\pi\sqd\cc {{\d S}\over{\d \t\uu i}}
\een{AppearanceOfMarginalsInTheActionNormalization}
so from this we infer
\bbb
g\ll{i\bar{j}} = |x-y|\uu 8\cc \langle C\ll i(x) C\ll{\bar{j}}
(y) \rangle\ .
\een{InferredEquationForRelationshipBetweenNormalizationsOfZamolodchikovMetricAndCMarginals}
This implies another relationship among normalizations: Comparing our equation \rr{InferredEquationForRelationshipBetweenNormalizationsOfZamolodchikovMetricAndCMarginals} with equation \aleq{1.8} of~\cite{Gerchkovitz:2016gxx}
we infer the normalization relationship
\bbb
O\ll i \cc \bigg |\ll{\hbox{eq.~\aleq{1.8}}} =    C\ll i\cc 
\een{InferredEqualityBetweenNoncalligraphicOAndCapitalC} 
The relationship between the K\"ahler potential and the correlators (as normalized by~\cite{Gerchkovitz:2016gxx}) is known from~\cite{Gerchkovitz:2016gxx} 's eq. \aleq{3.4} which reads
\bbb
G\ll 2 = 16\cc \pp\ll\t\pp\ll{\bar{\t}}\cc {\rm log}[Z]\ .
\eee
One also has
\bbb
{\rm log}[Z]  = {1\over{3\times 2\uu{12}}}\cc K \bigg |\ll{\hbox{eq.~\aleq{1.4}}} \ .
\een{NormalizationOfTheKahlerPotentialRelativeToLogOfZAsDefinedOperationallyAndStatedExplicitlyInFootnote16OfTheCBCPaper}
As the authors of ~\cite{Gerchkovitz:2016gxx} note in their footnote \aleq{16}, the normalization in eq. \aleq{1.4} of~\cite{Gerchkovitz:2016gxx} should be ignored for purposes of reading~\cite{Gerchkovitz:2016gxx}, as refers to a normalization from the previous papers
\cite{Gerchkovitz:2014gta} and~\cite{Gomis:2014woa}.  The operational definition of the relationship of the K\"ahler potential to the log of the partition function in
\cite{Gerchkovitz:2016gxx}, is actually \rr{NormalizationOfTheKahlerPotentialRelativeToLogOfZAsDefinedOperationallyAndStatedExplicitlyInFootnote16OfTheCBCPaper}.

  From \aleq{1.5} and \aleq{3.4} we have
\bbb
|x-y|\uu 4\langle \co\ls\t(x) \cob\ls{\bar{\t}}(y)\rangle = G\ll 2 =  16\cc \pp\ll\t\bar{\pp}\ll{\bar{\t}}\cc {\rm log}[Z]
\een{LotsOfNormalizationConventionsInCBCPaper}
and the equality between the first term and the last
is consistent (setting \bbd{r\to 1} anyway) with our result \rr{GeneralConventionIndependentFormulaeForAntipodalSphereCorrelatorsVC} for the
Weyl transformation between the computation of two-point functions in the \bbd{\IR\uu 4} and \bbd{S\uu 4} conformal frames, for operators with \bbd{\D = 2}.
Then we can rewrite this as
\bbb
\pp\ll\t\bar{\pp}\ll{\bar{\t}}\cc {\rm log}[Z]  = {1\over{16}}\cc
 |x-y|\uu 4\langle \co\ls\t(x) \cob\ls{\bar{\t}}(y)\rangle  
 = {1\over{16}}\cc  G\ll 2   
\een{LotsOfNormalizationConventionsInCBCPaperReversed}

Then the higher \bbd{G\ll{2n}} can only be normalized as the correlators of higher powers of the same operator insertions, with higher powers of the same normalization
factors,
\bbb
G\ll{2n} = |x-y|\uu {4n}\langle [\co\ls\t(x)]\uu n [\cob\ls{\bar{\t}}(y)]\uu n\rangle
\een{InferredDefinitionOfGSub2nInCBCPaper}

Next we explain how these relate to the normalization
of the Zamolodchikov metric \bbd{g\lrm{I\bar{J}}} and the (differently normalized as we shall see) capital \bbd{G} metric
\bbd{G\lrm{I\bar{J}}}.  Ref.~\cite{Gerchkovitz:2016gxx} is consistent about its normalization \rr{NormalizationOfTheKahlerPotentialRelativeToLogOfZAsDefinedOperationallyAndStatedExplicitlyInFootnote16OfTheCBCPaper} of the K\"ahler potential outside of
their \aleq{1.4} (see their footnote \aleq{16} -- they say \rr{NormalizationOfTheKahlerPotentialRelativeToLogOfZAsDefinedOperationallyAndStatedExplicitlyInFootnote16OfTheCBCPaper} is correct, and 
that their own eq. \aleq{1.4} was cut and pasted from~\cite{Gerchkovitz:2014gta} and~\cite{Gomis:2014woa} without matching normalizations consistently) then we have
\bbb
{\rm log}[Z]= {1\over{3\times 2\uu{12}}}\cc K \bigg |\ll{\hbox{eq.~\aleq{1.4}}} \ ,
\eee
and so
\bbb
g\lrm{I\bar{J}}  
 = K\ll{,I\bar{J}} \bigg |\ll{\hbox{eq.~\aleq{1.4}}}  = 3\times 2\uu{12}\cc \pp\ll{\t\uu I} \bar{\pp}\ll{\bar{\t}\uu J}\cc {\rm log}[Z]
 \xxnn
 = 
3\times 2\uu{12}\cc  \langle \co\ll{\t\uu I}(N) \cob\ll{\bar{\t}\uu J}(S) \rangle\ll{S\uu 4}  = 3\times 2\uu 8\cc
 |x-y|\uu 4\langle \co\ll{\t\uu I}(x) \cob\ll{\bar{\t}\uu J}(y)\rangle  
 = 3\times 2\uu 8\cc  G\ll 2   
\een{UsingLotsOfNormalizationConventionsInCBCPaperReversed}
with the subscripts \bbd{I\bar{J}} implicit in the last one, \bbd{G\ll 2}.

That equation fixes the relationship of the normalizations
of all the correlators to \bbd{K} and to the normalization
of the Zamolodchikov metric \bbd{g\ll{I\bar{J}}}.  Next
we fix the normalization of the capital-\bbd{G} metric \bbd{G\lrm{I\bar{J}},} the one defined by two-point functions of
the chiral primaries as opposed to the two-point functions of the marginal
operators.  

Then in \aleq{1.19} of~\cite{Gerchkovitz:2016gxx} we have
\bbb
 G\lrm{I\bar{J}} = G\ll{2~~ {\rm I\bar{J}}} = |x-y|\uu 4\cc \langle \co\lrm I(x) \cob\lrm J(y) \rangle 
 \xxnn
 = 16\cc \pp\ll{\t\uu I} \bar{\pp}\ll{\bar{\t}\uu J}\cc {\rm log}[Z] = {1\over{3\times 2\uu 8}}\cc g\lrm{I\bar{J}}  
 = {1\over{3\times 2\uu 8}}\cc  K\ll{,I\bar{J}} \bigg |\ll{\hbox{eq.~\aleq{1.4}}} 
\eee
We can also relate these to the two-point functions of the marginals \bbd{O\ll I} and \bbd{C\ll I} (which are just the same up to a phase,
see eq. \rr{InferredEqualityBetweenNoncalligraphicOAndCapitalC}) using  \rr{ZamolodchikovMetricEqualsMarginalTwoPointFunction},\rr{InferredEquationForRelationshipBetweenNormalizationsOfZamolodchikovMetricAndCMarginals} and we have
\bbb
 G\lrm{I\bar{J}} = G\ll{2~~ {\rm I\bar{J}}} = |x-y|\uu 4\cc \langle \co\lrm I(x) \cob\lrm J(y) \rangle 
 \xxnn
 = 16\cc \pp\ll{\t\uu I} \bar{\pp}\ll{\bar{\t}\uu J}\cc {\rm log}[Z] = {1\over{3\times 2\uu 8}}\cc g\lrm{I\bar{J}}  
 = {1\over{3\times 2\uu 8}}\cc  K\ll{,I\bar{J}} \bigg |\ll{\hbox{eq.~\aleq{1.4}}} 
 \xxnn
 = {1\over{3\times 2\uu 8}}\cc  |x-y|\uu 8\cc \langle C\ll i(x) C\ll{\bar{j}}
(y) \rangle =  {1\over{3\times 2\uu 8}}\cc  |x-y|\uu 8\cc \langle O\ll i(x) O\ll{\bar{j}}
(y) \rangle
\eee

So if we wish to ignore all other objects and just compare two-point functions directly,
\bbb
\bbsk
 |x-y|\uu 4\cc \langle \co\lrm I(x) \cob\lrm J(y) \rangle = 
 {1\over{3\times 2\uu 8}}\cc  |x-y|\uu 8\cc \langle C\ll i(x) C\ll{\bar{j}}
(y) \rangle =  {1\over{3\times 2\uu 8}}\cc  |x-y|\uu 8\cc \langle O\ll i(x) O\ll{\bar{j}}
(y) \rangle \llsk
\een{DirectRelationshipsBetweenNormalizationsOfCorrelatorsInCBCPaper}
The coefficients of proportionality among the various two-point functions are summarized in table~\ref{TwoPointFunctionCouplingNormalizationTable}.

\begin{center}
\begin{tabular}{ |c||c|c|c| } 
 \hline\hline
\backslashbox{denominator}{numerator} &  $  {{G\ll 2 = G\ll{\t\bar{\t}} =  }\atop{ \langle\co\cob\rangle\ll{}}}$ &  $   \pp\ll\t\pp\ll{\bar{\t}}\cc {\rm log}[Z] $&  $  {  {g\ll{\t\bar{\t}}  =   \langle C\overline{C}\rangle\ll{}  }\atop  { = \langle O\overline{O}\rangle\ll{}  }   }$   \\ \hline\hline
 $  {{G\ll 2 = G\ll{\t \bar{\t}}= }\atop{ \langle\co\cob\rangle\ll{}}} $ &   $1 $ & $ {1\over{16}} $ &   $ 3\times 2\uu {+8}  $    \\ \hline
$   \pp\ll\t\pp\ll{\bar{\t}}\cc {\rm log}[Z] $ & $    16 $          & $ 1 $&  $ 3\times 2\uu {+12}  $   \\ \hline
$  {  {     g\ll{\t\bar{\t}}  =  \langle C\overline{C}\rangle\ll{}  }\atop  { = \langle O\overline{O}\rangle\ll{}  }   }$       & $ 3\uu{-1}\times 2\uu{-8} $    & $ 3\uu{-1}\times 2\uu{-12}  $& $  1 $    \\ \hline
\end{tabular}
\captionof{table}{Relationships among normalizations of two-point functions in ref. ~\cite{Gerchkovitz:2016gxx}. In this table, the notation \bbd{\langle [{\rm operator}]\ll 1 \cc [{\rm operator}]\ll 2 \rangle} denotes
the correlator\\
 \bbd{ |x - y|\uu{2\D\lrm{operators}} \cc \langle [{\rm operator}]\ll 1(x) \cc [{\rm operator}]\ll 2(y) \rangle}.
The caligraphic \bbd{\co } are BPS scalar chiral primary operators of dimension \bbd{2} and the \bbd{C} and noncaligraphic \bbd{O} are the complex marginal operators
in the same \bbd{{\cal N} = 2} superconformal supermultiplet, with vanishing \bbd{R-}charge and dimension \bbd{4}.    }\label{TwoPointFunctionCouplingNormalizationTable}
\end{center}

\subsubsection{Relationship of \bbd{\IR\uu D} two-point functions with \bbd{S\uu D} two-point functions}

In everything above, we have given normalizations for two-point functions on flat space rather than
on the sphere.  The relationship between two-point functions on flat space and two-point functions at antipodal points on \bbd{S\uu D} of radius \bbd{R} is
\bbb
\langle \co\ll \D(N)\co\ll \D(S) \rangle\bigg |\ll{S\uu D} =   (2R)\uu{-2\D}\cc |x-y|\uu{2\D}\cc \langle \co\ll \D(x) \co\ll\D(y) \rangle \cc \bigg |\ll{\IR\uu D} \ ,
\xxn{GeneralConventionIndependentFormulaeForAntipodalSphereCorrelatorsVA}
\langle \co\ll \D(x) \co\ll\D(y) \rangle \cc \bigg |\ll{\IR\uu D} =  (2R)\uu{+2\D}\cc |x-y|\uu{-2\D}\cc\langle \co\ll \D(N)\co\ll \D(S) \rangle\bigg |\ll{S\uu D} \ .
\een{GeneralConventionIndependentFormulaeForAntipodalSphereCorrelatorsVB}
Or, the relationship can be summed up most intuitively and symmetrically by
\bbb
(2R)\uu{2\D}\cc \langle \co\ll \D(N) \co\ll\D(S) \rangle \cc \bigg |\ll{S\uu D} =  |x-y|\uu{2\D}\cc\langle \co\ll \D(x)\co\ll \D(y) \rangle\bigg |\ll{\IR\uu D} 
\een{GeneralConventionIndependentFormulaeForAntipodalSphereCorrelatorsVC}
This has nothing to do with the normalization of the operators themselves; it just follows from the application of the conformal transformation to the correlator, using
the standard metric on the round \bbd{S\uu D} of radius \bbd{R} as described in Euclidean coordinates on the conformally equivalent \bbd{\IR\uu D}:
\bbb
ds\sqd \cc\bigg |\ll{S\uu D} = {{4 R\uu 4}\over{(x\ll a\sqd + R\sqd)\sqd}}\cc ds\sqd\cc \bigg |\ll{\IR\uu D}\ .
\eee

So under the change of conformal frame, the metric transforms under the Weyl transformation as
\bbb
g\ll{\bullet\bullet} \bigg |\ll{S\uu D} =  \bigg [ \cc {4\cc R\uu 4\over{(R\sqd + r\sqd)\sqd}} \cc \bigg ]\uu{+1} g\ll{\bullet\bullet}\bigg |\ll{\IR\uu D}\ ,
\eee
a primary operator of dimension \bbd{\D} transforms as
\bbb
\co\ll \D \cc \bigg |\ll{S\uu D} = 
\bigg [ \cc {4\cc R\uu 4\over{(R\sqd + r\sqd)\sqd}} \cc \bigg ]\uu{-\D / 2}\cc \co\ll \D \cc \bigg |\ll{\IR\uu D} 
\eee

\subsubsection{Summary of various normalizations of the K\"ahler potential on \bbd{{\cal N}=2} superconformal theory space
\WarningOut{\shg{Make sure and keep this one, since we refer to it in sec.~\ref{AFuncHoloCouplingTransformation} in the body of the paper.}}
}\label{SummaryOfNormalizationsOfTheKahlerPotential}

The K\"ahler potential and Zamolodchikov metric are normalized differently in most of~\cite{Gerchkovitz:2016gxx} than in~\cite{Gerchkovitz:2014gta, Gomis:2014woa} and
in equation \aleq{1.4} of~\cite{Gerchkovitz:2016gxx};
here we are using the normalization for \bbd{K} that appears in~\cite{Gerchkovitz:2014gta, Gomis:2014woa}, and also in eq. \aleq{1.4} of~\cite{Gerchkovitz:2016gxx}.  It is related to the normalization appearing in most of
\cite{Gerchkovitz:2016gxx} by 
\bbb
K\bigg |\lrm{this~paper} = K\biggl |\ll{ {\hbox{ refs.~\cite{Gerchkovitz:2014gta, Gomis:2014woa} and }} \atop {\hbox{eq. \aleq{1.4} of~\cite{Gerchkovitz:2016gxx}}}}  = 
2\uu{- 10}\cc K\biggl |\ll{ {\hbox{ref. ~\cite{Gerchkovitz:2016gxx}, except}\atop{\hbox{for their eq. \aleq{1.4}}}}} = 12\cc {\rm log}[Z\ll{S\uu 4}]\ .
\eee
See footnote \alfoo{16} of ~\cite{Gerchkovitz:2016gxx}.

\subsubsection{The unnormalized partition function with insertions, \bbd{Z\ll n} and its connected version \bbd{q\ll n}}

The relationship with \bbd{G\ll{2n}} is
\bbb
G\ll{2n} = 2\uu{4n} {{e\uu{q\ll n}}\over{Z\ll{S\uu 4}}} = 2\uu{4n} {{e\uu{q\ll n}}\over{Z\ll{S\uu 4}}} 
\xxnn
 e\uu{q\ll n} = 2\uu{-4n}\cc Z\ll {S\uu 4}\cc G\ll{2n}\ .
\eee
where (from our inferred definition of \bbd{G\ll {2n}} in~\cite{Gerchkovitz:2016gxx})
\bbb
G\ll{2n} = |x-y|\uu {4n}\langle [\co\ls\t(x)]\uu n [\cob\ls{\bar{\t}}(y)]\uu n\rangle
\een{InferredDefinitionOfGSub2nInCBCPaperRecap}
So technically speaking,
\bbb
e\uu{q\ll n} = [{1\over{16}}\cc |x - y|\uu 4 ]\uu n \times [{\rm partition~function~with~additional~source~action~} S\lrm{source}]
\xxnn
S\lrm{source} \equiv -n\cc {\rm log}[\co(x)] - n\cc {\rm log}[\cob(y)]
\eee
Then we can further decompose these as
\bbb
e\uu{q\ll n} = Z\ll n = Z\ll n\uprm{EFT} Z\ll n\uprm{mmp} = \exp{q\ll n\uprm{EFT} + q\ll n\uprm{mmp}}\ .
\eee

\subsubsection{The Oviedo quotient \bbd{F\uprm{Oviedo}}}

The "Oviedo quotient" of \movie~ is something the authors of \movie~ do not even give a name to other than an unqualified "F".  This ratio is of central importance in \movie~ but many other
objects in the adjacent literature including the present paper
are denoted by "F" with various subscripts and superscripts,
so we give the ratio "F" of \movie~ a name,
\bbb
F\uprm{Oviedo}[n,g] = {{G\ll {2n}\urm{SQCD}}\over{G\ll{2n}\uu{{\cal N} = 4}}}\ ,
\eee
defined in their \aleq{2.22}. They also define
\bbb
F\uprm{Oviedo}\ll\infty[\ddsc] \equiv \DDSL F\uprm{Oviedo}[n,g] 
\eee
and
\bbb
{\cal F}\uprm{Oviedo} [\ddsc]\equiv {\rm Log}\bigg (\cc F\uprm{Oviedo}\ll\infty[\ddsc]\cc \bigg )
\eee

\subsubsection{The \bbd{\Delta C\ll 1} function}

First \mgkt define a symbol \bbd{\Delta G} that is just the same as the Oviedo quotient:
\bbb
\Delta G\bigg |\lrm{\smgkt} = F\uprm{Oviedo}
\een{GKDDeltaGDefinition}
Then they define
\bbb
C \equiv {\rm log}[G]  = \sum\ll{k\geq 0}\cc n\uu{1-k}\cc C\ll k[\ddsc]\ ,
\eee
\bbb
\Delta C = {\rm log}[(\Delta G)\lrm{\smgkt}] = {\rm log}[F\uprm{Oviedo}]  = \sum\ll{k\geq 0}\cc n\uu{1-k}\cc \Delta C\ll k[\ddsc]\ ,
\een{DeltaCDefinitionAndExpansion}
so that\footnote{These values are quoted from page 29 of~\cite{Grassi:2019txd}, below their eq. \aleq{4.9}.}
\bbb
C\ll 0\upp{{\cal N} = 4} = 2\cc {\rm log}[\ddsc] - 2\ ,
\xxnn
C\ll 1\upp{{\cal N} = 4} =  {3\over 2}\cc {\rm log}[\ddsc] + {\rm log}[2\pi]\ ,
\eee
and the additional contributions \bbd{\Delta C} for the case of SQCD, start at order \bbd{n\uu 0} in the double-scaling limit:
\bbb
\Delta C\ll 0 = 0
\xxnn
\Delta C\ll 1 =  {\cal F}\uprm{Oviedo}[\ddsc]\ .
\eee
and the main object of study of \mgkt's matrix model is \bbd{\Delta C\ll 1}.

\subsubsection{Ref.~\cite{Grassi:2019txd}'s worldline-instantont function \bbd{F\uprm{inst}}}

The function \bbd{F\uprm{inst}} in ref.~\cite{Grassi:2019txd} is defined as related to their \bbd{\Delta C\ll 1} by~\cite{Grassi:2019txd}'s eq. \aleq{4.20}:
\bbb
\Delta C\ll 1 = 12\cc {\rm log}[\g\lrm G] - 1 - {1\over 3}\cc {\rm log}[2] - 16\cc \ddsc\cc {\rm log}[2] + \hh\cc {\rm log}[\ddsc] + F\uprm{inst}[\ddsc]\ .
\eee

In terms of our own MMP function, the worldline-instanton function of~\cite{Grassi:2019txd} is exactly the double-scaling limit of the MMP piece of \bbd{q\ll n},
given in our eq. \rr{MatchingTheDoubleScalingLimitOfOurMMPFunctionWithTheWorldlineInstantonFunctionInGKT} and recapped here:
\bbb
F\uprm{inst}[\ddsc] =  \DDSL q\uprm{mmp}\ll n  = \DDSL\big [ q\ll n - q\ll n\uprm{eft} \big ]
\een{MatchingTheDoubleScalingLimitOfOurMMPFunctionWithTheWorldlineInstantonFunctionInGKTRecap}

\section{\blue{Appendix:} Weak-coupling expansions}\label{WeakCouplingExpansions}

\subsection{Weak-coupling expansion of the relationship between the UV coupling \bbd{\t} and the IR coupling \bbd{\s}}\label{WeakCouplingRelationshipBetweenTauAndSigma}

Using
\bbd{q\equiv e^{2 \pi i \tau} = \lambda(\sigma)}between \bbd{\s} and the UV coupling
\bbd{\t}. and defining
\bbb
y =e^{\pi i \sigma} 
\eee
we can expand the modular lambda function \bbd{\l(\sigma)} at large \bbd{{\rm Im}(\s),} giving
\begin{equation}
  \lambda(\sigma) \simeq 16 y - 128 y^2 + 704 y^3 - 3072 y^4 + 11488 y^5 + \dots
\label{MLWCE1}\end{equation}
so that the relationship between the couplings is
\bbb
\t 
=  { \s\over 2} - {{2i}\over \pi}\cc {\tt log}[2] + {i\over \pi}\cc \bigg [ \cc
4\cc y - 6 \cc y\sqd + {{16}\over 3}\cc y\uu 3 -3\cc  y\uu 4 + {{24}\over 5}\cc  y\uu 5
-8\cc y\uu 6
\llsk
\xxnn
  + {{32}\over 7}\cc y\uu 7 - {3\over 2}\cc  y\uu 8  + {{52}\over 9}\cc y\uu 9 - {{36}\over 5}\cc y\uu{10} + O(y\uu{11})  \cc \bigg ]\ .
\een{MLWCE2}
and
\bbb
\s = 2\t + {{4i}\over \pi}\cc {\tt log}[2]
- {i\over\pi} \cc \biggl [ \cc
 {q\over 2} + {{13}\over{64}}\cc q\sqd + {{23}\over{192}}\cc q\uu 3
+ {{2,701}\over{ 32,768}}\cc  q\uu 4 + {{5,057}\over{ 81,920}}\cc q\uu 5 + {{76,715}\over {1,572,864}}\cc q\uu 6
+ {{146,749}\over{ 3,670,016}}\cc  q\uu 7
\llsk
\xxnn
 + {{144,644,749}\over{4,294,967,296}} \cc  q\uu 8 + {{279,805,685}\over{ 9,663,676,416}} \cc q\uu 9+ {{4,346,533,901}\over{ 171,798,691,840}}\cc  q\uu{10} + O(q\uu{11})
 \cc \biggl ]
\een{MLWCE3}

We will mostly only need the relationships in the zero-instanton
approximation:
\bbb
\t 
\simeq  { \s\over 2} - {{2i}\over \pi}\cc {\tt log}[2] \ , 
\llsk\llsk \s \simeq 2\t + {{4i}\over \pi}\cc {\tt log}[2]\ .
\een{ZeroInstantonApproximationToRelationBetweenCouplings}

\subsection{Weak-coupling expansion of some modular forms as functions of \bbd{\t}}\label{DoubleScalingExpansionForPrefactor}

First we expand the Dedekind eta function \bbd{\eta(\s)}, which is a modular form of weight \bbd{(+\hh,0)}.  Using
\bbb
y = \exp{\pi i \s} = {1\over{16}}\cc q\ ,
\een{MLWCE5}
and the product formula for the eta function,
\bbb
\eta(\s) = \exp{{{\pi i}\over{12}}}\cc\prod\ll{m\geq 1}\cc (1 - \exp{2\pi i m \s})\ ,
\eee
we have
\bbb
\eta(\s) = y\uu{+{1\over{12}}} + O(y\uu{+{{25}\over{12}}}) = 2\uu{-{1\over 3}}\cc q\uu{+{1\over{12}}} + O(q\uu{+{{25}\over{12}}}) \ .
\eee
Its absolute value \bbd{|\eta(\s)|} is a modular form of weights \bbd{(+{1\over 4},+{1\over 4}),} whose weak-coupling expansion is
\bbb
|\eta(\s)| = 2\uu{-{1\over 3}}\cc |q|\uu{+{1\over{12}}}+ O(|q|\uu{+{{25}\over{12}}})\ ,
\eee
and its p\bbd{\uth} power, a modular form of weights \bbd{(+{p\over 4},+{p\over 4}),} has the weak-coupling expansion.
Specifically for \bbd{p=-8}we have
\bbb
|\eta(\s)|\uu{-8} =2\uu{+{8\over 3}}\cc |q|\uu{-{2\over 3}}+ O(|q|\uu{+{4\over 3}})\ .
\een{WeakCouplingExpansionOfAbsEtaOfSigmaToTheMinusEight}
which is a nonholomorphic modular form of weight \bbd{(-2,-2)}.
Next, using \rr{ZeroInstantonApproximationToRelationBetweenCouplings} we expand the imaginary part of \bbd{\s,} which is a nonholomorphic modular form of weights \bbd{(-1,-1)}:
\bbb
{\rm Im}[\s] = 2\cc {\rm Im}[\t]+ {4\over \pi}\cc {\tt log}[2] + O(|q|)\ ,
\een{MLWCE4}
and its p\bbd{\uth} power is a nonholomorphic modular form of weights \bbd{(-p,-p)}, with the weak-coupling expansion
\bbb
[{\rm Im}(\s)]\uu p = 2\uu p \cc  [{\rm Im}(\t) + {2\over \pi}\cc {\tt log}[2]]\uu p + O(|q|\cc [{\rm Im}(\t)]\uu{p-1})\ ,
\eee
and specifically for \bbd{p=-2} we have
\bbb
{1\over{[{\rm Im}(\s)]\sqd }} = {1\over 4}\cc  [{\rm Im}(\t) + {2\over \pi}\cc {\tt log}[2]]\uu{-2} + O({{|q|}\over[{\rm Im}(\s)]\uu 3 })\ ,
\een{WeakCouplingExpansionOfImSigmaToTheMinusTwo}
which is a modular form of weight \bbd{(+2,+2)}.

Finally we combine \rr{WeakCouplingExpansionOfAbsEtaOfSigmaToTheMinusEight} with \rr{WeakCouplingExpansionOfImSigmaToTheMinusTwo} to get the modular-invariant combination \bbd{{1\over{|\eta(\s)|\uu 8\cc [{\rm Im}(\s)]\sqd }} ,} which has the weak-coupling
expansion
\bbb
{1\over{|\eta(\s)|\uu 8\cc [{\rm Im}(\s)]\sqd }} =  {{2\uu{+{2\over 3}}\cc |q|\uu{-{2\over 3}}}\over{  [{\rm Im}(\t) + {2\over \pi}\cc {\tt log}[2]]\sqd  }} + O ({{ |q|\uu{+{1\over 3}} }\over{ [{\rm Im}[\t]]\sqd}})\ ,
\een{ResultForSixteenTimesExponentiatedACoefficient}

\subsection{Some weak-coupling/large-\bbd{n} expansions at fixed double-scaling parameter \bbd{\ddsc = {n\over{4\pi\cc {\rm Im}[\t]}}}}

Now we also would like to expand some quantities at weak-coupling simultaneously taking large \bbd{n = \JJM/2} while holding fixed the\footnote{In this paper we will always be using this definition  \bbd{\ddsc = {n\over{4\pi\cc {\rm Im}[\t]}}}, which agrees with the normalization of \bbd{\ddsc} as defined
in the second half of~\cite{Grassi:2019txd} and differs by a factor of \bbd{(4\pi)} from the normalization of \bbd{\ddsc} as defined
in the first half of~\cite{Grassi:2019txd}, and by a factor of \bbd{(4\pi)\sqd} from the the normalization of \bbd{\ddsc} as defined
in~\cite{Bourget:2018obm}.  See sec.~\ref{ThreeLambdaNormalizationsWhyWhyWhyWhyWhyGodWhy} of the Appendix.}
double-scaling parameter \bbd{\ddsc = {n\over{4\pi\cc {\rm Im}[\t]}}}.

Specifically we want to expand the prefactor in eq. \rr{MMPToOviedoQuotientInTermsOfFactor}.

First, use
\bbb
\exp{A\ls\t} \simeq {1\over {16}}\cc [{\rm lm}(\t) + {2\over \pi}\cc {\rm log}[2]]\uu{-2}
\eee
which means
\bbb
\exp{- nA\ls\t} \simeq 2\uu {4n}\cc  [{\rm lm}(\t) + {2\over \pi}\cc {\rm log}[2] ]\uu{+2n}
\eee
so
\bbb
\PREFAK \equiv \exp{- n A\ls{\t} - \tilde{B}} \cc {1\over{\G(2n + {5\over 2})}}\cc 2\uu{- 4n}\cc G\ll {2n}\uprm{{\cal N} = 4}
\xxnn
 \simeq
 \exp{ - \tilde{B}} \cc {1\over{\G(2n + {5\over 2})}}\cc 2\uu{- 4n}\cc G\ll {2n}\uprm{{\cal N} = 4} \cc 2\uu {4n}\cc  [{\rm lm}(\t) + {2\over \pi}\cc {\rm log}[2] ]\uu{+2n}
 \xxnn
 = \exp{ - \tilde{B}} \cc {1\over{\G(2n + {5\over 2})}}\cc G\ll {2n}\uprm{{\cal N} = 4} \cc   [{\rm lm}(\t) + {2\over \pi}\cc {\rm log}[2] ]\uu{+2n} 
 \xxnn
 = \exp{ - \tilde{B}} \cc {1\over{\G(2n + {5\over 2})}}\cc  [{\rm Im}[\t]]\uu{-2n}\cc \G(2n+2)  \cc   [{\rm lm}(\t) + {2\over \pi}\cc {\rm log}[2] ]\uu{+2n} 
  \xxnn
 = \exp{ - \tilde{B}} \cc {1\over{\G(2n + {5\over 2})}}\cc \G(2n+2)   \cc  [1 + {2\over \pi}\cc {\rm log}[2] \cc {1\over{{\rm lm}(\t)}} ]\uu{+2n} 
  \xxnn
 = \exp{ - \tilde{B}} \cc \bigg [ \cc 2\uu{-\hh}\cc n\uu{-\hh} + O(n\uu{-{3\over 2}}) \cc \bigg ]\cc  [1 + {2\over \pi}\cc {\rm log}[2] \cc {1\over{{\rm lm}(\t)}} ]\uu{+2n} 
 \xxnn
 = \exp{ - \tilde{B}} \cc \bigg [ \cc 2\uu{-\hh}\cc n\uu{-\hh} + O(n\uu{-{3\over 2}}) \cc \bigg ]\cc \exp{2n\cc {\rm log} [1 + {2\over \pi}\cc {\rm log}[2] \cc {1\over{{\rm lm}(\t)}} ]} 
  \xxnn
 = \exp{ - \tilde{B}} \cc \bigg [ \cc 2\uu{-\hh}\cc n\uu{-\hh} + O(n\uu{-{3\over 2}}) \cc \bigg ]\cc \exp{2n\cc   {2\over \pi}\cc {\rm log}[2] \cc {1\over{{\rm lm}(\t)}}  + O({n\over{[{\rm Im}(\t)]\sqd}})}
 \xxnn
 = \exp{ - \tilde{B}} \cc \bigg [ \cc 2\uu{-\hh}\cc n\uu{-\hh} + O(n\uu{-{3\over 2}}) \cc \bigg ]\cc \exp{ {4\over \pi}\cc {\rm log}[2] \cc {n\over{{\rm lm}(\t)}}}\cc
 \biggl [ \cc 1 +  O({n\over{[{\rm Im}(\t)]\sqd}}) \cc \biggl ]
\eee
Now, we know from eq. \rr{FinalResultForSchemeIndependentExpTildeBInTermsOfUnknownConstantBoldfaceC} that 
\bbb
\exp{\tilde{B}}\simeq \exp{\tilde{b}} \cc [{\rm Im}[\t]]\uu{-\hh}\ ,
\eee
for some coupling-independent constant \bbd{\exp{\tilde{b}}} that we do \emm{not} know how to calculate {\it ab initio.}

So it seems we have
\bbb
\PREFAK \simeq \exp{ - \tilde{B}} \cc 2\uu{-\hh}\cc n\uu{-\hh} \cc \exp{ {4\over \pi}\cc {\rm log}[2] \cc {n\over{{\rm lm}(\t)}}}
 \xxnn
  \simeq \exp{-\tilde{b}} \cc [{\rm Im}[\t]]\uu{+\hh}\cc 2\uu{-\hh}\cc n\uu{-\hh} \cc \exp{ {4\over \pi}\cc {\rm log}[2] \cc {n\over{{\rm lm}(\t)}}}\
  \xxnn 
  \simeq  \exp{-\tilde{b}}\cc \big [ \cc 2n / {\rm Im}[\t]\big ]\uu{-\hh} \cc \exp{ {4\over \pi}\cc {\rm log}[2] \cc {n\over{{\rm lm}(\t)}}}\ .
\eee
Now use the expression \bbd{\ddsc = {n\over{4\pi \cc {\rm Im}[\t]}}} for the double-scaling parameter~\cite{Bourget:2018obm, Grassi:2019txd},
to write
\bbb
 {{2n}\over{{\rm Im}[\t]}}= 8\pi \ddsc \llsk \llsk {4\over\pi}\cc {\rm log}(2)\cc {n\over{{\rm Im}[\t]}} = 16\cc {\rm log}(2)\cc\ddsc
 \ ,
\eee
and
\bbb
[ \cc 2n / {\rm Im}[\t]\big ]\uu{-\hh} = [8\pi\ddsc]\uu{-\hh}\llsk\llsk \exp{ {4\over \pi}\cc {\rm log}[2] \cc {n\over{{\rm lm}(\t)}}} = \exp{16\cc {\rm log}(2)\cc\ddsc }\ ,
\eee
and
\bbb
\PREFAK \simeq {{\exp{-\tilde{b}}}\over{\sqrt{8\pi \ddsc}}}\cc \exp{16\cc {\rm log}(2)\cc\ddsc }\
\een{FinalResultForDoubleScalingLimitOfPrefactorWithSimeqSymbol}
where \bbd{\tilde{b}} is a scheme-independent constant we have so far determined only numerically and the omitted terms are of order \bbd{n\uu{-1}}
at fixed \bbd{\ddsc.}  So we can write
\bbb
\PREFAK = {{\exp{-\tilde{b}}}\over{\sqrt{8\pi \ddsc}}}\cc \exp{16\cc {\rm log}(2)\cc\ddsc } + O(n\uu{-1})~{\rm at~fixed~}\ddsc\ ,
\een{FinalResultForDoubleScalingLimitOfPrefactorWithOrderOfOmittedTerms}
and/or
\bbb
\DDSL\cc \PREFAK = {{\exp{-\tilde{b}}}\over{\sqrt{8\pi \ddsc}}}\cc \exp{16\cc {\rm log}(2)\cc\ddsc } \ .
\een{FinalResultForDoubleScalingLimitOfPrefactorAsLimit}

\subsection{Weak-coupling expansion of the sphere partition function}\label{ZS4WeakCouplingExpansionSec}

Here we give the weak-coupling expansions of the \bbd{S\uu 4} partition functions for SQCD as computed in the Pestun-Nekrasov scheme and AGT scheme for the Euler-density
counterterm, which in the paper we have denoted \bbd{Z\lrm{{Pestun-}\atop{Nekrasov}}} and \bbd{Z\lrm{AGT}} respectively.

From ref.~\cite{Gerchkovitz:2016gxx} we have
\bbb
Z\lrm{{Pestun-}\atop{Nekrasov}}  = {1\over{4\pi\cc ({\rm Im}[\t])\uu{{3\over 2}}}}\cc \bigg [ \cc 1
- {{45\cc \zeta(3)}\over{16\pi\sqd\cc  ({\rm Im}[\t])\sqd} } + {{525\cc \zeta(5)}\over{64\pi\uu 3\cc  ({\rm Im}[\t])\uu 3 }} + O(({\rm Im}[\t])\uu{-4})\cc \bigg ] 
\xxnn
+ ({\rm gauge~instantons})\ ,
\een{ExplicitFormulaForSpherePartitionFunctionInAppendix}

From eqs. \aleq{3.9}, \aleq{4.4}-\aleq{4.7} of ref.\cite{Alday:2009aq} we have
\bbb
Z\lrm{AGT} = |1 - q|\uu{-4}\cc Z\lrm{{Pestun-}\atop{Nekrasov}} \ ,
\eee
where we have used the values \bbd{m\ll 0 = m\ll 1 = 1, Q=2} for the parameters \bbd{m\ll {0,1}, Q}, which correspond to conformal SQCD on the round \bbd{S\uu 4}.

\section{\blue{Appendix:} Modular forms and their anharmonic-group counterparts}\label{ModularFormsAndAnharmonicForms}

\subsection{Modular tensor calculus and the \bbd{\eta-}function as a compensator}

So as we all know, a holomorphic modular form of weight \bbd{k} is a function \bbd{F(\s)} of \bbd{\s} satisfying the
functional equation
\bbb
F(\s\pr) = (c\s + d)\uu k\cc F(\s)\ , 
\eee
for \bbd{\s\pr \equiv {{a\s + b}\over{c\s + d}}} for integers \bbd{a,b,c,d\in\IZ} satisfying \bbd{ad-bc=1}.  The modular group
is generated by the elements \bbd{T:\s\mapsto \s + 1} and \bbd{S:\s\to - {1\over\s}}, so we can specify the modular transformation
properties of any object by how it transforms under \bbd{S} and \bbd{T}.  So a holomorphic modular form \bbd{\O\ll k} of weight \bbd{k} transforms as 
\bbb
\s\to\s + 1:\llsk \O\ll k\to \O\ll k\ ,
\xxnn
\s\to - {1\over \s}:\llsk \O\ll k\to \s\uu k\cc \O\ll k\ .
\eee

The transformation property of a holomorphic modular form
is defined so that the holomorphic derivative with respect to \bbd{\s} transforms as a modular form of weight \bbd{+2}:
\bbb
\s\to\s + 1:\llsk \pp\ll \s\to \pp\ll \s\ ,
\xxnn
\s\to - {1\over \s}:\llsk \pp\ll\s \to \s\sqd \cc \pp\ll \s\ .
\eee

  The \bbd{\eta-}function
doesn't transform quite as a modular form, but as almost a modular form of weight \bbd{+\hh,} except some extra phases,
\bbb
\eta(\s + 1) = \exp{{{\pi i}\over{12}}}\cc \eta(\s)\ , \llsk
\eta(- {1\over\s}) = (- i \s)\uu{+\hh}\cc \eta(\s)\ .
\een{EtaFunctionTransformations}
The fact that the additional factors are pure phases, means that \bbd{|\eta(\s)|} transforms as a nonholomorphic
modular form of weights \bbd{(+{1\over 4}, +{1\over 4})} and the fact that the phases lie in \bbd{ {{\pi}\over {12}}\IZ} means
that \bbd{\eta\uu{24}(\s)} transforms as a modular form of weight \bbd{(+{12},0).}  This means that 
\bbd{|\eta(\s)|} and \bbd{\eta\uu{24}(\s)} can play the role of nonholomorphic and holomorphic "compensators" for nontrivial
modular transformation laws.

\subsection{Anharmonic tensor calculus}

\subsubsection{Transformation of \bbd{\t-}derivatives under the anharmonic group}\label{TransformationOfLittleTauDerivatitesUnderAnharmonicGroup}

Under the generators of the modular group, the holomorphic coordinates \bbd{q \equiv e\uu{2\pi i \t} = \l(\s)} transforms according to \rr{ModularTransformationsOfq}, which we recap here:
\bbb
S: \llsk q\to 1-q\ ,
\llsk\llsk
T: \llsk q\to {q\over{q-1}}\ .
\een{ModularTransformationsOfqRecap}

Consider the transformation law of the holomorphic \bbd{\t-}derivative
\bbb
\pp\ll\t = {q\over{2\pi i}}\cc \pp\ll q
\eee
under the anharmonic group.  First consider the transformation under \bbd{q\to 1-q,} which is
\bbb
q\to 1-q\ , \llsk \pp\ll\t \to {{1-q}\over q}\pp\ll\t\ .
\eee
And of course under \bbd{q\to +{1\over q}} we have 
\bbb
q\to +{1\over q}\ , \llsk \t\to -\t\  \ .
\eee

\subsubsection{Tensor calculus of the anharmonic group}
The correlation function \bbd{G\ll 2 = |x-y|\uu 4 \langle \co\ls\t(x) \cob\ls{\bar{\t}}(y)\rangle} transforms as the product of a \bbd{\t} derivative and a \bbd{\bar{\t}} derivative:
\bbb
q\to 1-q\ , \llsk G\ll 2\to {{|1-q|\sqd}\over{|q|\sqd}}\cc G\ll 2\ , 
\xxnn
q\to + {1\over q}\ , \llsk G\ll 2\to G\ll 2\ ,
\eee
and, since \bbd{\exp{A\ls{\t}}} transforms the same way as \bbd{G\ll 2,} so
\bbb
q\to 1-q\ , \llsk \exp{A\ls{\t}} \to  {{|1-q|\sqd}\over{|q|\sqd}}\cc \exp{A\ls{\t}}
\xxnn
q \to + {1\over q}\ , \llsk \exp{A\ls{\t}}  \to \exp{A\ls{\t}} \ ,
\eee
So in general we will say that an "anharmonic form" \bbd{\phi\ll{(w,\tilde{w})}} of "anharmonic weights" \bbd{(w,\tilde{w})} is any object that transforms as
\bbb
q\to 1-q\ , \llsk  \phi\ll{(w,\tilde{w})} \to \big ( {{1-q}\over q} \big )\uu{w/2}\cc \big ( {{1-\qb}\over \qb} \big )\uu{\tilde{w}/2}\cc \phi\ll{(w,\tilde{w})} \ ,
\xxnn
q \to + {1\over q}\ , \llsk  \phi\ll{(w,\tilde{w})} \to (-1)\uu{(w - \tilde{w})/2}\cc \phi\ll{(w,\tilde{w})} \ .
\eee
The phases are unambiguous so long as \bbd{w - \tilde{w}} is an even integer,
which is always the case we shall consider. In particular for \bbd{w = \tilde{w}}
we have
\bbb
q\to 1-q\ , \llsk  \phi\ll{(w,w)} \to \big | {{1-q}\over q} \big |\uu w\cc \phi\ll{(w,w)} \ ,
\xxnn
q \to + {1\over q}\ , \llsk  \phi\ll{(w,w)} \to  \phi\ll{(w,w)} \ .
\een{AnharmonicFormNonchiralWeightsWOnBothSides}
The derivative \bbd{\pp\ll\t} transforms as an anharmonic form of weights \bbd{(2,0)} and \bbd{\pp\ll\tb} transforms as an
anharmonic form of weight \bbd{(0,2)} and the laplacian \bbd{\pp\ll\t\pp\ll\tb} transforms as an anharmonic form
of weight \bbd{(2,2).}

\subsubsection{The anharmonic compensator \bbd{a\ll 6\sqd(q)}}

So just as \bbd{\eta\uu{24}(\s)} is a modular form of weights \bbd{(+12, 0)} whose \bbd{-k} power can take a modular form
of weights \bbd{(12k,0)} to a modular invariant function whose logarithm has the same laplacian, and 
\bbd{|\eta(\s)|} is a modular form of weights \bbd{(+{1\over 4}, +{1\over 4})} whose \bbd{-4k} power can take a modular form
of equal weights \bbd{(k,k)} to a modular invariant function whose logarithm has the same laplacian, 
we can try to construct holomorphic and nonchiral compensators playing the same role for anharmonic forms.

Start with the quantity
\bbb
a\ll 6(q) \equiv {q\over{(1-q)\sqd}}\ .
\eee
Under anharmonic transformations we have
\bbb
q\to 1-q\ , \llsk  a\ll 6 \to \big ( {{1-q}\over q} \big )\uu 3\cc  a\ll 6 \ ,
\xxnn
q \to + {1\over q}\ , \llsk  a\ll 6 \to +a\ll 6 \ .
\eee
The \bbd{q\to 1-q} transformation is that of an anharmonic form of weight \bbd{+6,} but the \bbd{q\to +{1\over q}}
transformation is not: An anharmonic form of weight \bbd{w} equal to \bbd{2} mod \bbd{4} would transform
under \bbd{q\to + {1\over q}} with a \bbd{-} sign rather than a \bbd{+} sign; in order to find an object with a covariant transformation law, we can simply square \bbd{a\ll 6} and we find that
\bbd{a\ll 6\sqd} does indeed transform as an anharmonic form of weight \bbd{+12.}

So we can shift the weights of any holomorphic anharmonic form by any integer multiple of \bbd{12}, by multiplying
it by \bbd{a\ll 6\sqd,} just as we could shift the weights of any holomorphic modular form by an integer multiple of 
\bbd{12}, by multiplying it by a power of \bbd{\eta\uu{24}(\s).}

We can also shift the anharmonic weights by \emm{any} amount that is the same for both the holomorphic and 
antiholomorphic transformations.  If \bbd{\phi\ll{(w,\tilde{w})}} is an anharmonic form of weights \bbd{(w,\tilde{w})} then
\bbb
\phi\ll{(w+\ell ,\tilde{w}+\ell)} \equiv |a\ll 6|\uu{\ell/6}\cc \phi\ll{(w,\tilde{w})}  = \big | {q\over{(1-q)\sqd}} \big |\uu{{\ell}\over 6}
\cc \cc \phi\ll{(w,\tilde{w})} 
\eee
is an anharmonic form of weights \bbd{(w+\ell ,\tilde{w}+\ell)} for any real \bbd{\ell} (or any complex \bbd{\ell} for that matter).

\subsection{Bridge between anharmonic and modular tensor calculi}

So the function \bbd{\ModToAnharbridge  \equiv {1\over{2\pi i}}\cc {{\l\pr(\s)}\over{\l(\s)}} = {1\over{2\pi i}}\cc{{d({\rm log}[\l(\s)])}\over{d\s}} ={1\over{2\pi i}}\cc {{d({\rm log}[q])}\over{d\s}} = {{d\t}\over{d\s}}} transforms as a modular form of weight \bbd{(2,0)} times an anharmonic
form of weight \bbd{(-2,0)}.  So \bbd{\ModToAnharbridge } is a bridge between modular forms and anharmonic forms.  In particular, if
\bbd{\phi\ll{(w,\tilde{w})}} is an anharmonic form of weights \bbd{(w,\tilde{w}),} then 
\bbb
\Upsilon\ll{(w,\tilde{w})} \equiv \ModToAnharbridge \uu{w/2} \bar{\ModToAnharbridge }\uu{\tilde{w}/2} \cc \phi\ll{(w,\tilde{w})}
\eee
transforms as a modular form of weight \bbd{(w,\tilde{w})} with no additional transformation.  Likewise
if \bbd{\Upsilon\ll{(w,\tilde{w})}} is a modular form of weight \bbd{(w,\tilde{w})}, then 
\bbb
\phi\ll{(w,\tilde{w})} \equiv \ModToAnharbridge \uu{-w/2} \bar{\ModToAnharbridge }\uu{-\tilde{w}/2} \cc \Upsilon\ll{(w,\tilde{w})}
\eee
transforms as an anharmonic form of weight \bbd{(w,\tilde{w})} with no additional transformation. 

\subsection{An identity on the derivative of the modular lambda function}\label{IdentityOnTheDerivativeOfTheModularLambdaFunction}

From a bridge and a compensator of each kind, we can construct the invariant combination
\bbb
\r \equiv a\ll 6\uu{-2}\cc \ModToAnharbridge \uu{-6}\cc \eta\uu{24}
\eee
which would seem to transform trivially under all modular and/or anharmonic transformations.  That is, it is completely invariant under the
modular group.  \bbd{\r} is invariant under \bbd{\s\to\s + 1} and under
\bbd{\s\to - {1\over \s}} which means it is invariant under the full modular group \bbd{SL(2,\IZ).}  It is also
single-valued, nonsingular, and holomorphic.  Since it is also modular invariant, that means it must be an entire function
of the klein \bbd{J}-invariant which we are calling \bbd{\k}.  Specifically, by checking weak coupling asymptotics we can see it asymptotes to \bbd{+{1\over 4}} at infinity.  So by Liouville's theorem in the \bbd{k-}plane (where \bbd{k} is the Kleinian invariant \bbd{J[\s]}) it must be equal to 
\bbd{+{1\over 4}} identically everywhere.

So we find
\bbb
\r = a\ll 6\uu{-2}\times \ModToAnharbridge \uu{-6} \times \eta\uu{24}  = {1\over 4}\ .
\eee
Or, written as an identity for the derivative of the modular \bbd{\l(\s)} function, we have
\bbb
[\l\pr(\s)]\uu{+6} = -256\pi\uu 6\times \l\uu 4(\s) \times (1 - \l(\s))\uu 4\times \eta\uu{24}(\s)\ .
\een{IdentityForDerivativeOfLambdaFunction}

Taking the cube root of the absolute value, and dividing by \bbd{4\pi\sqd |\l(\s)|\sqd} we find
\bbb
\big |\cc {{d\t}\over{d\s}}\cc \big |\sqd = {1\over{4\pi\sqd}}\cc \big |\cc {{d\cc {\rm Log}[q]}\over{d\s}}\cc \big |\sqd 
= 2\uu{+{2\over 3}}\cc  |\l(\s)|\uu{-{2\over 3}}\times |1 - \l(\s)|\uu{+{4\over 3}} \times |\eta(\s)|\uu 8\ .
\een{IdentityForTauToSigmaJacobian}
Using this identity we can write the expression for the exponentiated \bbd{A-}coefficient in the \bbd{\t-}frame as
\bbb
\bbsk
\exp{A\ls\t} = \big |\cc {{\l\pr(\s)}\over{\l(\s)}}\cc \big |\uu{-2}\cc \exp{A\ls\s} = 2\uu{-{{14}\over 3}}\cc  |\l(\s)|\uu{+{2\over 3}}\times |1 - \l(\s)|\uu{-{4\over 3}} \times |\eta(\s)|\uu {-8} \times {1\over{[ {\rm Im}(\s)]\sqd}}
\llsk
\een{SimplerFormulaForExponentialOfACoefficientInTauFrame}

\subsection{Uniqueness theorem for modular invariant harmonic functions}\label{UniquenessTheoremProof}

Suppose \bbd{f[\s,\bar{\s}]} is a real-valued, smooth, harmonic function that is \bbd{SL(2,\IZ)} invariant with \bbd{\s} transforming in the usual way by fractional linear transformations,
and suppose \bbd{f} is bounded by some power \bbd{{\rm Im}[\s]\uu p} as \bbd{{\rm Im}[\s]\to \infty}  Since \bbd{f} is modular-invariant, it can be written as a function of
the Klein invariant \bbd{k \equiv J[\s]}, which holomorphically maps the fundamental domain of \bbd{SL(2,\IZ)} in the UHP to the complex plane.  Harmonic functions
map holomorphically to harmonic functions, so 
\bbb
f[\s,\sb] = g[k, \bar{k}]\ , \llsk\llsk \pp\ll k \bar{\pp}\ll{\bar{k}}g[k,\bar{k}] = 0\ .
\eee
The function \bbd{k = J[\s]} grows as \bbd{|\exp{-2\pi i \s}| = \exp{+ 2\pi {\rm Im}[\s]}} at large \bbd{{\rm Im}[\s]} so a function bounded  by some power \bbd{{\rm Im}[\s]\uu p} is bounded by
\bbd{ ({\rm log}|k| / (2\pi))\uu p} .  Then \bbd{{\bar{\pp}\ll{\bar{k}}}\cc g[k,\bar{k}]} is homomorphic in \bbd{k}, nonsingular, and bounded by \bbd{{ ({\rm log}|k| )\uu{p-1}\over |k|}} as \bbd{|k|\to\infty}.
By Liouville's theorem,a bounded entire function is constant, so \bbd{f[\s,\bar{\s}]} must be constant if it is modular invariant and grows no faster than polynomially with \bbd{{\rm Im}[\s]}.

This same theorem automatically applies to harmonic functions of \bbd{q = e\uu{2\pi i \t} = \l(\s)} that are invariant under the anharmonic group and
are bounded by a power of \bbd{{\rm Im}[\t]} at weak coupling.  Any such
function is a modular invariant function of \bbd{\s} that is bounded by a power of \bbd{{\rm Im}[\s]} at weak coupling, and so must necessarily be a constant.

\newpage

\bibliographystyle{JHEP}
\bibliography{references,nospire}

\providecommand{\href}[2]{#2}\begingroup\raggedright\begin{thebibliography}{10}

\bibitem{Alday:2009aq}
L.~F. Alday, D.~Gaiotto and Y.~Tachikawa, \emph{{Liouville Correlation
  Functions from Four-dimensional Gauge Theories}},
  \href{https://doi.org/10.1007/s11005-010-0369-5}{\emph{Lett. Math. Phys.}
  {\bfseries 91} (2010) 167--197},
  [\href{https://arxiv.org/abs/0906.3219}{{\ttfamily 0906.3219}}].

\bibitem{Grassi:2019txd}
A.~Grassi, Z.~Komargodski and L.~Tizzano, \emph{{Extremal Correlators and
  Random Matrix Theory}},  \href{https://arxiv.org/abs/1908.10306}{{\ttfamily
  1908.10306}}.

\bibitem{Hellerman:2017sur}
S.~Hellerman and S.~Maeda, \emph{{On the Large $R$-charge Expansion in
  ${\mathcal N} = 2$ Superconformal Field Theories}},
  \href{https://doi.org/10.1007/JHEP12(2017)135}{\emph{JHEP} {\bfseries 12}
  (2017) 135}, [\href{https://arxiv.org/abs/1710.07336}{{\ttfamily
  1710.07336}}].

\bibitem{Hellerman:2018xpi}
S.~Hellerman, S.~Maeda, D.~Orlando, S.~Reffert and M.~Watanabe,
  \emph{{Universal correlation functions in rank 1 SCFTs}},
  \href{https://doi.org/10.1007/JHEP12(2019)047}{\emph{JHEP} {\bfseries 12}
  (2019) 047}, [\href{https://arxiv.org/abs/1804.01535}{{\ttfamily
  1804.01535}}].

\bibitem{Hellerman:2020sqj}
S.~Hellerman, S.~Maeda, D.~Orlando, S.~Reffert and M.~Watanabe,
  \emph{{S-duality and correlation functions at large R-charge}},
  \href{https://arxiv.org/abs/2005.03021}{{\ttfamily 2005.03021}}.

\bibitem{Berenstein:2003gb}
D.~E. Berenstein, J.~M. Maldacena and H.~S. Nastase, \emph{{Strings in flat
  space and pp waves from N=4 Super Yang Mills}},
  \href{https://doi.org/10.1063/1.1524550}{\emph{AIP Conf. Proc.} {\bfseries
  646} (2002) 3--14}.

\bibitem{Alday:2007qf}
L.~F. Alday, G.~Arutyunov, M.~K. Benna, B.~Eden and I.~R. Klebanov, \emph{{On
  the Strong Coupling Scaling Dimension of High Spin Operators}},
  \href{https://doi.org/10.1088/1126-6708/2007/04/082}{\emph{JHEP} {\bfseries
  04} (2007) 082}, [\href{https://arxiv.org/abs/hep-th/0702028}{{\ttfamily
  hep-th/0702028}}].

\bibitem{Alday:2007mf}
L.~F. Alday and J.~M. Maldacena, \emph{{Comments on operators with large
  spin}}, \href{https://doi.org/10.1088/1126-6708/2007/11/019}{\emph{JHEP}
  {\bfseries 11} (2007) 019},
  [\href{https://arxiv.org/abs/0708.0672}{{\ttfamily 0708.0672}}].

\bibitem{Alday:2013cwa}
L.~F. Alday and A.~Bissi, \emph{{Higher-spin correlators}},
  \href{https://doi.org/10.1007/JHEP10(2013)202}{\emph{JHEP} {\bfseries 10}
  (2013) 202}, [\href{https://arxiv.org/abs/1305.4604}{{\ttfamily 1305.4604}}].

\bibitem{Hellerman:2013kba}
S.~Hellerman and I.~Swanson, \emph{{String Theory of the Regge Intercept}},
  \href{https://doi.org/10.1103/PhysRevLett.114.111601}{\emph{Phys. Rev. Lett.}
  {\bfseries 114} (2015) 111601},
  [\href{https://arxiv.org/abs/1312.0999}{{\ttfamily 1312.0999}}].

\bibitem{Alday:2015eya}
L.~F. Alday, A.~Bissi and T.~Lukowski, \emph{{Large spin systematics in CFT}},
  \href{https://doi.org/10.1007/JHEP11(2015)101}{\emph{JHEP} {\bfseries 11}
  (2015) 101}, [\href{https://arxiv.org/abs/1502.07707}{{\ttfamily
  1502.07707}}].

\bibitem{Fitzpatrick:2012yx}
A.~L. Fitzpatrick, J.~Kaplan, D.~Poland and D.~Simmons-Duffin, \emph{{The
  Analytic Bootstrap and AdS Superhorizon Locality}},
  \href{https://doi.org/10.1007/JHEP12(2013)004}{\emph{JHEP} {\bfseries 12}
  (2013) 004}, [\href{https://arxiv.org/abs/1212.3616}{{\ttfamily 1212.3616}}].

\bibitem{Komargodski:2012ek}
Z.~Komargodski and A.~Zhiboedov, \emph{{Convexity and Liberation at Large
  Spin}}, \href{https://doi.org/10.1007/JHEP11(2013)140}{\emph{JHEP} {\bfseries
  11} (2013) 140}, [\href{https://arxiv.org/abs/1212.4103}{{\ttfamily
  1212.4103}}].

\bibitem{Caron-Huot:2016icg}
S.~Caron-Huot, Z.~Komargodski, A.~Sever and A.~Zhiboedov, \emph{{Strings from
  Massive Higher Spins: The Asymptotic Uniqueness of the Veneziano Amplitude}},
  \href{https://doi.org/10.1007/JHEP10(2017)026}{\emph{JHEP} {\bfseries 10}
  (2017) 026}, [\href{https://arxiv.org/abs/1607.04253}{{\ttfamily
  1607.04253}}].

\bibitem{Son:1995wz}
D.~T. Son, \emph{{Semiclassical approach for multiparticle production in scalar
  theories}}, \href{https://doi.org/10.1016/0550-3213(96)00386-0}{\emph{Nucl.
  Phys. B} {\bfseries 477} (1996) 378--406},
  [\href{https://arxiv.org/abs/hep-ph/9505338}{{\ttfamily hep-ph/9505338}}].

\bibitem{Srednicki}
M.~Srednicki, \emph{Chaos and quantum thermalization},
  \href{https://doi.org/10.1103/PhysRevE.50.888}{\emph{Physical Review E}
  {\bfseries 50} (1994) 888}, [\href{https://arxiv.org/abs/9403051}{{\ttfamily
  9403051}}].

\bibitem{Deutsch}
J.~M. Deutsch, \emph{Quantum statistical mechanics in a closed system},
  {\emph{Physical Review A} {\bfseries 43} (1991) 2046}.

\bibitem{Cardy:1986ie}
J.~L. Cardy, \emph{{Operator Content of Two-Dimensional Conformally Invariant
  Theories}}, \href{https://doi.org/10.1016/0550-3213(86)90552-3}{\emph{Nucl.
  Phys. B} {\bfseries 270} (1986) 186--204}.

\bibitem{Hartman:2014oaa}
T.~Hartman, C.~A. Keller and B.~Stoica, \emph{{Universal Spectrum of 2d
  Conformal Field Theory in the Large c Limit}},
  \href{https://doi.org/10.1007/JHEP09(2014)118}{\emph{JHEP} {\bfseries 09}
  (2014) 118}, [\href{https://arxiv.org/abs/1405.5137}{{\ttfamily 1405.5137}}].

\bibitem{Delacretaz:2020nit}
L.~V. Delacretaz, \emph{{Heavy Operators and Hydrodynamic Tails}},
  \href{https://doi.org/10.21468/SciPostPhys.9.3.034}{\emph{SciPost Phys.}
  {\bfseries 9} (2020) 034},
  [\href{https://arxiv.org/abs/2006.01139}{{\ttfamily 2006.01139}}].

\bibitem{Mukhametzhanov:2020swe}
B.~Mukhametzhanov and S.~Pal, \emph{{Beurling-Selberg Extremization and Modular
  Bootstrap at High Energies}},
  \href{https://doi.org/10.21468/SciPostPhys.8.6.088}{\emph{SciPost Phys.}
  {\bfseries 8} (2020) 088},
  [\href{https://arxiv.org/abs/2003.14316}{{\ttfamily 2003.14316}}].

\bibitem{Bachas:1991fd}
C.~Bachas, \emph{{A Proof of exponential suppression of high-energy transitions
  in the anharmonic oscillator}},
  \href{https://doi.org/10.1016/0550-3213(92)90304-T}{\emph{Nucl. Phys. B}
  {\bfseries 377} (1992) 622--648}.

\bibitem{Libanov:1994ug}
M.~V. Libanov, V.~A. Rubakov, D.~T. Son and S.~V. Troitsky,
  \emph{{Exponentiation of multiparticle amplitudes in scalar theories}},
  \href{https://doi.org/10.1103/PhysRevD.50.7553}{\emph{Phys. Rev. D}
  {\bfseries 50} (1994) 7553--7569},
  [\href{https://arxiv.org/abs/hep-ph/9407381}{{\ttfamily hep-ph/9407381}}].

\bibitem{Jaeckel:2018ipq}
J.~Jaeckel and S.~Schenk, \emph{{Exploring High Multiplicity Amplitudes in
  Quantum Mechanics}},
  \href{https://doi.org/10.1103/PhysRevD.98.096007}{\emph{Phys. Rev. D}
  {\bfseries 98} (2018) 096007},
  [\href{https://arxiv.org/abs/1806.01857}{{\ttfamily 1806.01857}}].

\bibitem{Monin:2018cbi}
A.~Monin, \emph{{Inconsistencies of higgsplosion}},
  \href{https://arxiv.org/abs/1808.05810}{{\ttfamily 1808.05810}}.

\bibitem{Khoze:2018mey}
V.~V. Khoze and J.~Reiness, \emph{{Review of the semiclassical formalism for
  multiparticle production at high energies}},
  \href{https://doi.org/10.1016/j.physrep.2019.06.004}{\emph{Phys. Rept. C}
  {\bfseries 822} (2019) 1--52},
  [\href{https://arxiv.org/abs/1810.01722}{{\ttfamily 1810.01722}}].

\bibitem{Dine:2020ybn}
M.~Dine, H.~H. Patel and J.~F. Ulbricht, \emph{{Behavior of Cross Sections for
  Large Numbers of Particles}},
  \href{https://arxiv.org/abs/2002.12449}{{\ttfamily 2002.12449}}.

\bibitem{Hellerman:2015nra}
S.~Hellerman, D.~Orlando, S.~Reffert and M.~Watanabe, \emph{{On the CFT
  Operator Spectrum at Large Global Charge}},
  \href{https://doi.org/10.1007/JHEP12(2015)071}{\emph{JHEP} {\bfseries 12}
  (2015) 071}, [\href{https://arxiv.org/abs/1505.01537}{{\ttfamily
  1505.01537}}].

\bibitem{Monin:2016jmo}
A.~Monin, D.~Pirtskhalava, R.~Rattazzi and F.~K. Seibold, \emph{{Semiclassics,
  Goldstone Bosons and CFT data}},
  \href{https://doi.org/10.1007/JHEP06(2017)011}{\emph{JHEP} {\bfseries 06}
  (2017) 011}, [\href{https://arxiv.org/abs/1611.02912}{{\ttfamily
  1611.02912}}].

\bibitem{Cuomo:2017vzg}
G.~Cuomo, A.~de~la Fuente, A.~Monin, D.~Pirtskhalava and R.~Rattazzi,
  \emph{{Rotating superfluids and spinning charged operators in conformal field
  theory}}, \href{https://doi.org/10.1103/PhysRevD.97.045012}{\emph{Phys. Rev.
  D} {\bfseries 97} (2018) 045012},
  [\href{https://arxiv.org/abs/1711.02108}{{\ttfamily 1711.02108}}].

\bibitem{Sharon:2020mjs}
A.~Sharon and M.~Watanabe, \emph{{Transition of Large $R$-Charge Operators on a
  Conformal Manifold}},
  \href{https://doi.org/10.1007/JHEP01(2021)068}{\emph{JHEP} {\bfseries 01}
  (2021) 068}, [\href{https://arxiv.org/abs/2008.01106}{{\ttfamily
  2008.01106}}].

\bibitem{Gaume:2020bmp}
L.~A. Gaum\'e, D.~Orlando and S.~Reffert, \emph{{Selected Topics in the Large
  Quantum Number Expansion}},
  \href{https://arxiv.org/abs/2008.03308}{{\ttfamily 2008.03308}}.

\bibitem{Cuomo:2020rgt}
G.~Cuomo, \emph{{A note on the large charge expansion in 4d CFT}},
  \href{https://doi.org/10.1016/j.physletb.2020.136014}{\emph{Phys. Lett. B}
  {\bfseries 812} (2021) 136014},
  [\href{https://arxiv.org/abs/2010.00407}{{\ttfamily 2010.00407}}].

\bibitem{Orlando:2019hte}
D.~Orlando, S.~Reffert and F.~Sannino, \emph{{A safe CFT at large charge}},
  \href{https://doi.org/10.1007/JHEP08(2019)164}{\emph{JHEP} {\bfseries 08}
  (2019) 164}, [\href{https://arxiv.org/abs/1905.00026}{{\ttfamily
  1905.00026}}].

\bibitem{Cuomo:2019ejv}
G.~Cuomo, \emph{{Superfluids, vortices and spinning charged operators in 4d
  CFT}}, \href{https://doi.org/10.1007/JHEP02(2020)119}{\emph{JHEP} {\bfseries
  02} (2020) 119}, [\href{https://arxiv.org/abs/1906.07283}{{\ttfamily
  1906.07283}}].

\bibitem{Orlando:2020yii}
D.~Orlando, S.~Reffert and F.~Sannino, \emph{{Charging the Conformal Window}},
  \href{https://arxiv.org/abs/2003.08396}{{\ttfamily 2003.08396}}.

\bibitem{Alvarez-Gaume:2016vff}
L.~Alvarez-Gaume, O.~Loukas, D.~Orlando and S.~Reffert, \emph{{Compensating
  strong coupling with large charge}},
  \href{https://doi.org/10.1007/JHEP04(2017)059}{\emph{JHEP} {\bfseries 04}
  (2017) 059}, [\href{https://arxiv.org/abs/1610.04495}{{\ttfamily
  1610.04495}}].

\bibitem{Hellerman:2017veg}
S.~Hellerman, S.~Maeda and M.~Watanabe, \emph{{Operator Dimensions from
  Moduli}}, \href{https://doi.org/10.1007/JHEP10(2017)089}{\emph{JHEP}
  {\bfseries 10} (2017) 089},
  [\href{https://arxiv.org/abs/1706.05743}{{\ttfamily 1706.05743}}].

\bibitem{Hellerman:2017efx}
S.~Hellerman, N.~Kobayashi, S.~Maeda and M.~Watanabe, \emph{{A Note on
  Inhomogeneous Ground States at Large Global Charge}},
  \href{https://doi.org/10.1007/JHEP10(2019)038}{\emph{JHEP} {\bfseries 10}
  (2019) 038}, [\href{https://arxiv.org/abs/1705.05825}{{\ttfamily
  1705.05825}}].

\bibitem{Hellerman:2018sjf}
S.~Hellerman, N.~Kobayashi, S.~Maeda and M.~Watanabe, \emph{{Observables in
  Inhomogeneous Ground States at Large Global Charge}},
  \href{https://arxiv.org/abs/1804.06495}{{\ttfamily 1804.06495}}.

\bibitem{Jafferis:2017zna}
D.~Jafferis, B.~Mukhametzhanov and A.~Zhiboedov, \emph{{Conformal Bootstrap At
  Large Charge}}, \href{https://doi.org/10.1007/JHEP05(2018)043}{\emph{JHEP}
  {\bfseries 05} (2018) 043},
  [\href{https://arxiv.org/abs/1710.11161}{{\ttfamily 1710.11161}}].

\bibitem{Alvarez-Gaume:2019biu}
L.~Alvarez-Gaume, D.~Orlando and S.~Reffert, \emph{{Large charge at large N}},
  \href{https://doi.org/10.1007/JHEP12(2019)142}{\emph{JHEP} {\bfseries 12}
  (2019) 142}, [\href{https://arxiv.org/abs/1909.02571}{{\ttfamily
  1909.02571}}].

\bibitem{Kumar:2018nkf}
S.~P. Kumar, D.~Roychowdhury and S.~Stratiev, \emph{{Roton-phonon excitations
  in Chern-Simons matter theory at finite density}},
  \href{https://doi.org/10.1007/JHEP12(2018)116}{\emph{JHEP} {\bfseries 12}
  (2018) 116}, [\href{https://arxiv.org/abs/1806.06976}{{\ttfamily
  1806.06976}}].

\bibitem{Watanabe:2019adh}
M.~Watanabe, \emph{{Chern-Simons-Matter Theories at Large Global Charge}},
  \href{https://arxiv.org/abs/1904.09815}{{\ttfamily 1904.09815}}.

\bibitem{Arias-Tamargo:2019xld}
G.~Arias-Tamargo, D.~Rodriguez-Gomez and J.~G. Russo, \emph{{The large charge
  limit of scalar field theories and the Wilson-Fisher fixed point at
  $\epsilon=0$}}, \href{https://doi.org/10.1007/JHEP10(2019)201}{\emph{JHEP}
  {\bfseries 10} (2019) 201},
  [\href{https://arxiv.org/abs/1908.11347}{{\ttfamily 1908.11347}}].

\bibitem{Arias-Tamargo:2019kfr}
G.~Arias-Tamargo, D.~Rodriguez-Gomez and J.~G. Russo, \emph{{Correlation
  functions in scalar field theory at large charge}},
  \href{https://doi.org/10.1007/JHEP01(2020)171}{\emph{JHEP} {\bfseries 01}
  (2020) 171}, [\href{https://arxiv.org/abs/1912.01623}{{\ttfamily
  1912.01623}}].

\bibitem{Arias-Tamargo:2020fow}
G.~Arias-Tamargo, D.~Rodriguez-Gomez and J.~G. Russo, \emph{{On the UV
  completion of the $O(N)$ model in $6-\epsilon$ dimensions: a stable
  large-charge sector}},
  \href{https://doi.org/10.1007/JHEP09(2020)064}{\emph{JHEP} {\bfseries 09}
  (2020) 064}, [\href{https://arxiv.org/abs/2003.13772}{{\ttfamily
  2003.13772}}].

\bibitem{Badel:2019oxl}
G.~Badel, G.~Cuomo, A.~Monin and R.~Rattazzi, \emph{{The Epsilon Expansion
  Meets Semiclassics}},
  \href{https://doi.org/10.1007/JHEP11(2019)110}{\emph{JHEP} {\bfseries 11}
  (2019) 110}, [\href{https://arxiv.org/abs/1909.01269}{{\ttfamily
  1909.01269}}].

\bibitem{Badel:2019khk}
G.~Badel, G.~Cuomo, A.~Monin and R.~Rattazzi, \emph{{Feynman diagrams and the
  large charge expansion in $3-\varepsilon$ dimensions}},
  \href{https://doi.org/10.1016/j.physletb.2020.135202}{\emph{Phys. Lett. B}
  {\bfseries 802} (2020) 135202},
  [\href{https://arxiv.org/abs/1911.08505}{{\ttfamily 1911.08505}}].

\bibitem{Giombi:2020enj}
S.~Giombi and J.~Hyman, \emph{{On the Large Charge Sector in the Critical
  $O(N)$ Model at Large $N$}},
  \href{https://arxiv.org/abs/2011.11622}{{\ttfamily 2011.11622}}.

\bibitem{Antipin:2021akb}
O.~Antipin, J.~Bersini, F.~Sannino, Z.-W. Wang and C.~Zhang, \emph{{Untangling
  scaling dimensions of fixed charge operators in Higgs Theories}},
  \href{https://arxiv.org/abs/2102.04390}{{\ttfamily 2102.04390}}.

\bibitem{Cuomo:2021qws}
G.~Cuomo, L.~V. Delacretaz and U.~Mehta, \emph{{Large Charge Sector of 3d
  Parity-Violating CFTs}},  \href{https://arxiv.org/abs/2102.05046}{{\ttfamily
  2102.05046}}.

\bibitem{Komargodski:2021zzy}
Z.~Komargodski, M.~Mezei, S.~Pal and A.~Raviv-Moshe, \emph{{Spontaneously
  Broken Boosts in CFTs}},  \href{https://arxiv.org/abs/2102.12583}{{\ttfamily
  2102.12583}}.

\bibitem{Cuomo:2021ygt}
G.~Cuomo, \emph{{The OPE meets semiclassics}},
  \href{https://arxiv.org/abs/2103.01331}{{\ttfamily 2103.01331}}.

\bibitem{Kravec:2018qnu}
S.~M. Kravec and S.~Pal, \emph{{Nonrelativistic Conformal Field Theories in the
  Large Charge Sector}},
  \href{https://doi.org/10.1007/JHEP02(2019)008}{\emph{JHEP} {\bfseries 02}
  (2019) 008}, [\href{https://arxiv.org/abs/1809.08188}{{\ttfamily
  1809.08188}}].

\bibitem{Favrod:2018xov}
S.~Favrod, D.~Orlando and S.~Reffert, \emph{{The large-charge expansion for
  Schr\"odinger systems}},
  \href{https://doi.org/10.1007/JHEP12(2018)052}{\emph{JHEP} {\bfseries 12}
  (2018) 052}, [\href{https://arxiv.org/abs/1809.06371}{{\ttfamily
  1809.06371}}].

\bibitem{Kravec:2019djc}
S.~M. Kravec and S.~Pal, \emph{{The Spinful Large Charge Sector of
  Non-Relativistic CFTs: From Phonons to Vortex Crystals}},
  \href{https://doi.org/10.1007/JHEP05(2019)194}{\emph{JHEP} {\bfseries 05}
  (2019) 194}, [\href{https://arxiv.org/abs/1904.05462}{{\ttfamily
  1904.05462}}].

\bibitem{Orlando:2020idm}
D.~Orlando, V.~Pellizzani and S.~Reffert, \emph{{Near-Schr\"odinger dynamics at
  large charge}},  \href{https://arxiv.org/abs/2010.07942}{{\ttfamily
  2010.07942}}.

\bibitem{Hellerman:2020eff}
S.~Hellerman and I.~Swanson, \emph{{Droplet-Edge Operators in Nonrelativistic
  Conformal Field Theories}},
  \href{https://arxiv.org/abs/2010.07967}{{\ttfamily 2010.07967}}.

\bibitem{Jack:2021ypd}
I.~Jack and D.~R.~T. Jones, \emph{{Anomalous dimensions at large charge in d=4
  O(N) theory}},  \href{https://arxiv.org/abs/2101.09820}{{\ttfamily
  2101.09820}}.

\bibitem{Papadodimas:2009eu}
K.~Papadodimas, \emph{{Topological Anti-Topological Fusion in Four-Dimensional
  Superconformal Field Theories}},
  \href{https://doi.org/10.1007/JHEP08(2010)118}{\emph{JHEP} {\bfseries 08}
  (2010) 118}, [\href{https://arxiv.org/abs/0910.4963}{{\ttfamily 0910.4963}}].

\bibitem{Baggio:2014sna}
M.~Baggio, V.~Niarchos and K.~Papadodimas, \emph{{Exact correlation functions
  in $SU(2) \mathcal N=2$ superconformal QCD}},
  \href{https://doi.org/10.1103/PhysRevLett.113.251601}{\emph{Phys. Rev. Lett.}
  {\bfseries 113} (2014) 251601},
  [\href{https://arxiv.org/abs/1409.4217}{{\ttfamily 1409.4217}}].

\bibitem{Baggio:2014ioa}
M.~Baggio, V.~Niarchos and K.~Papadodimas, \emph{{tt$^{*}$ equations,
  localization and exact chiral rings in 4d $ \mathcal{N} $ =2 SCFTs}},
  \href{https://doi.org/10.1007/JHEP02(2015)122}{\emph{JHEP} {\bfseries 02}
  (2015) 122}, [\href{https://arxiv.org/abs/1409.4212}{{\ttfamily 1409.4212}}].

\bibitem{Argyres:1995jj}
P.~C. Argyres and M.~R. Douglas, \emph{{New phenomena in SU(3) supersymmetric
  gauge theory}},
  \href{https://doi.org/10.1016/0550-3213(95)00281-V}{\emph{Nucl. Phys. B}
  {\bfseries 448} (1995) 93--126},
  [\href{https://arxiv.org/abs/hep-th/9505062}{{\ttfamily hep-th/9505062}}].

\bibitem{Argyres:1995xn}
P.~C. Argyres, M.~R. Plesser, N.~Seiberg and E.~Witten, \emph{{New N=2
  superconformal field theories in four-dimensions}},
  \href{https://doi.org/10.1016/0550-3213(95)00671-0}{\emph{Nucl. Phys. B}
  {\bfseries 461} (1996) 71--84},
  [\href{https://arxiv.org/abs/hep-th/9511154}{{\ttfamily hep-th/9511154}}].

\bibitem{Xie:2012hs}
D.~Xie, \emph{{General Argyres-Douglas Theory}},
  \href{https://doi.org/10.1007/JHEP01(2013)100}{\emph{JHEP} {\bfseries 01}
  (2013) 100}, [\href{https://arxiv.org/abs/1204.2270}{{\ttfamily 1204.2270}}].

\bibitem{Anselmi:1997ys}
D.~Anselmi, J.~Erlich, D.~Z. Freedman and A.~A. Johansen, \emph{{Positivity
  constraints on anomalies in supersymmetric gauge theories}},
  \href{https://doi.org/10.1103/PhysRevD.57.7570}{\emph{Phys. Rev. D}
  {\bfseries 57} (1998) 7570--7588},
  [\href{https://arxiv.org/abs/hep-th/9711035}{{\ttfamily hep-th/9711035}}].

\bibitem{Nekrasov:2002qd}
N.~A. Nekrasov, \emph{{Seiberg-Witten prepotential from instanton counting}},
  \href{https://doi.org/10.4310/ATMP.2003.v7.n5.a4}{\emph{Adv. Theor. Math.
  Phys.} {\bfseries 7} (2003) 831--864},
  [\href{https://arxiv.org/abs/hep-th/0206161}{{\ttfamily hep-th/0206161}}].

\bibitem{Pestun:2007rz}
V.~Pestun, \emph{{Localization of gauge theory on a four-sphere and
  supersymmetric Wilson loops}},
  \href{https://doi.org/10.1007/s00220-012-1485-0}{\emph{Commun. Math. Phys.}
  {\bfseries 313} (2012) 71--129},
  [\href{https://arxiv.org/abs/0712.2824}{{\ttfamily 0712.2824}}].

\bibitem{Gerchkovitz:2016gxx}
E.~Gerchkovitz, J.~Gomis, N.~Ishtiaque, A.~Karasik, Z.~Komargodski and S.~S.
  Pufu, \emph{{Correlation Functions of Coulomb Branch Operators}},
  \href{https://doi.org/10.1007/JHEP01(2017)103}{\emph{JHEP} {\bfseries 01}
  (2017) 103}, [\href{https://arxiv.org/abs/1602.05971}{{\ttfamily
  1602.05971}}].

\bibitem{Gaiotto:2009we}
D.~Gaiotto, \emph{{N=2 dualities}},
  \href{https://doi.org/10.1007/JHEP08(2012)034}{\emph{JHEP} {\bfseries 08}
  (2012) 034}, [\href{https://arxiv.org/abs/0904.2715}{{\ttfamily 0904.2715}}].

\bibitem{Grimm:2007tm}
T.~W. Grimm, A.~Klemm, M.~Marino and M.~Weiss, \emph{{Direct Integration of the
  Topological String}},
  \href{https://doi.org/10.1088/1126-6708/2007/08/058}{\emph{JHEP} {\bfseries
  08} (2007) 058}, [\href{https://arxiv.org/abs/hep-th/0702187}{{\ttfamily
  hep-th/0702187}}].

\bibitem{Dorey:1996bn}
N.~Dorey, V.~V. Khoze and M.~P. Mattis, \emph{{On N=2 supersymmetric QCD with
  four flavors}},
  \href{https://doi.org/10.1016/S0550-3213(97)00132-6}{\emph{Nucl. Phys. B}
  {\bfseries 492} (1997) 607--622},
  [\href{https://arxiv.org/abs/hep-th/9611016}{{\ttfamily hep-th/9611016}}].

\bibitem{Billo:2013fi}
M.~Billo, M.~Frau, L.~Gallot, A.~Lerda and I.~Pesando, \emph{{Deformed N=2
  theories, generalized recursion relations and S-duality}},
  \href{https://doi.org/10.1007/JHEP04(2013)039}{\emph{JHEP} {\bfseries 04}
  (2013) 039}, [\href{https://arxiv.org/abs/1302.0686}{{\ttfamily 1302.0686}}].

\bibitem{Bourget:2018obm}
A.~Bourget, D.~Rodriguez-Gomez and J.~G. Russo, \emph{{A limit for large
  $R$-charge correlators in $\mathcal{N}=2$ theories}},
  \href{https://doi.org/10.1007/JHEP05(2018)074}{\emph{JHEP} {\bfseries 05}
  (2018) 074}, [\href{https://arxiv.org/abs/1803.00580}{{\ttfamily
  1803.00580}}].

\bibitem{Seiberg:1990eb}
N.~Seiberg, \emph{{Notes on quantum Liouville theory and quantum gravity}},
  \href{https://doi.org/10.1143/PTPS.102.319}{\emph{Prog. Theor. Phys. Suppl.}
  {\bfseries 102} (1990) 319--349}.

\bibitem{Gerchkovitz:2014gta}
E.~Gerchkovitz, J.~Gomis and Z.~Komargodski, \emph{{Sphere Partition Functions
  and the Zamolodchikov Metric}},
  \href{https://doi.org/10.1007/JHEP11(2014)001}{\emph{JHEP} {\bfseries 11}
  (2014) 001}, [\href{https://arxiv.org/abs/1405.7271}{{\ttfamily 1405.7271}}].

\bibitem{Gomis:2014woa}
J.~Gomis and N.~Ishtiaque, \emph{{K\"ahler potential and ambiguities in 4d $
  \mathcal{N} $ = 2 SCFTs}},
  \href{https://doi.org/10.1007/JHEP04(2015)169}{\emph{JHEP} {\bfseries 04}
  (2015) 169}, [\href{https://arxiv.org/abs/1409.5325}{{\ttfamily 1409.5325}}].

\bibitem{LeFloch:2020uop}
B.~Le~Floch, \emph{{A slow review of the AGT correspondence}},
  \href{https://arxiv.org/abs/2006.14025}{{\ttfamily 2006.14025}}.

\bibitem{Okuda:2010ke}
T.~Okuda and V.~Pestun, \emph{{On the instantons and the hypermultiplet mass of
  N=2* super Yang-Mills on $S^{4}$}},
  \href{https://doi.org/10.1007/JHEP03(2012)017}{\emph{JHEP} {\bfseries 03}
  (2012) 017}, [\href{https://arxiv.org/abs/1004.1222}{{\ttfamily 1004.1222}}].

\bibitem{Poland:2018epd}
D.~Poland, S.~Rychkov and A.~Vichi, \emph{{The Conformal Bootstrap: Theory,
  Numerical Techniques, and Applications}},
  \href{https://doi.org/10.1103/RevModPhys.91.015002}{\emph{Rev. Mod. Phys.}
  {\bfseries 91} (2019) 015002},
  [\href{https://arxiv.org/abs/1805.04405}{{\ttfamily 1805.04405}}].

\bibitem{Seiberg:1999xz}
N.~Seiberg and E.~Witten, \emph{{The D1 / D5 system and singular CFT}},
  \href{https://doi.org/10.1088/1126-6708/1999/04/017}{\emph{JHEP} {\bfseries
  04} (1999) 017}, [\href{https://arxiv.org/abs/hep-th/9903224}{{\ttfamily
  hep-th/9903224}}].

\bibitem{Watanabe:2019pdh}
M.~Watanabe, \emph{{Accessing Large Global Charge via the
  $\epsilon$-Expansion}},  \href{https://arxiv.org/abs/1909.01337}{{\ttfamily
  1909.01337}}.

\bibitem{Dondi:2021buw}
N.~Dondi, I.~Kalogerakis, D.~Orlando and S.~Reffert, \emph{{Resurgence of the
  large-charge expansion}},  \href{https://arxiv.org/abs/2102.12488}{{\ttfamily
  2102.12488}}.

\bibitem{richter2000semiclassical}
K.~Richter and K.~Richter, \emph{Semiclassical theory of mesoscopic quantum
  systems}, vol.~11.
\newblock Springer Berlin, 2000.

\bibitem{HellermanExponentialCorrectionsToAppear}
S.~Hellerman, \emph{{On The Exponentially Small Corrections to \bbd{{\cal N} =
  2} Superconformal Correlators At Large R-Charge}}, {\emph{to appear} }.

\bibitem{Tachikawa:2013kta}
Y.~Tachikawa, \emph{{N=2 supersymmetric dynamics for pedestrians}}, vol.~890.
\newblock 2014,
  \href{https://doi.org/10.1007/978-3-319-08822-8}{10.1007/978-3-319-08822-8}.

\bibitem{Seiberg:1994rs}
N.~Seiberg and E.~Witten, \emph{{Electric - magnetic duality, monopole
  condensation, and confinement in N=2 supersymmetric Yang-Mills theory}},
  \href{https://doi.org/10.1016/0550-3213(94)90124-4}{\emph{Nucl. Phys. B}
  {\bfseries 426} (1994) 19--52},
  [\href{https://arxiv.org/abs/hep-th/9407087}{{\ttfamily hep-th/9407087}}].

\bibitem{Seiberg:1994aj}
N.~Seiberg and E.~Witten, \emph{{Monopoles, duality and chiral symmetry
  breaking in N=2 supersymmetric QCD}},
  \href{https://doi.org/10.1016/0550-3213(94)90214-3}{\emph{Nucl. Phys. B}
  {\bfseries 431} (1994) 484--550},
  [\href{https://arxiv.org/abs/hep-th/9408099}{{\ttfamily hep-th/9408099}}].

\end{thebibliography}\endgroup
\end{document}